\documentclass[11pt]{article}
\usepackage[a4paper,margin=1in]{geometry}

\usepackage{fullpage}
\usepackage{amsmath,amssymb,amsthm,mathtools}
\usepackage{microtype}
\usepackage{cite}
\usepackage[hidelinks]{hyperref}
\usepackage{xcolor}
\usepackage{booktabs,tabularx,array,float}
\usepackage[table]{xcolor}

\usepackage[dvipsnames]{xcolor}

\newcommand{\U}{\mathsf{U}}
\newcommand{\hyb}{\mathcal{H}}
\newcommand{\chyb}[1]{\Phi_{\hyb_{#1}}}
\newcommand{\pf}{\mathsf{pf}}
\newcommand{\Epq}{E_{p,q}}
\newcommand{\Kpq}{K_{p,q}^{\mathrm{PF}}}
\newcommand{\qcK}{\overline{\Phi}_K}
\newcommand{\qcE}{\overline{\Phi}_E}
\newcommand{\pibij}{\Pi^{\mathrm{bij}}}
\newcommand{\cD}{\mathcal{D}}
\newcommand{\nEspace}[1]{\mathcal K_{#1}^{\mathrm{nE}}}
\newcommand{\abs}[1]{\lvert#1\rvert}
\newcommand{\norm}[1]{\lVert#1\rVert}
\newcommand{\tr}{\operatorname{Tr}}
\newcommand{\ket}[1]{\lvert#1\rangle}

\newcommand{\Splus}{S_{+}}
\newcommand{\Sminus}{S_{-}}

\theoremstyle{definition}
\newtheorem{definition}{Definition}[section]
\newtheorem{lemma}[definition]{Lemma}

\newtheorem{theorem}[definition]{Theorem}
\newcounter{algorithm}

\title{Birthday-Scale Non-Adaptive Security for Strong Pseudorandom Unitaries}
\author{%
  Zvika Brakerski, Lior Ozer, and Shahar Silberstein\\[0.4em]
  \small Weizmann Institute of Science\\
  \small\texttt{\{zvika.brakerski, lior.ozer, shahar.silberstein\}@weizmann.ac.il}
}
\date{}

\begin{document}

\maketitle

\begin{abstract}
Pseudorandom unitaries (PRUs) are efficiently implementable unitary ensembles that are computationally indistinguishable from Haar-random unitaries, providing efficient substitutes for Haar-randomness in quantum information and cryptography. Strong PRUs require this indistinguishability to hold even when the adversary is given access to all of $U$, $U^\dagger$, $U^*$, and $U^T$. 

In this work, we prove birthday-scale non-adaptive security for the DPFC (Clifford-Phase-Permutation-Clifford) strong PRU candidate. For an $N$-dimensional unitary, allowing arbitrary entangled inputs and coherent superpositions of query types, we show that DPFC can withstand $t \approx \sqrt{N}$ queries. This matches the best known query complexity even for PRUs that are not strong, compared with $t \approx N^\alpha$ for $\alpha < 1/2$ in previous strong-PRU analyses. 

Our proof utilizes two techniques from previous works: the EPR-sector decomposition of Schuster et al. (arXiv 2025), previously used for unitary designs but not for PRUs, and the path-recording framework of Ma and Huang (STOC 2025), previously used for adaptive security analysis.

We complement the non-adaptive security result with an upper bound in the adaptive setting, showing that DPFC cannot remain secure for $t \gg \sqrt{N}$ queries.

\end{abstract}

\newpage

\tableofcontents
\clearpage

\section{Introduction}
\label{sec:introduction}

\newcommand{\poly}{\mathrm{poly}}
\newcommand{\polylog}{\mathrm{polylog}}

Pseudorandom Unitaries (PRU) \cite{JLS18} is an ensemble of efficiently sampleable and computable unitaries that is nonetheless computationally indistinguishable from Haar random unitaries. More precisely, an efficient adversary that is given query access to either a PRU or a Haar random unitary, will only have a negligible advantage in distinguishing the two cases. Haar random unitaries are considered as a good proxy for various physical phenomena such as thermalization and black-hole information mixing \cite{YK17} and therefore the ability to model their behavior in a computationally efficient manner is important if we wish to simulate such systems. PRUs also found applications in quantum cryptography \cite{GJMZ23, PRV26, HBMKB25}.

The first constructions of PRUs only appeared in the literature recently \cite{MPSY24, MH24}, and, as of yet, several additional candidate constructions with various properties have been introduced \cite{SMLBH25, LQSYZ26, GKT26}. A common theme that is shared between \emph{all} candidates with claimed cryptographic security is the reliance on the notion of ROM-PRU, explicitly defined later in ~\cite{BY26}. Essentially, a construction is presented in the \emph{random oracle model} (i.e.\ using a random classical function as an oracle), where a distinguishing bound was shown for all bounded-query algorithms, even using computationally unbounded resources. Then, the random functions are replaced with post-quantum secure pseudorandom functions to obtain cryptographic security. Since the second step is standard, our works (and all other works in the area) focus on the task of building a useful ROM-PRU, namely one where even when the number of distinguisher queries $t$ is super-polynomial, the distinguishing gap remains negligible.

Indeed, in all known cases, there is a tradeoff between $t$ and the distinguishing gap. For example, in the ``base case'' of \cite{MH24}, the distinguishing gap is roughly $t^2/N$, where $N$ is the dimension of the Hilbert space on which the unitary is defined (i.e.\ $N=2^n$ where $n$ is the number of qubits).
We refer to this bound of security as the \emph{birthday bound}, since it allows obtaining a meaningful result for values of $t$ that approach $\sqrt{N}$.\footnote{A meticulous reader would notice that we go from claiming $t^2/N$ security to arguing that the scheme is secure for $t \approx \sqrt{N}$ queries. This is of course not immediate since setting $t$ to such a value would only guarantee constant security rather than negligible. We purposely neglect this distinction, partly in order to simplify exposition, and partly because security can be amplified by repetition \cite{MPSY24}, so constant security can be amplified to any desired security level by at the cost of complicating the construction. In the same vain, we view the difference between $t^2/N$ and $t/\sqrt{N}$ security as immaterial for the purposes of this paper.} As we discuss below, the birthday scale arises naturally in analyses based on collisions among $t$ random elements of $[N]$, and it is the best security scale currently established for efficiently implementable PRU constructions. 

Given a construction as above, we can simultaneously set $t$ to be superpolynomial in $n$ and the distinguishing gap to be negligible in $n$, which provides adequate security, so long as $n$ is polynomially related to the security parameter. There are cases, though, where the specific relation between $t$ and the distinguishing gap is more important. In particular, one might not want to be restricted to $t = \poly(n)=\polylog(N)$, and consider a setting where even for $t = \poly(N)$ (i.e.\ $t$ is exponential in $n$), we have a security guarantee. The most extreme version of this requirement is known as \emph{scalable security} \cite{BS20,BY26}. 

In a parallel and independent work, Huang, Ramkumar, and Wright~\cite{HRW26} show how to go beyond the birthday scale and obtain security against $t \approx N$ queries to the unitary, using a polynomial number of queries to the underlying random oracle, but using inefficiently implementable unitaries in-between. They also show that such $t$ is essentially optimal, namely that there is no random oracle construction making polynomially many queries that achieves $t > N^{1+\gamma}$ for any $\gamma$. Nevertheless, as we pointed out, the birthday scale still stands as the best known security scale for efficiently implementable constructions.

So far, we only considered a setting where a distinguisher gets query access to run $U$ itself. However, there are settings where the security requirements are more stringent in terms of the type of queries that the distinguisher can make. Specifically, one may wish to consider adversaries that can apply the sampled unitary $U$ along with its inverse $U^\dagger$, complex conjugate $U^*$, and transpose $U^T$. PRUs that remain secure even when all such queries $U, U^\dagger, U^*, U^T$ are allowed are known as \emph{strong} PRUs \cite{MH24,SMLBH25}.\footnote{The notion presented in \cite{MH24} only allowed $U$ and $U^\dagger$.} We may (and indeed do in this work) even consider settings where the distinguisher is allowed to make a query in coherent superposition of the four types. The importance of strong PRU (of the type described here) was established in the literature \cite{Z25,TWZ25} with applications to the security of cryptographic objects such as commitment schemes, and the hardness of Hayden-Preskill decoding~\cite{YK17}.

Concretely, our work builds on the seminal ``PFC-style'' PRU candidates first introduced in \cite{MPSY24}. They considered a family of unitaries that consists of a random $n$-qubit Clifford circuit (C), followed by a random $n$-qubit phase gate (F),\footnote{In the original PFC construction, the phase is binary, but we may and indeed require higher-order phases. Specifically, strong PRUs rely on ternary phase.} and finally a random permutation over $\{0,1\}^n$ (P). This construction was then adopted by \cite{MH24}. As mentioned above, this construction is shown to be secure all the way up to the birthday bound, and the analysis (in both \cite{MPSY24,MH24}) relies on the probability of collisions in a $t$-tuple being sufficiently low. At an extremely high level, the Clifford $C$ ``scrambles'' the input, and the $PF$ part is shown to behave like a PRU for a random input with no collisions. It was shown in \cite{BY26} that an attack on PFC with $t = O(\sqrt{N})$ queries exists, even using non-adaptive queries, which rules out the possibility of an improved analysis that would significantly change this status.

For strong PRU, the PFC construction is slightly modified to ``DPFC''. An additional Clifford (D) is applied after the permutation. This modification is essential in order to support even just $U^\dagger$ queries. This setting had been analyzed for $U^\dagger$ queries in \cite{MH24} and for general strong queries in \cite{SMLBH25} but, perhaps surprisingly, these analyses were unable to match the birthday bound and were only shown to be secure up to $t \approx N^\alpha$ for some small constant $\alpha$ which varies, depending on the variant of the construction and proof, between $1/16$ and $1/32$.\footnote{It may be possible to tweak the analysis to slightly improve $\alpha$, but approaching $1/2$ does not seem to follow naively.} Furthermore, the PFC upper bound from \cite{BY26} does not seem to hold for DPFC, so one could possibly hope for improved performance even beyond the birthday bound.

In this work, we provide both lower and upper bounds for the behavior of DPFC under strong queries. On one hand, we show a \emph{non-adaptive security proof at the birthday bound}. Showing that DPFC remains secure up to $t \approx \sqrt{N}$ strong queries, in the non-adaptive model. That is, if all queries are provided ahead of time. To complement this result, we present a $t = O(\sqrt{N})$ attack on DPFC in the strong query model (in particular $U, U^\dagger$ are needed). This attack, however, is \emph{adaptive}.

We note that the previous $N^\alpha$ results were proven in an adaptive model, so our results come close but do not completely close the gap between the positive and negative analyses of DPFC. However, we do show that the birthday bound is the most we can aspire to under DPFC (without modifications), and it is actually achievable at least for non-adaptive queries.

\subsection{Our Results}
\label{sec:informal-our-results}

As we outlined above, the DPFC ensemble on $n$ qubits is the product of $4$ unitary random variables $C, F, P, D$. The \emph{endpoints} $C,D$ are random $n$-qubit Clifford unitaries. The permutation operator $P$ is a random permutation over the $N=2^n$ elements of the computational basis. The phase function $F$ is a diagonal unitary, whose entries (in our case) are random ternary roots of unity. Applying them in order on an $n$-qubit input implements the unitary $U=DPFC$ which is the subject of investigation in this work.

We notice like all previous analyses, we do not require the full Clifford property of $C,D$. Instead, we only require that they are independently sampled from distributions known as $2$-unitary designs ($2$-designs for short). These distributions behave identically to Haar random, if only $2$ queries are made. This property is satisfied by random Clifford unitaries \cite{DCEL09}, which are also efficiently implementable. Our analysis is in an idealized model where $P, F$ are sampled using a random permutation and function (which can be replaced by cryptographic objects as explained above).

In terms of adversarial model, our adversary is computationally unbounded and is tasked with distinguishing DPFC from Haar random. It does so by producing $t$ nonadaptive queries, where in order to achieve birthday scale security we require that if $t \ll \sqrt{N}$, the distinguishing advantage is small (see below for a more formal expression). More explicitly, the adversary prepares $t$ input registers, each of dimension $N$, it also prepares a ``routing register'' which indicates, for each $i \in [t]$ whether the $i$th query is a query for \(U, U^{T},U^{*}\) or \(U^{\dagger}\). In addition, the adversary may have its own state register. All of these registers may be arbitrarily entangled. After the register is prepared, the unitaries \(U, U^{T},U^{*},U^{\dagger}\) are properly applied, and the output is processed by the adversary to decide whether $U$ was drawn from DPFC or Haar.

We can now state our claim of birthday-scale nonadaptive security for DPFC against all four query types, including a coherent superposition of query patterns.

\begin{theorem}[DPFC nonadaptive strong security, informal, see Theorem~\ref{thm:main-four-query-dpfc-security}]
\label{thm:informal-dpfc-non-adaptive-security}
    A non-adaptive adversary, making \(t\) parallel queries to any coherent superposition of query patterns to \(U, U^{T}, U^{*}, U^{\dagger}\), has a distinguishing advantage of at most \(O(t/\sqrt{N})\) between a DPFC unitary and a Haar-random unitary.
\end{theorem}

At the heart of our proof is a technique that uses path recording oracles \cite{MH24}, which have so far been used for \emph{adaptive} security analysis, as an intermediate hybrid for comparing the two ensembles. We show that in the non-adaptive setting this technique provides additional power when combined with the \emph{EPR-sector decomposition} of \cite{SMLBH25}, which is a tool that has up to now been used to study unitary designs. 
This enables us to separate out invariant EPR sectors that complicate the analysis when the adversary can use all four query types, while preserving coherence between sectors, allowing us to recover the birthday-scale bound.

To complement our results, we describe an algorithm that makes $O(\sqrt{N})$ \emph{adaptive} queries to \(U, U^{\dagger}\) and reaches a constant distinguishing advantage between the two ensembles, when the chosen endpoints for DPFC are Cliffords. In the adaptive setting, the adversary can make queries to \(U, U^{\dagger}\) that depend on the outputs of previous queries.

\begin{theorem}[Adaptive attack on DPFC, informal, see Theorem~\ref{thm:cba-main}]
    \label{thm:informal-addaptive-attack-on-DPFC}
    For sufficiently large \(N = 2^n\), there exists an adaptive adversary that can distinguish the DPFC ensemble of size \(N\) with \(n\)-qubit Clifford endpoints from Haar-random unitaries with constant distinguishing advantage using \(O(\sqrt{N})\) queries to the sampled unitary \(U\) and its inverse, \(U^\dagger\).
\end{theorem}

In terms of techniques, our algorithm deviates significantly from the forward-only PFC attack of~\cite{BY26} and the constant-query inverse attack on PFC described in~\cite{MPSY24}.
Using the fact that Cliffords normalize the Pauli group, we identify a hidden fraction of \(\sim1/N\) Pauli operators \(Q\) whose image under the adjoint action of \(U\), \(UQU^\dagger\), for \(U\) sampled from Clifford-endpoint DPFC, have only mutually commuting Paulis in their Pauli expansion.
In contrast, we show that when \(U\) is Haar-random, two Paulis in the expansion above commute with probability close to \(1/2\).
Then, we use an amplitude amplification to identify these Paulis with high probability and apply a commutation test to random Paulis in the expansion.  

We believe that our results provide a new perspective for studying the non-adaptive hardness of unitary families that follow the popular PFC / DPFC structure.
While previous non-adaptive analyses used primarily representation-theoretic techniques, we show that the path recording tool combined with the EPR-sector decomposition can be used to compare the channels induced by parallel queries in the strong setting. Our attack complements this result by showing that the birthday scale limits the adaptive security of the Clifford-DPFC. 

Together, these results give an overall picture of the security of  Clifford-endpoint DPFC: nonadaptive security guarantees together with adaptive security limitations. 
A significant question that remains open is whether it is possible to achieve the birthday-scale security or even better, in the strong adaptive setting, with an efficiently implementable construction.  
Achieving adaptive security beyond this scale requires overcoming the structure exploited by our attack.

\subsection{Technical overview}
\label{sec:technical-overview}
\subsubsection{DPFC nonadaptive strong security (Theorem~\ref{thm:informal-dpfc-non-adaptive-security})}
We consider an adversary making \(t\) parallel queries, each of which may be forward, inverse, conjugate, or transpose.
Since the queries are parallel, their ordering does not matter, up to a permutation of registers. 
It therefore suffices to consider a fixed query pattern of the form
\begin{align}
    \U_{a,b,c,d}(U)
    :=
    U^{\otimes a}
    \otimes (U^T)^{\otimes b}
    \otimes (U^*)^{\otimes c}
    \otimes (U^\dagger)^{\otimes d},
    \label{eq:abcd_query_pattern}
\end{align}
where \(t=a+b+c+d\).
The adversary may prepare an arbitrary joint input state, possibly entangled with an external reference system.
For a unitary ensemble, averaging the action of
\(\U_{a,b,c,d}(U)\) over the sampled unitary defines the \emph{four-type twirl} (Definition~\ref{def:parallel-four-query-twirl}).
We measure its distance from the corresponding Haar twirl in \emph{diamond norm}, thereby accounting for all such inputs.\footnote{Recall that we allow the adversary to control the identity of the queries by an additional routing register, thus in actuality we have a superposition of \(\U_{a,b,c,d}(U)\) calls with different $a,b,c,d$ values (that sum to $t$). This turns out to not be crucial, and we handle it in the end.}

There are two approaches in the literature for analyzing pseudorandom unitaries. The first, by \cite{MPSY24} relies on representation theory, since $U^{\otimes t}$ is a representation of the unitary group. This approach does not seem applicable in the strong PRU setting since tensor products involving both $U$ and $U^\dagger$ do not define a representation. The second approach was initiated in \cite{MH24} and relies on a technique called \emph{path recording} (more details will follow below). Whereas the introduction of path recording was tied to its ability to handle \emph{adaptive queries}, in this work we show how to utilize it in the nonadaptive setting.

We think of the \(DPF_3C\) ensemble as a composition of three layers: two layers of randomly and independently chosen endpoint unitaries C and D, sampled from an exact 2-design family, and a middle layer sandwiched in-between that consists of the PF (permutation-function) operators. In particular, PF contains a random ternary phase function, followed by a random permutation on the $n$-qubit computational basis elements.%

Similarly to \cite{MH24,SMLBH25} we use the path recorder within a hybrid argument. We consider a hybrid where the PF layer is replaced with our path recorder operator. We call our recorder the \emph{parallel recorder} since it relies on all queries being provided in parallel (or equivalently in a nonadaptive manner).

Let us briefly explain how the hybrid method works here, which is an adaptation of the recorder blueprint from \cite{MH24} and followup works. Our first hybrid $\hyb_0$ samples $D, P, F, C$ from their prescribed distributions. In $\hyb_1$ we replace PF with the path recorder (in our case, the parallel recorder). This recorder is a pure operator (i.e.\ not sampled from a distribution) but it works over a larger ambient space, which also contains a ``database'' register in addition to the regular input-output register of PF. The premise is that if the database register is traced out, then $\hyb_1$ produces a similar output distribution as $\hyb_0$, under specific conditions on $C, D$ that we outline. The error term here will be of birthday-bound scale. 
Note that in both $\hyb_0, \hyb_1$, the ``endpoints'' $C,D$ are sampled from their prescribed distributions and applied on the input-output register. In $\hyb_2$, however, $C,D$ do not act on the input-output register, but rather on the database register. In our case, we show that $\hyb_1\to\hyb_2$ incurs no loss at all. Now, in $\hyb_2$, we notice that the database register is traced out in the end, which means that the output in $\hyb_2$ does not depend on $C,D$ at all. The conclusion is that whenever $C,D$ satisfy the conditions for the $\hyb_0\to\hyb_1$ transition, their particular distributions do not change the resulting output state, up to a small statistical distance. Finally, we show that any unitary $2$-design distribution on $C,D$ would suffice for $\hyb_0\to\hyb_1$. Therefore, DPFC with Clifford endpoints is statistically close to DPFC with Haar random endpoints. Since the latter is itself Haar random, it follows that the former is close to Haar-random too.

In what follows, we start by explaining the parallel recorder functionality when only ``regular'' (forward) queries and inverse queries are allowed. Then, we address the additional challenges introduced by conjugate queries. Finally, we show how the two analyses combine to establish security against all four query types.

\paragraph{Analyzing inverse queries using the parallel recorder.}
We consider the setting where only $U, U^\dagger$ (parallel) queries are allowed. Recall that our goal, as per the hybrid description above, is to replace the PF layer with a path recording operator.

Consider \(t=p+q\) parallel queries to the permutation-phase layer PF, consisting of \(p\) forward queries and \(q\) inverse queries.
For a fixed permutation \(\pi\) and phase function \(f\), a forward query maps \(|x\rangle\) to \(\omega_3^{f(x)}|\pi(x)\rangle\), whereas an inverse query maps \(|y\rangle\) to \(\omega_3^{-f(\pi^{-1}(y))}|\pi^{-1}(y)\rangle\).
To analyze the averaged action of $t$ queries to the PF layer, we consider its purification \(\Kpq\), which keeps the random choices \(\pi,f\) coherently in an auxiliary environment. 
For a computational-basis input
\(|\boldsymbol{x}\rangle|\boldsymbol{y}\rangle\), where
\(\boldsymbol{x}=(x_1,\ldots,x_p)\) and
\(\boldsymbol{y}=(y_1,\ldots,y_q)\), we can write
\begin{align}
\Kpq
|\boldsymbol{x}\rangle|\boldsymbol{y}\rangle
\nonumber 
&=
\frac{1}{\sqrt{N!\,3^N}}
\sum_{\pi\in S_N}
\sum_{f:[N]\to\mathbb Z_3}
\omega_3^{
\sum_{i=1}^p f(x_i)
-
\sum_{j=1}^q f(\pi^{-1}(y_j))
}
|\pi(\boldsymbol{x})\rangle
|\pi^{-1}(\boldsymbol{y})\rangle
|\pi,f\rangle
\nonumber\\
&=
\frac{1}{N^{t/2}}
\sum_{\substack{
\boldsymbol{u}\in[N]^p\\
\boldsymbol{v}\in[N]^q
}}
w_{\Splus,\Sminus}\,
|\boldsymbol{u}\rangle|\boldsymbol{v}\rangle
|\pf_{\Splus,\Sminus}\rangle,
\label{eq:overview-pf-purification}
\end{align}
where 
\begin{align*}
\Splus=\{(x_i,u_i):i\in[p]\},
\qquad
\Sminus=\{(v_j,y_j):j\in[q]\}~,
\end{align*}
and
\begin{align*}
|\pf_{\Splus,\Sminus}\rangle
\propto
\sum_{\substack{
\pi\in S_N\\
\pi\supseteq \Splus\uplus \Sminus
}}
\sum_{f:[N]\to\mathbb Z_3}
\omega_3^{
\sum_{i=1}^p f(x_i)
-
\sum_{j=1}^q f(v_j)
}
|\pi,f\rangle
\end{align*}
is a normalized state (so long as there exists $\pi$ s.t.\ \(\pi\supseteq \Splus\uplus \Sminus\), i.e.\ \(\pi\) satisfies every input-output constraint in \(\Splus\uplus \Sminus\)). Note that  $\Splus$ contains pairs of the form $(x,u)$ (i.e.\ the input $x$ appears first), whereas $\Sminus$ contains pairs of the form $(v,y)$ (i.e.\ the input $y$ to the inverse query appears second). This makes sense since a $U^\dagger$ query both reverses the input-output order of the unitary and also flips the sign of the $F$ component.

This decomposition groups the random permutation-function choices according to the input-output size-\(t\) relations they produce.
Each environment state \(|\pf_{\Splus,\Sminus}\rangle\) is a coherent superposition over all permutations consistent with the relations and all phase functions, weighted by their corresponding phases.
For a fixed input, different output strings correspond to orthogonal environment states, since no permutation can satisfy two different sets of output constraints.

In the first hybrid $\hyb_1$, we replace the PF layer with a linear map called the \emph{parallel recorder} \(\Epq\), which we introduce next. The parallel recorder acts on the input register, and produces output on both the input register and an additional new database register.
Although the map it defines is not necessarily trace preserving (and thus it is not an isometry), we show that, after averaging over the endpoint unitaries, this replacement incurs a diamond norm error of at most \(O(t/\sqrt{N})\). 

The parallel recorder sends each query input to a superposition over its possible outputs, and coherently stores the input-output relations in databases, based on the phase sign contribution.
For \(t=p+q\) parallel queries, with \(p\) forward inputs \(\boldsymbol{x}=(x_1,\ldots,x_p)\) and \(q\) inverse inputs
\(\boldsymbol{y}=(y_1,\ldots,y_q)\), the parallel recorder acts as
\begin{align}
    \Epq|\boldsymbol{x}\rangle|\boldsymbol{y}\rangle
    =
    \frac{1}{N^{t/2}}
    \sum_{\substack{
    \boldsymbol{u}\in[N]^p\\
    \boldsymbol{v}\in[N]^q
    }}
    w'_{\Splus,\Sminus}   \,|\boldsymbol{u}\rangle|\boldsymbol{v}\rangle
    |\Splus,\Sminus\rangle,
    \label{eq:forward_inverse_recorder}
\end{align}

where \(w'_{\Splus,\Sminus}\) is the weight arising from symmetrizing the databases (see Equation~\eqref{eq:parallel-four-query-recorder-basis-action} for details). The intuition behind the parallel recorder being able to simulate random permutations and phases comes from lazy sampling. 
Just as classical lazy sampling generates and stores only the information needed to answer the queries, without specifying the entire random permutation and phase function, when only a few queries are made, it suffices to store only the signed input-output relations associated with the queries and sum over all outcome possibilities.

The recorder $\Epq$ and the physical purification $\Kpq$ are both described by the same signed relations \(\Splus,\Sminus\), but with different coefficients.
The recorder allows arbitrary relations, including those inconsistent with any permutation, whereas the physical purification includes only relations compatible with the sampled permutation.
To compare the two maps, we focus on the \emph{bijective subspace} defined as follows. We say that $(\Splus, \Sminus)$ is bijective if all \(t\) domain labels are distinct and all \(t\) image labels are distinct, i.e., there are no domain or image collisions in $(\Splus, \Sminus)$. The bijective subspace for $\Epq$ (respectively $\Kpq$) is the span of all $\ket{\Splus,\Sminus}$ (respectively $\ket{\pf_{\Splus,\Sminus}}$) s.t.\ $(\Splus, \Sminus)$ is bijective. We also let $\pibij$ denote the projector onto this subspace (slightly overloading notation, we use the same operator for both $\Epq$ and $\Kpq$).

We note that on the bijective subspace, $\Epq$ and $\Kpq$ behave almost identically. Indeed, if $(\Splus, \Sminus)$ is bijective then both $w_{\Splus,\Sminus}$ and $w'_{\Splus,\Sminus}$ are constants (i.e.\ independent of $\Splus,\Sminus$). One can calculate that
$w'_{\Splus,\Sminus}=1$ and $w_{\Splus,\Sminus}= \sqrt{{N^t(N-t)!}/{N!}}=(1+O(t^2/N))$. Furthermore, in the bijective setting $\pf_{\Splus,\Sminus}$ uniquely specifies $(\Splus, \Sminus)$ itself. Therefore, in the bijective subspace, there is a well-defined isometry mapping $\ket{\pf_{\Splus,\Sminus}} \leftrightarrow \ket{\Splus,\Sminus}$.

Under this identification, the physical purification and the recorder match exactly on their components in the bijective subspace, up to a normalization factor $\alpha = \frac{w}{w'}=1+O(t^2/N)$.
This is the \emph{exact matching property} of Lemma~\ref{lem:exact-four-query-bijective-sector-matching}.
It provides the basis for our approximation: the two maps agree on bijective relations, and any difference stems from relations containing repeated domain or image labels.

Our strategy, therefore, for making the transition $\hyb_0 \to \hyb_1$ will be the following. It will be convenient to work with channels that are not necessarily trace preserving, but are completely positive (CP). Let us denote by $\qcK$, $\qcE$ the CP maps obtained by sampling $C,D$ from their prescribed distributions, and using them as endpoints for $\Kpq$, $\Epq$ respectively. (Whereas $\qcK$ is a CPTP quantum channel, $\qcE$ is not trace preserving, as we discuss below.) Then the channels applied in the hybrids, denoted $\chyb{0}, \chyb{1}$, exactly correspond, respectively to applying $\qcK,\qcE$ followed by tracing out the database register.
We established that $\qcK,\qcE$ are identical (up to small normalization) on their (respective) bijective subspaces:
\begin{align}
    \pibij \qcK(\rho) \pibij = \alpha^2  \pibij \qcE(\rho) \pibij~,
\end{align}
up to the mapping $\ket{\pf_{\Splus,\Sminus}} \leftrightarrow \ket{\Splus,\Sminus}$.

We are therefore left with the task of bounding the non-bijective weight of these channels. We use techniques similar to \cite{MH24} to show that the non-bijective weight in the output of $\qcE$ is at most $O(t^2/N)$:
\begin{align}
    \tr[(I-\pibij) \qcE(\rho)] \le O(t^2/N)~.
\end{align}
We would like to infer that $\tr[\pibij \qcE(\rho)] \ge 1- O(t^2/N)$, but this requires a delicate argument since $\qcE$ is not trace preserving. We therefore consider an intermediate ordered version of $\Epq$ which is an isometry (specifically, it saves $\Splus,\Sminus$ in order of appearance, rather than as multisets), show the collision bound there, and then transfer back. This ordered recorder is also identical, up to scaling, to $\Epq$. Once we establish that $\tr[\pibij \qcE(\rho)] \ge 1- O(t^2/N)$, it follows that $\tr[\pibij \qcK(\rho)] \ge 1- O(t^2/N)$ (the factor of $\alpha^2$ does not change the asymptotics), and $\tr[(I-\pibij)\qcK(\rho)] \le O(t^2/N)$. Combining all of these properties, we conclude that $\norm{\chyb{0} - \chyb{1}}_1 \le O(t/\sqrt{N})$, where the square root is due to cross terms between the bijective and non-bijective components.

The tool we use to bound the non-bijective weight also establishes the transition from $\hyb_1$ to $\hyb_2$. As mentioned above, $\hyb_1$ is equivalent, for all $C,D$, to a hybrid where $C,D$ act on the database register rather than the input register. This latter form, where $C,D$ act on the database, turns out to be useful for establishing the collision bound. This is the so-called \emph{unitary invariance} property (Lemma~\ref{lem:four-type-endpoint-covariance}):
\begin{align}
\left[
    \left(D^{\otimes p}\right)_U
    \otimes \left((C^\dagger)^{\otimes q}\right)_{\dagger}
    \otimes I_{\cD}
\right] \Epq
\left[
    \left(C^{\otimes p}\right)_U
    \otimes \left((D^\dagger)^{\otimes q}\right)_{\dagger}
\right]
=\left[I_{\mathcal A_{p,q}}\otimes\mathcal Q_{p,q}[C,D]\right]\Epq~,
\label{eq:overview-endpoint-covariance}
\end{align}
where $\cD$ is the database register and \(\mathcal Q_{p,q}[C,D]\) is a unitary acting only on $\cD$.
Consequently, tracing out the databases eliminates the dependence on \(C,D\), so the recorder induces the same map for every choice of endpoint unitaries (this is the transition $\hyb_1 \to \hyb_2$).

Let us briefly explain why this property holds. The derivation is similar to \cite{MH24} and in fact somewhat simpler and tighter in the non-adaptive setting (in the adaptive setting it only holds approximately). If we consider Equation~\eqref{eq:forward_inverse_recorder}, we see that the path recording operator essentially ``copies'' $\boldsymbol{x}, \boldsymbol{y}$ from the input register onto the database register. This means that the application of $C$ on the input to $U$ or $D^\dagger$ on the input to $U^\dagger$ might as well be applied after $\boldsymbol{x}, \boldsymbol{y}$ have been copied into the database, and not before. Likewise the output values $\boldsymbol{u}, \boldsymbol{v}$ also have a ``twin'' in the database register, so the application of the output endpoint $D$ or $C^\dagger$ respectively, can also be shifted to the database register (similarly to how applying a unitary on one end of an EPR pair is equivalent to applying its transpose on the other end). 

Formalizing this argument turns out to require some more care. This is since the database information is kept in an order-agnostic fashion, and the above argument requires that the order of queries is preserved (so that the ``EPR-pair argument'' can be carried through). This is handled (as in previous work) by considering an ordered version of the recording operator and connecting it to the actual recording operator by applying a symmetric-subspace projector.

Let us now go back to the task of bounding collisions, in order to (finally) establish the transition from $\hyb_0$ to $\hyb_1$. We show that this holds whenever $C,D$ are sampled from a $2$-design. To do this, we notice that the operator \(\mathcal Q_{p,q}[C,D]\) that is applied to the database register, amounts to applying a random Clifford (either $C, D^T, C^*, D^\dagger$) to each input and output element in the database. As analyzed in previous works, this means that the probability of landing outside the bijective subspace is similar to the probability of obtaining a collision when sampling $t$ random elements from a universe of size $N$, namely $O(t^2/N)$. As explained above, we need to apply this argument on the ordered version of the recorder in order to obtain our bound.

At this point, a subtlety from above creeps back in. We noted that $\Epq$ is not unitary, and it can therefore blow up the weights of some input vectors, in particular outside the bijective subspace. To handle this, we bound the spectral norm of the Gram operator \(G_{p,q}:=\Epq^\dagger \Epq\). In particular, in Lemma~\ref{lem:four-query-gram-upper-bound}, we show that
\begin{align}
G_{p,q}-I
\preceq
\left(e^{t^2/(2N)}-1\right)I~,
\end{align}
which means that the distortion can be at most $1+O(t^2/N)$. Interestingly, the bound on the Gram operator can also be viewed as some variant of collision counting. As we explained, $\Epq$ is an isometry on the bijective subspace, but it assigns more weight to an input vector the more collisions it has. Essentially, this is because the database register is symmetrized, and a vector with repetitions has a larger projection onto the symmetric subspace. We refer to Lemma~\ref{lem:four-query-gram-upper-bound} for the details.

This concludes both transitions $\hyb_0 \to \hyb_1$ and $\hyb_1 \to \hyb_2$. We note that extending our techniques to the adaptive setting seems to hit a barrier.
In the adaptive setting, an adversary may use the output of one query as the input to another, so applying \(U\) followed by \(U^\dagger\) must restore the original state.
An append-only recorder cannot capture this cancellation, requiring the adaptive recorder of~\cite{MH24} to support coherent erasure of previously recorded relations.
However, the erasure mechanism prevents the full recorder from satisfying exact unitary invariance, which instead holds only approximately, with an error of \(O(k/\sqrt N)\) for a database of size \(k\).
These errors accumulate over successive queries, preventing \cite{MH24}-style analysis from achieving birthday-scale security.
In our parallel setting, no such erasure is necessary, allowing us to retain exact unitary invariance and obtain the birthday-scale bound.

\paragraph{Analyzing conjugate queries via EPR sector decomposition.} When considering 
query patterns that include both \(U,U^*\) or \(U^T, U^\dagger\), we notice that contrary to the previous case, these so called \emph{mixed} queries introduce \emph{invariant subspaces}.
Specifically, an EPR state shared between a pair of mixed query registers:
\begin{align}
    |\Omega_N\rangle := 
    \frac{1}{\sqrt N}
    \sum_{x=1}^N |x,x\rangle,
\end{align}
satisfies for every unitary \(U\),
\begin{align}
    (U\otimes U^*)|\Omega_N\rangle
    =|\Omega_N\rangle, \quad (U^T\otimes U^\dagger)|\Omega_N\rangle
    =|\Omega_N\rangle~.
\end{align}
Such inputs should therefore pass through our unitary operation intact.

This property would create a fundamental problem for any attempt to naturally extend the parallel recorder in Equation~\eqref{eq:forward_inverse_recorder} for all four query types. This is because the parallel recorder works by creating entanglement between the query register and the state of the database, which is prohibitive if invariant subspaces exist. %
    
Let us nevertheless try to envision the trivial extension, bearing in mind that we must handle invariant subspaces differently down the line. To extend the parallel recorder, the roles of the database sign and orientation of input-output relations from before should be preserved.
Specifically, referring back to Equation~\eqref{eq:overview-pf-purification}, \(U^*\) queries with input $x$ are recorded in the negative database $\Sminus$, but they introduce pairs in the \emph{forward orientation}, i.e.\ of the form $(x,u)$, whereas \(U^T\) queries with input $y$ are recorded in the positive $\Splus$ database with \emph{inverse orientation} $(v,y)$.
Thus, the action of the extended parallel recorder on a computational basis input \( |\boldsymbol x\rangle_U
|\boldsymbol y\rangle_T
|\boldsymbol x'\rangle_*
|\boldsymbol y'\rangle_\dagger\) is given by 
\begin{align}
    & E_{a,b,c,d}
    \bigl(
    |\boldsymbol x\rangle_U
    |\boldsymbol y\rangle_T
    |\boldsymbol x'\rangle_*
    |\boldsymbol y'\rangle_\dagger
    \bigr)
    = \nonumber\\
    & \qquad
    \frac{1}{N^{t/2}}
    \sum_{\boldsymbol u,\boldsymbol v,\boldsymbol u',\boldsymbol v'}
    w'_{\Splus,\Sminus}\,
    |\boldsymbol u\rangle_U
    |\boldsymbol v\rangle_T
    |\boldsymbol u'\rangle_*
    |\boldsymbol v'\rangle_\dagger
    |\Splus,\Sminus\rangle,
    \label{eq:overview-full-parallel-four-query-recorder-basis-action}
\end{align}
where \(a,b,c,d\) correspond to the query pattern in Equation~\eqref{eq:abcd_query_pattern}, \(t = a + b + c + d\), and we extend the definitions of \(\Splus, \Sminus\) to 
\begin{align*}
    \Splus := \{(x_i,u_i): i \in [a]\} \uplus
    \{(v_i,y_i): i \in [b]\}, \quad
    \Sminus :=\{(x_i',u_i'): i\in [c]\} \uplus \{(v_i',y_i'): i\in[d]\}.
\end{align*}
As expected, for a shared EPR pair between the \(U,U^*\) registers, this parallel recorder fails and acts as 
\begin{align}
    E_{1,0,1,0} |\Omega_N\rangle = \frac{1}{N^{3/2}} \sum_{x,u,u' \in [N]} |u,u'\rangle
    |\{(x,u)\},\{(x,u')\}\rangle.
\end{align}
Tracing out the database register yields the maximally mixed state \(I/N^2\), while the action of the physical PF layers simply preserves \(|\Omega_N\rangle\). 
Hence, this direct extension cannot be used as is to approximate the PF layers.

Nevertheless, this extension should work when there are no EPR pairs in the input. One would therefore be tempted to check every pair of coordinates in the input (or, more sophisticatedly, write it in the Bell basis and separate the EPR component from the rest). Then, for the EPR part do not perform any operation, and use the extended oracle above for the remainder.
This approach does not work in general though, since two projectors on chosen EPR location subspaces are not orthogonal, so the entire Hilbert space cannot be decomposed straightforwardly into a direct sum of subspaces corresponding to possible EPR locations.

To handle the overlap between projectors, we adapt the EPR-sector decomposition used to analyze unitary designs in \cite{SMLBH25}. This decomposition can be viewed as a careful, working version of the naive decomposition attempt from above. To explain their decomposition, let us consider the setting where only $U, U^*$ queries are allowed, and we refer to ``invariant pairs'' when considering an EPR pair between a $U$ query slot and a $U^*$ query slot (namely, an EPR pair that should ``pass through'' the unitary oracle call).

Instead of projecting on each possible location of EPR pairs, the decomposition in \cite{SMLBH25} jointly orthogonalizes the subspaces corresponding to different possible placements of a \emph{fixed number} of invariant pairs, \(\ell\). 
The full Hilbert space is now a direct sum of sectors, each containing a fixed number $0\leq\ell\leq \min\{a,c\}$ of invariant pairs, occupying $2\ell$ query slots, and no EPR pairs on the other slots.
The $\ell$-th sector can then be further decomposed into a tensor product of the ``multiplicity register'', which specifies on which locations of the original coordinates the invariant pairs resided, and a ``data register'' containing the EPR-free information. 
Formally, they define partial isometries $I_\ell:\mathcal{H}^{\otimes (a+c)} \to \nEspace{(a-\ell,c-\ell)}\otimes \mathcal M_{\ell}^{(a,c)}$ that project onto the $\ell$-th sector, mapping the sector to the multiplicity space $\mathcal M_{\ell}^{(a,c)}$ and the \emph{no-EPR subspace} \(\nEspace{(a-\ell,c-\ell)}\). 
The no-EPR subspace consists of states whose projection on the subspace spanned by invariant pairs in any location is zero.
Indeed, this is where we would want to apply our naive path recorder. 

By construction, the maps \(I_{\ell}\) corresponding to different \(\ell\) numbers of pairs are orthogonal, so the direct sum over all possible \(\ell\) pairs, \(\mathcal{I}_{(a,c)}\) is a unitary transformation. 
Crucially, as shown in \cite{SMLBH25}, the action of the query operator in these new coordinates (i.e., $\mathcal{I} \U(U) \mathcal{I}^\dagger$, for any $U$) manifests as the same unitary operation, applied only to the residual EPR-free query registers in each EPR sector.
The multiplicity registers remain unchanged.
This property is referred to as \emph{unitary equivariance}.
This is a result of the fact that the subspaces associated with different EPR-pair placements are invariant under the query action. 
Therefore, the operators used in the aforementioned joint orthogonalization process: their orthogonal projectors, their complements, products, and positive square roots all commute with the query action.
Finally, the change of coordinates removes the entanglement problem we had before so the recorder definition can be extended.
We define the full parallel recorder \(\mathbf E_{(a,c)}\) by first applying \(\mathcal I_{(a,c)}\), then applying the map \(E_{(a-\ell,0,c-\ell,0)}\) from Equation~\eqref{eq:overview-full-parallel-four-query-recorder-basis-action} to the residual registers in each $\ell$ EPR sector separately $\bigoplus_{\ell}\left.E_{(a-\ell,0,c-\ell,0)}\right|_{\nEspace{(a-\ell,c-\ell)}}$. 
Now, the inputs to the recorder in each block do not contain any EPR pairs.
Thus, at a high-level, we first apply the unitary change of basis, then compare the recorder map with the physical DPFC channel in every sector (the physical channel as well acts on the reduced space only).

\paragraph{Sectorwise comparison and coherent assembly.}
So far, we have analyzed the parallel recorder that simulates the permutation–phase layers within each EPR sector. 
Returning to the hybrid sequence introduced at the beginning of this overview, we need to bound the errors in \(\hyb_0\to\hyb_1\), which replace the permutation–phase layers with the recorder, both for Clifford and Haar-random endpoints when mixed queries are allowed.
We show that the same approximation errors hold for a coherent superposition of different EPR sectors and query patterns.

Continuing only with $U,U^*$ queries for convenience, fix an EPR sector \(\ell\), with residual query pattern \((a-\ell, c-\ell)\) after EPR sector decomposition. 
The equivariance property mentioned above allows us to compare the physical action of the PF layers and the parallel recorder on the corresponding residual no-EPR subspace.
Recall that \(K^{\mathrm{PF}}\) from Equation~\eqref{eq:overview-pf-purification} (extended to $U,U^*$ and later to all query types) purifies the physical PF action. 
To compare this purification with the recorder, we map its purifying registers isometrically into the database space, using the identification of physical environment states with database states described earlier. 
This changes only the representation of the environment and leaves the physical channel unchanged. 
Either \(K^{\mathrm{PF}}\) or the recorder \(E_{(a-\ell,c-\ell)}\) is placed between the input $C^{\otimes (a-\ell)}\otimes (C^*)^{\otimes (c-\ell)}$ and output $D^{\otimes (a-\ell)}\otimes (D^*)^{\otimes (c-\ell)}$ layers. 
The approximation errors established above for \(U,U^\dagger\) queries apply here separately in each sector in the same way: there is an exact matching up to normalization on the bijective component, while the collision and norm estimates bound the remaining error, resulting in an approximation error of order \(O(t/\sqrt{N})\). 

Finally, applying the unitary transformation $\mathcal I_{(a,c)}$ allows us to express the input in EPR-sector coordinates. 
Extending the argument to superpositions of inputs from different sectors requires handling cross terms.
Orthogonality of the outputs in each sector then allows the largest sector to bound the full part outside the bijective subspace without separate treatment of cross terms, yielding the same $O(t/\sqrt{N})$ diamond-norm error.
Thus, as above, we get the same $O(t/\sqrt{N})$ diamond norm error.

To address all four query types, we actually apply this decomposition twice, once for $U,U^*$ registers and once for $U^T,U^\dagger$ registers. 
A sector is then labeled by the numbers $(\ell,m)$ of invariant pairs in each groups. 
The same process is carried out on each resulting no-EPR subspace, including unitary equivariance, the definition of the full recorder, the final bound.

The argument also accommodates coherent superpositions of query patterns. 
The proof compares the physical channel and the recorder map together with their endpoint and database registers, using the same representation of the physical environment across all query patterns and EPR sectors. 
The pattern and sector labels remain part of the output, so the total weight outside the bijective database subspace is again bounded by the largest blockwise estimate, $O(t^2/N)$.
Consequently, using the same techniques as above, the $O(t/\sqrt N)$ diamond-norm estimate holds for arbitrary coherent superpositions of both query patterns and EPR sectors.

\subsubsection{Adaptive attack on DPFC (Theorem~\ref{thm:informal-addaptive-attack-on-DPFC})}

Recall that our goal is to show an algorithm that uses $O(\sqrt{N})$ (adaptive) queries to $U, U^\dagger$ and distinguishes between DPFC and the Haar distribution with high probability. Our approach relies on $C,D$ being Cliffords, and in particular on the fact that the Clifford group normalizes the Pauli group. That is, if $Q$ is Pauli and $C$ is Clifford then $CQC^\dagger$ is Pauli.

Let \(U=DPFC\) with Clifford endpoints, and consider a Pauli $Q$ of the form $Q=Q_s = C^\dagger Z^s C$, for $s \in \{0,1\}^n$, where $Z^s = Z^{s_1}\otimes \cdots \otimes Z^{s_n}$. If we calculate $U Q_s U^\dagger$ we get
\begin{align}
    UQ_sU^{\dagger} &= (DPFC) (C^\dagger Z^s C) (C^\dagger F^\dagger P^{\dagger} D^{\dagger})
    \\ \nonumber &= DPZ^sP^{\dagger}D^{\dagger}~,
\end{align}
since the $C$ terms cancel out, and $F$ commutes with $Z^s$.

Let us consider the term $PZ^sP^{\dagger}$. This is a diagonal matrix, and therefore its Pauli basis decomposition\footnote{Recall that any square operator can be written as a sum of Pauli operators.} only contains diagonal elements, i.e.\ it is possible to write $PZ^sP^{\dagger} = \sum_{t\in\{0,1\}^n} r_a Z^a$ for some coefficients $\{\alpha_t\}_t$. This in turn means that the Pauli decompsition of $UQ_sU^{\dagger}=DPZ^sP^{\dagger}D^{\dagger}$ is equal to $\sum_{a\in\{0,1\}^n} r_a D Z^a D^\dagger$ (recall that $D$ is Clifford). We note that $D Z^a D^\dagger$ all commute with each other. 

The idea of our distinguisher is to find such a $Q_s$ and test whether the elements of its Pauli basis decomposition commute. How do we actually perform this test? And how likely is it to find such a Pauli? Let us start with the latter. Essentially by definition there are $N=2^n$ many such $Q_s$, out of $2^{2n}$ many $n$-qubit Paulis. Therefore, if we just sample $Q$ at random, the probability of it being in $\{Q_s\}_s$ is $\sim 1/N$. Once we have a candidate $Q$, we can test it using the following procedure. We recall that if $V$ is a unitary with Pauli decomposition $V=\sum_a r_a R_a$, then we can sample from the set $\{R_a\}_a$ with respective probabilities $\abs{r_a}^2$. This is done by applying $V$ to one end of an $n$-qubit EPR pair and measuring in the Bell basis. In our case, we let $V=UQU^\dagger$ and apply the above procedure twice to obtain $R_a, R_b$ with probability $\abs{r_a}^2 \abs{r_b}^2$. We accept if the two Paulis commute and reject otherwise.

If indeed $U=DPFC$, and if we were lucky to pick $Q=Q_s$, then our testing procedure accepts with probability $1$. In contrast, if $U$ is Haar random, then on average the test succeeds with probability close to $1/2$ (since $R_a, R_b$ behave almost like random Paulis). Using a concentration bound, it holds that with overwhelming probability over $U$, there does not exist nonidentity $Q$ that causes the test to accept with probability greater than $3/4$.

We therefore found a distinguishing test between DPFC and Haar. Sample $Q$, run the test a sufficient number of time to estimate its success probability, if this value is greater than $3/4$ declare DPFC, otherwise repeat. If after $\sim N$ attempts no suitable $Q$ was found, declare Haar.

This is a valid test but it requires $\tilde{O}(N)$ queries to $U, U^\dagger$, since each execution of the test requires a constant number of queries. We improve this to $O(\sqrt{N})$ using amplitude amplification. Notice that a naive application of amplitude amplification would lead to query complexity of $O(\sqrt{N}\log N)$, which can be reduced to $O(\sqrt{N})$  using techniques from \cite{HMW03}.

\subsection{Paper Organization}

In Section~\ref{sec:preliminaries} we introduce necessary definitions and the proof setting. Sections~\ref{sec:recorder-and-physical-environment} and~\ref{sec:four-query-endpoint-averaging} form the core of the proof, presenting the main constructions and the assembly of the final security bound. 
Section~\ref{sec:cba-attack} describes the protocol for the adaptive attack on the DPFC ensemble.
Lengthy proofs are deferred to the appendix. 

\subsection{Acknowledgments}

The authors are supported by the Horizon Europe Research and Innovation Program via ERC Project ACQUA (Grant 101087742).

\subsection{AI Disclosure}

This work involved extensive, iterative interaction with OpenAI’s models using ChatGPT: GPT-5.5 SOL, GPT-5.6 SOL, GPT-6 ASTRA, and GPT-6.1 SOL, for research discussions and manuscript preparation.
The authors proposed the idea of proving the nonadaptive security of the DPFC ensemble against all four query types and suggested the mathematical tools and techniques used in the paper.
The construction of the parallel recorder using the append maps of \cite{MH24}, as well as the adaptive attack, came about after many discussions between the authors and AI. 
The authors suggested the tool of EPR sector decomposition of \cite{SMLBH25} as a way to handle mixed queries in the extension to all four query types, as well as the assembly of query patterns.
AI was also used extensively to draft, review, and simplify the manuscript over multiple sessions. 
The authors take full responsibility for all mathematical claims, proofs, and conclusions.

\section{Preliminaries and Setup}
\label{sec:preliminaries}

Fix a dimension \(N\geq 1\), and let \(\mathcal H_N:=\mathbb C^N\). Write \(\mathrm U(N)\) for the group of unitary operators on \(\mathcal H_N\), and \(\mu_{\mathrm H}\) for its normalized Haar measure.

For finite-dimensional Hilbert spaces \(\mathcal H\) and \(\mathcal K\), let \(\mathsf L(\mathcal H,\mathcal K)\) denote the space of linear operators from \(\mathcal H\) to \(\mathcal K\), and abbreviate \(\mathsf L(\mathcal H):=\mathsf L(\mathcal H,\mathcal H)\). 

\subsection{Channels, Isometries, and Norms}

\begin{definition}[Quantum channel]
\label{def:quantum-channels}
A linear map \(\Phi:\mathsf L(\mathcal H)\to\mathsf L(\mathcal K)\) is completely positive if \(\Phi\otimes\operatorname{id}_{\mathcal R}\) maps positive operators to positive operators for every finite-dimensional reference space \(\mathcal R\). 
It is trace-preserving if \(\operatorname{Tr}(\Phi(X))=\operatorname{Tr}(X)\) for every
\(X\in\mathsf L(\mathcal H)\). 
A quantum channel is a completely positive, trace-preserving linear map.
\end{definition}

\begin{definition}[Adjoint action]
\label{def:adjoint-action}
For a linear operator \(A:\mathcal H\to\mathcal K\), define its adjoint action (or conjugation map) by
\begin{align}
    \operatorname{Ad}_A:
    \mathsf L(\mathcal H)\longrightarrow\mathsf L(\mathcal K),
    \qquad
    \operatorname{Ad}_A(X):=AXA^\dagger.
\end{align}
\end{definition}

\begin{definition}[Isometries and Stinespring dilations]
\label{def:isometries-and-stinespring}
A linear map \(V:\mathcal H\to\mathcal K\) is an isometry if \(V^\dagger V=I_{\mathcal H}\). 
A Stinespring dilation of a completely positive map \(\Phi:\mathsf L(\mathcal H)\to\mathsf L(\mathcal K)\) is a linear map \(V:\mathcal H\to\mathcal K\otimes\mathcal E\) satisfying
\begin{align}  \Phi
    =
    \operatorname{Tr}_{\mathcal E}
    \circ
    \operatorname{Ad}_V.
\end{align}
The space \(\mathcal E\) is called the environment.
For every \(X\in\mathsf L(\mathcal H)\),
\begin{align}
    \operatorname{Tr}(\Phi(X))
    =
    \operatorname{Tr}(V^\dagger V X).
\end{align}
Thus \(\Phi\) is trace-preserving exactly when \(V^\dagger V=I_{\mathcal H}\). In particular, a dilation of a quantum channel is an isometry.
\end{definition}

\begin{definition}[Vector, trace, Hilbert-Schmidt, operator, and diamond norms]
\label{def:norms}
For \(\lvert\psi\rangle\in\mathcal H\), define its Hilbert-space norm by
\begin{align}
    \bigl\|\lvert\psi\rangle\bigr\|_2
    :=
    \sqrt{\langle\psi\vert\psi\rangle}.
\end{align}
For finite-dimensional Hilbert spaces \(\mathcal H\) and \(\mathcal K\), and \(X\in\mathsf L(\mathcal H,\mathcal K)\), define
\begin{align}
\begin{aligned}
    \lVert X\rVert_1
    :=
    \operatorname{Tr}\sqrt{X^\dagger X}, \quad
    \lVert X\rVert_2
    :=
    \sqrt{\operatorname{Tr}(X^\dagger X)}, \quad
    \lVert X\rVert_\infty
    :=
    \sup_{\lvert\psi\rangle\neq 0}
    \frac{\lVert X\lvert\psi\rangle\rVert_2}
         {\lVert\lvert\psi\rangle\rVert_2}.
\end{aligned}
\end{align}
These are respectively the trace norm, the unnormalized Hilbert-Schmidt (Schatten \(2\)) norm, and the operator norm. In terms of the singular values of \(X\), they are the sum, the square root of the sum of squares, and the maximum. When \(X\) is Hermitian, the singular values are the absolute values of its eigenvalues.

Let \(\Lambda: \mathsf L(\mathcal H) \longrightarrow \mathsf L(\mathcal K)\) be a linear map. Its diamond norm is
\begin{align}
    \lVert\Lambda\rVert_\diamond
    :=
    \sup_{\mathcal R}\;
    \sup_{\substack{
        Z\in\mathsf L(\mathcal H\otimes\mathcal R)\\
        Z\neq 0
    }}
    \frac{
        \lVert
            (\Lambda\otimes\operatorname{id}_{\mathcal R})(Z)
        \rVert_1
    }{
        \lVert Z\rVert_1
    },
\end{align}
where the first supremum ranges over all finite-dimensional reference spaces \(\mathcal R\), and \(\operatorname{id}_{\mathcal R}\) is the identity channel on \(\mathsf L(\mathcal R)\). 
For the diamond norm in finite dimensions, it is equivalent to take only \(\mathcal R\simeq\mathcal H\), since a reference space of dimension \(\dim(\mathcal H)\) is sufficient.
\end{definition}

\begin{lemma}[Common-environment dilation bound]
\label{lem:common-environment-stinespring-bound}
Let \(V, W:\mathcal H\to\mathcal K\otimes\mathcal E\) be linear maps with finite-dimensional input and output spaces and a common, possibly infinite-dimensional, environment. 
Define the completely positive maps
\begin{align}
    \Phi_V
    :=
    \operatorname{Tr}_{\mathcal E}\circ\operatorname{Ad}_V,
    \qquad
    \Phi_W
    :=
    \operatorname{Tr}_{\mathcal E}\circ\operatorname{Ad}_W.
\end{align}
Then,
\begin{align}
    \left\|\Phi_V-\Phi_W\right\|_\diamond
    &\leq
    \bigl(\|V\|_\infty+\|W\|_\infty\bigr)
    \left\|V-W\right\|_\infty.
    \label{eq:four-query-common-environment-diamond-bound}
\end{align}
In particular, if \(V\) and \(W\) are isometries, then,
\begin{align}
    \left\|\Phi_V-\Phi_W\right\|_\diamond
    \leq
    2\left\|V-W\right\|_\infty.
\end{align}
\end{lemma}

This is the elementary direction of the continuity theorem in \cite{KSW08}.

\begin{proof}
    See Appendix~\ref{app:common-environment-stinespring-bound}.
\end{proof}

\subsection{Parallel Twirls with Four Query Types}

Fix nonnegative integers \(a,b,c,d\), write \(\theta:=(a,b,c,d)\) for the query pattern. Set
\begin{align}
    t:=a+b+c+d,
    \qquad
    \mathcal A_{a,b,c,d}
    :=
    \mathcal H_N^{\otimes a}
    \otimes
    \mathcal H_N^{\otimes b}
    \otimes
    \mathcal H_N^{\otimes c}
    \otimes
    \mathcal H_N^{\otimes d}.
\end{align}
The parameters \(a,b,c,d\) count the numbers of \(U,U^T,U^*,U^\dagger\) queries, respectively, and the tensor factors are placed in this \emph{canonical query order}:
\begin{align*}
    U,
    \qquad
    U^T,
    \qquad
    U^*,
    \qquad
    U^\dagger,
\end{align*}
where \(U^*\) denotes the entrywise complex conjugate of \(U\).

\begin{definition}[Parallel four-type twirl]
\label{def:parallel-four-query-twirl}
For \(U\in\mathrm U(N)\), define the fixed-pattern query operator
\begin{align}
    \U_{a,b,c,d}(U)
    :=
    U^{\otimes a}
    \otimes
    (U^T)^{\otimes b}
    \otimes
    (U^*)^{\otimes c}
    \otimes
    (U^\dagger)^{\otimes d}.
\end{align}
Let \(\nu\) be a probability measure on \(\mathrm U(N)\). The parallel four-type \((a,b,c,d)\)-twirl is the channel \(\Phi_\nu^{a,b,c,d}:\mathsf L(\mathcal A_{a,b,c,d})\to\mathsf L(\mathcal A_{a,b,c,d})\) characterized, for every finite-dimensional reference space \(\mathcal R\) and every \(X\in\mathsf L(\mathcal A_{a,b,c,d}\otimes\mathcal R)\), by
\begin{align}
    \bigl(
        \Phi_\nu^{a,b,c,d}
        \otimes\operatorname{id}_{\mathcal R}
    \bigr)(X)
    &:=
    \mathbb E_{U\sim\nu}
    \left[
        \bigl(\U_{a,b,c,d}(U)\otimes I_{\mathcal R}\bigr)
        X
        \bigl(\U_{a,b,c,d}(U)^\dagger
        \otimes I_{\mathcal R}\bigr)
    \right].
\end{align}
The type of every query slot is fixed in advance. Adaptive
interleaving is not included.

When \(U\sim\mu_{\mathrm H}\) is Haar distributed, abbreviate
\begin{align}
    \Phi_{\mathrm H}^{a,b,c,d}
    :=
    \Phi_{\mu_{\mathrm H}}^{a,b,c,d}.
\end{align}
\end{definition}

\begin{definition}[Coherent parallel query operator and twirl]
\label{def:coherent-query-pattern}
For a total number of queries \(t\geq0\), define
\begin{align}
    \Theta_t
    :=\{(a,b,c,d)\in\mathbb Z_{\geq0}^4:a+b+c+d=t\},
    \qquad
    \mathcal Q_t
    :=\operatorname{span}\{|\theta\rangle:\theta\in\Theta_t\},
\end{align}
where the pattern states \(|\theta\rangle\) form an orthonormal basis.
Identify each \(\mathcal A_{a,b,c,d}\), for \((a,b,c,d)\in\Theta_t\), with the query space \(\mathcal A_t:=\mathcal H_N^{\otimes t}\) in canonical query order.
Throughout, the superscript \([t]\) denotes coherent assembly over
all query patterns with a total of \(t\) queries.
For \(U\in\mathrm U(N)\), define the controlled query operator on
\(\mathcal Q_t\otimes\mathcal A_t\) by
\begin{align}
    \U^{[t]}(U)
    &:={}
    \sum_{\theta=(a,b,c,d)\in\Theta_t}
        |\theta\rangle\!\langle\theta|
        \otimes\U_{a,b,c,d}(U).
\end{align}
The pattern register may be entangled with the query registers and an arbitrary reference system. All branches use the same unitary \(U\), and all queries are applied in one parallel layer.

For a probability measure \(\nu\) on \(\mathrm U(N)\), define the
corresponding channel \(\Phi_\nu^{[t]}:\mathsf L(\mathcal Q_t\otimes\mathcal A_t)\to\mathsf L(\mathcal Q_t\otimes\mathcal A_t)\), for every finite-dimensional reference space \(\mathcal R\) and every
\(X\in\mathsf L(\mathcal Q_t\otimes\mathcal A_t\otimes\mathcal R)\),
\begin{align}
    \bigl(\Phi_\nu^{[t]}
        \otimes\operatorname{id}_{\mathcal R}\bigr)(X)
    &:=
    \mathbb E_{U\sim\nu}
    \left[
        \bigl(\U^{[t]}(U)
            \otimes I_{\mathcal R}\bigr)
        X
        \bigl(\U^{[t]}(U)^\dagger
            \otimes I_{\mathcal R}\bigr)
    \right].
\end{align}
Abbreviate \(\Phi_{\mathrm H}^{[t]} :=\Phi_{\mu_{\mathrm H}}^{[t]}\).
\end{definition}

An arbitrary superposition of query type strings can be put in canonical query order by coherently sorting the query registers, retaining the original ordering in an auxiliary register, and undoing the sorting after the query layer. 
These permutations are independent of \(U\).

\subsection{Unitary Designs and the DPFC Ensemble}

\begin{definition}[Exact unitary \(t\)-design]
\label{def:exact-unitary-k-design}
Let \(t\geq 1\) be an integer. A probability measure \(\mathfrak D\) on \(\mathrm U(N)\) is an exact unitary \(t\)-design if, for every \(X\in\mathsf L(\mathcal H_N^{\otimes t})\),
\begin{align}
    \mathbb E_{C\sim\mathfrak D}
    \left[C^{\otimes t}X(C^\dagger)^{\otimes t}\right]
    =
    \mathbb E_{C\sim\mu_{\mathrm H}}
    \left[C^{\otimes t}X(C^\dagger)^{\otimes t}\right].
\end{align}
\end{definition}

The following definition is adapted from the strong additive-error notion in \cite{SMLBH25}, extending it to all four types of parallel queries.

\begin{definition}[Strong and ordinary approximate unitary \(t\)-designs]
\label{def:approximate-unitary-k-design}
Let \(t\geq1\) be an integer and let \(\varepsilon\geq0\).
A probability measure \(\mathfrak D\) on \(\mathrm U(N)\) is an
\emph{\(\varepsilon\)-approximate nonadaptive strong unitary
\(t\)-design} if, for every \(a,b,c,d\in\mathbb Z_{\geq0}\) satisfying \(a+b+c+d\leq t\),
\begin{align}
    \left\|
        \Phi_{\mathfrak D}^{a,b,c,d}
        -\Phi_{\mathrm H}^{a,b,c,d}
    \right\|_\diamond
    \leq\varepsilon.
\end{align}
Here \(a,b,c,d\) count parallel queries to \(U,U^T,U^*,U^\dagger\), respectively, as in Definition~\ref{def:parallel-four-query-twirl}.

Restricting to forward queries, \(b=c=d=0\), gives the ordinary notion: \(\mathfrak D\) is an \emph{\(\varepsilon\)-approximate unitary \(t\)-design} if, for every \(0\leq a\leq t\),
\begin{align}
    \left\|
        \Phi_{\mathfrak D}^{a,0,0,0}
        -\Phi_{\mathrm H}^{a,0,0,0}
    \right\|_\diamond
    \leq\varepsilon.
\end{align}
For the ordinary notion, \(\varepsilon=0\) recovers Definition~\ref{def:exact-unitary-k-design}.
\end{definition}

Following \cite{MH24}, define the DPFC ensemble.

\begin{definition}[DPFC ensemble]
\label{def:dpfc-ensemble}
Let \(\mathfrak D\) be an exact unitary \(2\)-design on \(\mathrm U(N)\).
Write \([N]:=\{1,\ldots,N\}\), let \(S_N\) be the symmetric group on
\([N]\), and set
\(
    \omega_3:=e^{2\pi i/3}
\). 
For \(\pi\in S_N\) and \(f:[N]\to\mathbb Z_3\), define
\begin{align}
    P_\pi\lvert x\rangle:=\lvert\pi(x)\rangle,
    \qquad
    F_f\lvert x\rangle:=\omega_3^{f(x)}\lvert x\rangle.
\end{align}
The \(\mathsf{DPFC}(\mathfrak D)\) ensemble is the distribution of \(U:=DP_\pi F_fC\) obtained by sampling all four choices independently:
\begin{align*}
    C,D\sim\mathfrak D,
    \qquad
    \pi\sim S_N,
    \qquad
    f\sim \mathbb Z_3^N.
\end{align*}
\end{definition}

Abbreviate the permutation-phase layer by
\begin{align}
    M_{\pi,f}:=P_\pi F_f,
    \qquad
    U=DM_{\pi,f}C.
\end{align}

\subsection{Pauli operators and Clifford conjugation}
\label{subsec:cba-paulis}

The adaptive attack in Section~\ref{sec:cba-attack} exploits the fact that the Clifford group introduced in this subsection is a unitary \(2\)-design and normalizes the Pauli group (defined in this subsection as well). 
This part is needed only for that attack and may otherwise be skipped.

\begin{definition}[Hermitian Pauli representatives]
\label{def:cba-pauli-representatives}
Let \(N=2^n\). Index the Hermitian Pauli representatives by
\(a=(u,v)\in\mathbb F_2^{2n}\):
\begin{align}
    P_a=i^{\sum_{j=1}^n u_jv_j}X^uZ^v,
    \qquad P_0=I.
\end{align}
The \(n\)-qubit Pauli group is
\begin{align}
    \mathcal P_n:=\bigl\{\omega P_a:
        a\in\mathbb F_2^{2n},\ \omega\in\{\pm1,\pm i\}\bigr\}.
\end{align}
\end{definition}

\begin{definition}[Symplectic form]
\label{def:cba-symplectic-form}
For \(a=(u,v),b=(w,z)\in\mathbb F_2^{2n}\), define the symplectic form by
\begin{align}
    [a,b]:=u\cdot z+v\cdot w\in\mathbb F_2.
\end{align}
\end{definition}

The Pauli elements satisfy
\begin{align}
    P_a^\dagger=P_a,\qquad P_a^2=I,\qquad
    \operatorname{Tr}(P_aP_b)=N\delta_{a,b},\qquad
    P_aP_b=(-1)^{[a,b]}P_bP_a.
    \label{eq:cba-paulis}
\end{align}
Thus, the \(N^2\) Pauli representatives form an orthogonal basis
of \(\mathsf L(\mathcal H_N)\), and two Pauli representatives \(P_a,P_b\) commute exactly
when \([a,b]=0\).

\begin{definition}[Clifford group]
\label{def:clifford-group}
The \(n\)-qubit \emph{Clifford group} is the normalizer of \(\mathcal P_n\) in \(\mathrm U(N)\)~\cite{Gottesman98}:
\begin{align}
    \operatorname{Clifford}(n)
    :=\bigl\{V\in\mathrm U(N):
        V\mathcal P_nV^\dagger=\mathcal P_n\bigr\}.
\end{align}
\end{definition}

Since conjugation preserves commutation, a conjugation by Clifford maps any mutually commuting family of Paulis to another such family. 
In~\cite{DCEL09} it is proved that uniformly sampling Clifford unitaries modulo global phase gives an exact unitary \(2\)-design. 
The Clifford ensemble provides a standard nontrivial example of a unitary \(2\) design beyond the trivial Haar measure on \(\mathrm U(N)\), but the non-adaptive DPFC security proof is not specific to Cliffords. 

\subsection{EPR-Sector Decomposition}

\begin{definition}[Normalized EPR state]
\label{def:normalized-epr-state}
On \(\mathcal H_N\otimes\mathcal H_N\), define the normalized EPR state and its rank-one projector by
\begin{align}
    \lvert\Omega_N\rangle
    :=
    \frac{1}{\sqrt N}\sum_{x=1}^N\lvert x,x\rangle,
    \qquad
    \Phi_N:=\lvert\Omega_N\rangle\!\langle\Omega_N\rvert.
\end{align}
\end{definition}

Mixed query actions (formed by a unitary tensored by its complex conjugate) act trivially on EPR pairs for all unitaries.
Indeed, for every \(V\in\mathrm U(N)\),
\begin{align}
    (V\otimes V^*)\lvert\Omega_N\rangle
    &=\lvert\Omega_N\rangle,
    &
    (V^T\otimes V^\dagger)\lvert\Omega_N\rangle
    &=\lvert\Omega_N\rangle.
    \label{eq:epr-invariance}
\end{align}
In particular, applying a mixed query action on an EPR component sampled from any unitary ensemble reveals no information about the original distribution.

As mentioned in Section~\ref{sec:technical-overview}, the comparison between Haar and DPFC uses an intermediate map that simulates the PF layers, called the parallel recorder, which is introduced formally in Section~\ref{sec:recorder-and-physical-environment}. 
This map creates entanglement between the input registers and auxiliary registers which will later be called database registers. 
Such entanglement fails to approximate the PF layers when the input contains an EPR pair: the physical actions \(U\otimes U^*\) and \(U^T\otimes U^\dagger\) preserve this pair as a pure state, whereas discarding the recorder’s database can make it mixed. 

A natural approach to address this problem is to project onto subspaces corresponding to different placements of EPR pairs and handle each subspace separately. 
However, these subspaces overlap, so their projectors do not yield an orthogonal decomposition of the query space. 
The decomposition introduced in this subsection resolves this overlap by jointly orthogonalizing subspaces corresponding to different possible placements of a fixed number of pairs.
This separates the invariant EPR components from the remaining no-EPR registers, allowing to apply the recorder only to the remaining registers.
The comparison then reduces to a comparison in each residual query space.

To formalize this separation, first characterize the subspace with no EPR component between registers carrying conjugate query actions~\cite{SMLBH25}. 

\begin{definition}[No-EPR subspace]
\label{def:no-epr-subspace}
For integers \(u,v\geq 0\), set the mixed-query space \(\mathcal H_N^{\otimes u}\otimes\mathcal H_N^{\otimes v}\).
Let \(\Phi_{i,j}\) denote the EPR projector of Definition~\ref{def:normalized-epr-state} acting on the \(i\)-th factor of the first tensor power and the \(j\)-th factor of the second tensor power, and as the identity on all remaining factors. 
The \emph{no-EPR subspace} is defined by
\begin{align}
    \nEspace{u,v}
    :=
    \bigcap_{i=1}^u\bigcap_{j=1}^v\ker\Phi_{i,j}
    \subseteq \mathcal H_N^{\otimes u}\otimes\mathcal H_N^{\otimes v}.
\end{align}
It consists of the states having zero projection onto an EPR pair for every choice of one factor from the first tensor power and one factor from the second.
Denote its orthogonal projector by \(\Pi_{u,v}^{\mathrm{nE}}\).
If \(u=0\) or \(v=0\), there are no pairs on which to impose an EPR constraint. 
In other words, \(\nEspace{u,0} =  H_N^{\otimes u}\), \(\nEspace{0,v} =  H_N^{\otimes v}\).
\end{definition}

Definition~\ref{def:no-epr-subspace} concerns one mixed tensor product \(V^{\otimes u}\otimes(V^*)^{\otimes v}\). 
Its indices \(i,j\) do not range over all four query blocks simultaneously. 
The definition is applied twice: \((u,v)=(a,c)\) for forward and conjugate queries and \((u,v)=(b,d)\) for transpose and inverse queries.
There are no restrictions between any other two query blocks.

The following orthogonal decomposition of the mixed-query space into EPR-sector subspaces, with the uncontracted registers in each sector lying in a no-EPR subspace, was shown in~\cite{SMLBH25}.

\begin{lemma}[Orthogonal EPR-sector coordinates]
\label{lemma:orthogonal-epr-sector-coordinates}
Let \(u,v\geq0\), and suppose that \(N\geq u+v\). For every \(0\leq\ell\leq\min\{u,v\}\), there exist a projector \(P_\ell^{(u,v)}\), a multiplicity space \(\mathcal M_\ell^{(u,v)}\), and a partial isometry
\begin{align}
    I_\ell^{(u,v)}:
    \mathcal H_N^{\otimes u}\otimes\mathcal H_N^{\otimes v}
    \longrightarrow
    \nEspace{u-\ell,v-\ell}
    \otimes\mathcal M_\ell^{(u,v)}
\end{align}
such that
\begin{align}
    P_\ell^{(u,v)}P_{\ell'}^{(u,v)}
    &=
    \delta_{\ell,\ell'}P_\ell^{(u,v)},
    \label{eq:epr-sector-orthogonality}\\
    \bigl(I_\ell^{(u,v)}\bigr)^\dagger I_\ell^{(u,v)}
    &=P_\ell^{(u,v)},
    \label{eq:epr-sector-initial-space}\\
    I_\ell^{(u,v)}\bigl(I_\ell^{(u,v)}\bigr)^\dagger
    &=
    \Pi_{u-\ell,v-\ell}^{\mathrm{nE}}
    \otimes I_{\mathcal M_\ell^{(u,v)}},
    \label{eq:epr-sector-final-space}\\
    \sum_{\ell=0}^{\min\{u,v\}}P_\ell^{(u,v)}
    &=I.
    \label{eq:epr-sector-resolution}
\end{align}
Moreover, for every \(V\in\mathrm U(N)\),
\begin{align}
    I_\ell^{(u,v)}
    \bigl(V^{\otimes u}\otimes(V^*)^{\otimes v}\bigr)
    &=
    \left(
        V^{\otimes(u-\ell)}
        \otimes(V^*)^{\otimes(v-\ell)}
        \otimes I_{\mathcal M_\ell^{(u,v)}}
    \right)
    I_\ell^{(u,v)},
    \label{eq:epr-sector-intertwining}
\end{align}
and
\begin{align}
    \dim\mathcal M_\ell^{(u,v)}
    =
    \binom{u}{\ell}\binom{v}{\ell}\ell!.
\end{align}
\end{lemma}

\begin{proof}
    See ~\cite{SMLBH25}.
\end{proof}

Next, the decomposition is extended to all four query types, and is applied separately to the \(U, U^*\) registers and the \(U^T, U^\dagger\) registers. 
The resulting sectors are indexed by two integers, counting the EPR pairs contracted in each group. 
Abbreviate
\begin{align}
    \U_\theta(U)
    :=
    \U_{a,b,c,d}(U).
\end{align}
Assume from this point onward that \(N\geq t=a+b+c+d\), so that Lemma~\ref{lemma:orthogonal-epr-sector-coordinates} applies to both mixed tensor products below and all later factorials are defined.
Define the sector-index set
\begin{align}
    \Lambda_\theta
    :=
    \left\{
        (\ell,m):
        0\leq\ell\leq\min\{a,c\},\quad
        0\leq m\leq\min\{b,d\}
    \right\}.
\end{align}

Denote
\begin{align}
    a_\lambda:=a-\ell,\qquad
    b_\lambda:=b-m,\qquad
    c_\lambda:=c-\ell,\qquad
    d_\lambda:=d-m.
\end{align}

For \(\lambda=(\ell,m)\in\Lambda_\theta\), define the residual
pattern
\begin{align}
    \theta_\lambda
    :=
    (a_\lambda, b_\lambda, c_\lambda, d_\lambda),
\end{align}
and abbreviate
\begin{align}
    \U_{\theta_\lambda}(U)
    :=
    \U_{a_\lambda, b_\lambda, c_\lambda, d_\lambda}(U),\quad \mathcal A_{\theta_\lambda}
    :=
    \mathcal A_{a_\lambda, b_\lambda, c_\lambda, d_\lambda}.
\end{align}

Set
\begin{align}
    p_\lambda:=a_\lambda+b_\lambda,
    \qquad
    q_\lambda:=c_\lambda+d_\lambda,
    \qquad
    t_\lambda:=p_\lambda+q_\lambda.
\end{align}

\begin{definition}[Residual no-EPR subspace]
\label{def:four-query-residual-no-epr-sector}
All operators below are specified by the query blocks on which they act. The residual no-EPR subspace is defined by
\begin{align}
    \nEspace{\theta_\lambda}
    &:={}
    \left(
        \bigcap_{i=1}^{a-\ell}
        \bigcap_{k=1}^{c-\ell}
        \ker\Phi_{i,k}^{U,U^*}
    \right)
    \cap
    \left(
        \bigcap_{j=1}^{b-m}
        \bigcap_{r=1}^{d-m}
        \ker\Phi_{j,r}^{U^T,U^\dagger}
    \right)
    \subseteq\mathcal A_{\theta_\lambda}.
\end{align}
The operators \(\Pi_{a-\ell,c-\ell}^{U,U^*}\) and \(\Pi_{b-m,d-m}^{U^T,U^\dagger}\) act on disjoint register pairs, and hence commute. 
Therefore,
\begin{align}
    \Pi_{\theta_\lambda}^{\mathrm{nE}}
    &:={}
    \Pi_{a-\ell,c-\ell}^{U,U^*}
    \Pi_{b-m,d-m}^{U^T,U^\dagger}
\end{align}
is the orthogonal projector onto
\(\nEspace{\theta_\lambda}\).
\end{definition}

Applying Lemma~\ref{lemma:orthogonal-epr-sector-coordinates} separately to the \(U,U^*\) and \(U^T,U^\dagger\) blocks gives the following sector projectors and coordinate maps for the full query space.

\begin{definition}[Double EPR-sector coordinate maps]
\label{def:four-query-epr-sector-coordinate-maps}
Regard \(P_\ell^{(a,c)}\) and \(P_m^{(b,d)}\) from Lemma~\ref{lemma:orthogonal-epr-sector-coordinates} as acting on the \(U,U^*\) and \(U^T,U^\dagger\) blocks, respectively, with identity operators on the remaining blocks understood. 
Set
\begin{align}
    P_\lambda
    &:={}
    P_\ell^{(a,c)}P_m^{(b,d)},
\end{align}
\begin{align}
    \mathcal M_\lambda
    &:={}
    \mathcal M_\ell^{(a,c)}
    \otimes
    \mathcal M_m^{(b,d)}.
    \label{eq:double-sector-multiplicity-space}
\end{align}
The partial isometry \(I_\lambda\) is defined directly in canonical query order. 
Fix orthonormal bases \(\{|r\rangle\}_r\) and \(\{|s\rangle\}_s\) of \(\mathcal M_\ell^{(a,c)}\) and \(\mathcal M_m^{(b,d)}\), respectively.
For arbitrary block vectors, expand the two maps as
\begin{align}
    I_\ell^{(a,c)}|x\rangle_U|z\rangle_*
    &=\sum_r |\psi_r(x,z)\rangle_{U,*}\otimes|r\rangle,\\
    I_m^{(b,d)}|y\rangle_T|w\rangle_\dagger
    &=\sum_s |\varphi_s(y,w)\rangle_{T,\dagger}\otimes|s\rangle.
\end{align}
Here
\begin{align*}
    |\psi_r(x,z)\rangle\in\nEspace{a-\ell,c-\ell},
    \qquad
    |\varphi_s(y,w)\rangle\in\nEspace{b-m,d-m}
\end{align*}
are generally unnormalized vectors.
Define
\begin{align*}
    I_\lambda:
    \mathcal A_{a,b,c,d}
    \longrightarrow
    \nEspace{\theta_\lambda}
    \otimes\mathcal M_\lambda
\end{align*}
by linearly extending the rule
\begin{align}
    I_\lambda
    \bigl(
        |x\rangle_U|y\rangle_T|z\rangle_*|w\rangle_\dagger
    \bigr)
    &:=
    \sum_{r,s}
    |\psi_r(x,z)\rangle_{U,*}
    \otimes|\varphi_s(y,w)\rangle_{T,\dagger}
    \otimes|r\rangle\otimes|s\rangle.
\end{align}
Thus \(I_\lambda\) applies \(I_\ell^{(a,c)}\) only to the \(U,U^*\) registers and \(I_m^{(b,d)}\) only to the \(U^T,U^\dagger\) registers. 
Register subscripts specify the placement of each factor.
The residual query registers are ordered as \(U,U^T,U^*,U^\dagger\), followed by the two multiplicity registers.
The rule extends linearly to arbitrary inputs, including inputs entangled across the four query blocks.
\end{definition}

\begin{lemma}[Exact double EPR-sector coordinates]
\label{prop:exact-double-epr-sector-coordinates}
The projectors and partial isometries above satisfy
\begin{align}
    P_\lambda P_{\lambda'}
    &=
    \delta_{\lambda,\lambda'}P_\lambda
    \quad
    \text{for all }\lambda,\lambda'\in\Lambda_\theta,
    \label{eq:double-sector-orthogonality}\\
    I_\lambda^\dagger I_\lambda
    &=P_\lambda,
    \quad
    I_\lambda I_\lambda^\dagger
    =
    \Pi_{\theta_\lambda}^{\mathrm{nE}}
    \otimes I_{\mathcal M_\lambda},
    \label{eq:double-sector-isometry-relations}\\
    \sum_{\lambda\in\Lambda_\theta}P_\lambda
    &=I_{\mathcal A_{a,b,c,d}}.
    \label{eq:double-sector-resolution}
\end{align}
Thus the initial spaces of the \(I_\lambda\) are mutually orthogonal.
Their final spaces are placed in mutually orthogonal summands of \(\widetilde{\mathcal A}_\theta\) below.
For every \(U\in\mathrm U(N)\), they also satisfy the exact equivariance identity
\begin{align}
    I_\lambda\U_\theta(U)
    &=
    \left(
        \U_{\theta_\lambda}(U)
        \otimes I_{\mathcal M_\lambda}
    \right)I_\lambda.
    \label{eq:double-sector-intertwining}
\end{align}
Moreover,
\begin{align}
    \dim\mathcal M_\lambda
    =
    \binom a\ell\binom c\ell\ell!
    \binom b m\binom d m m!.
\end{align}
Consequently,
\begin{align}
    \mathcal I_\theta
    &:={}
    \bigoplus_{\lambda\in\Lambda_\theta}I_\lambda
    :\mathcal A_{a,b,c,d}
    \longrightarrow
    \widetilde{\mathcal A}_\theta,
    \label{eq:global-sector-unitary}
\end{align}
\begin{align*}
    \widetilde{\mathcal A}_\theta
    &:={}
    \bigoplus_{\lambda\in\Lambda_\theta}
    \left(
        \nEspace{\theta_\lambda}
        \otimes\mathcal M_\lambda
    \right),
\end{align*}
is unitary.
\end{lemma}

\begin{proof}
    See Appendix~\ref{app:proof-double-epr-sector-coordinates}.
\end{proof}

Equation~\eqref{eq:double-sector-intertwining} shows that, within each EPR sector, the query acts only on the residual no-EPR registers.
For an arbitrary input operator \(X\) in the original query coordinates, its matrix in sector coordinates is
\begin{align}
    X_{\lambda,\lambda'}:=I_\lambda X I_{\lambda'}^\dagger.
\end{align}
For every \(U\), changing to sector coordinates and substituting~\eqref{eq:double-sector-intertwining} gives
\begin{align}
    \begin{aligned}
    \U_\theta(U)X\U_\theta(U)^\dagger
    &=
    \mathcal I_\theta^\dagger
    \bigl[\mathcal I_\theta\U_\theta(U)X\U_\theta(U)^\dagger\mathcal I_\theta^\dagger\bigr]
    \mathcal I_\theta
    \\
    &=
    \sum_{\lambda,\lambda'\in\Lambda_\theta}
    I_\lambda^\dagger
    \bigl(\U_{\theta_\lambda}(U)\otimes I_{\mathcal M_\lambda}\bigr)
    X_{\lambda,\lambda'}
    \bigl(\U_{\theta_{\lambda'}}(U)^\dagger\otimes I_{\mathcal M_{\lambda'}}\bigr)
    I_{\lambda'}.
    \end{aligned}
\end{align}
The residual query operators are understood on their no-EPR subspaces.
Averaging this identity over the sampled unitary describes the physical twirl on every block, including \(\lambda\neq\lambda'\).
The sum includes all off-diagonal terms \(\lambda\neq\lambda'\), so the unitary change of coordinates \(\mathcal I_\theta\) introduces no dephasing.
The multiplicity registers are untouched by the query action.

\subsection{Database Registers}
\label{sec:symmetric-relation-databases}
\label{sec:bijective-database-sectors}

The comparison between Haar and DPFC will use an auxiliary map that simulates the PF layers, called the parallel recorder, introduced in Section~\ref{sec:recorder-and-physical-environment}.
This subsection defines the auxiliary database registers used by parallel recorder, in which it stores pairs \((x,y)\), representing candidate constraints \(\pi(x)=y\) on the permutation. 
Both ordered and symmetric representations of these records will be used in the analysis.

More precisely, an edge is an ordered pair \(e=(x,y)\in[N]^2\). The coordinates \(x\) and \(y\) are called its \emph{domain label} and \emph{image label}, respectively. 
An edge register is
\(
    \mathcal E=\mathcal H_N\otimes\mathcal H_N,
\)
with \(|x,y\rangle_{\mathcal E}\) storing these labels in registers \(\mathcal E_{\mathrm X}\) and \(\mathcal E_{\mathrm Y}\), respectively. 
For \(s\geq0\), denote the ordered database space
\begin{align}
    \mathcal O_s:=\mathcal E^{\otimes s},
    \qquad
    \mathcal O_0:=\operatorname{span}\{|\varnothing\rangle\}.
\end{align}
The relation states below follow \cite{MH24}.

\begin{definition}[Relation databases]
\label{def:oriented-relation-databases}
A relation \(S\) of length \(k\) is a multiset of \(k\) edges.
Let \(m_e(S)\) denote the number of occurrences of \(e\) in \(S\), and let \(\mathcal R_k\) be the set of all length-\(k\) relations.
Writing \(S=\{e_1,\ldots,e_k\}\), with repetitions included, define the normalized relation state
\begin{align}
    |S\rangle
    &:=
    \frac{1}{\sqrt{k!\prod_e m_e(S)!}}
    \sum_{\sigma\in S_k}
    |e_{\sigma(1)},\ldots,e_{\sigma(k)}\rangle.
    \label{eq:oriented-normalized-relation-state}
\end{align}
The sum contains \(k!/\prod_e m_e(S)!\) distinct basis strings, each repeated \(\prod_e m_e(S)!\) times. 
Its squared norm is therefore \(k!\prod_e m_e(S)!\), so the prefactor in \eqref{eq:oriented-normalized-relation-state} makes \(|S\rangle\) a unit vector.
For any local dimension \(d\), let \(\Pi_{\mathrm{sym}}^{d,k}\) denote the orthogonal projector onto \(\operatorname{Sym}^k(\mathbb C^d)\), the subspace fixed by every permutation of the \(k\) tensor factors. 
In particular,
\begin{align}
    \Pi_{\mathrm{sym}}^{N^2,k}
    =\sum_{S\in\mathcal R_k}|S\rangle\langle S|.
\end{align}
The length-\(k\) database space is given by
\begin{align}
    \mathcal D_k
    :=
    \operatorname{Im}\bigl(\Pi_{\mathrm{sym}}^{N^2,k}\bigr)
    =
    \operatorname{Sym}^k
    \bigl(\mathcal H_N\otimes\mathcal H_N\bigr),
    \qquad
    \mathcal D_0
    :=
    \operatorname{span}\{|\varnothing\rangle\}.
\end{align}
The space \(\mathcal D_k\) is a subspace of \(\mathcal O_k\). 
The later construction will use two variable-length databases to distinguish positive phase contributions from the \(U,U^T\) queries and negative phase contributions from the \(U^*,U^\dagger\) queries:
\begin{align}
    \mathcal D_+
    :=
    \bigoplus_{k\geq0}\mathcal D_k,
    \qquad
    \mathcal D_-
    :=
    \bigoplus_{k\geq0}\mathcal D_k.
\end{align}
For a query pattern \(\theta=(a,b,c,d)\), the \(+\) and \(-\) databases therefore contain \(a+b\) and \(c+d\) records, respectively.
\end{definition}

Distinct relations have disjoint computational-basis support. 
Hence \(\{|S\rangle:S\in\mathcal R_k\}\) is an orthonormal basis of \(\mathcal D_k\).

The following definition is adapted from the bijective relations and projectors of \cite{MH24}. 
A \emph{collision} occurs when two records, within either database or across the two databases, share a domain label or share an image label.
When there are no collisions, the records specify a bijection between the recorded domain labels and the recorded image labels.
The subspace spanned by the corresponding database states is called the \emph{bijective subspace}.
On this subspace, the later comparison matches the recorder map to the physical permutation-phase layer, up to a normalization factor.

Here and below, \(\uplus\) denotes multiset union.
The query pattern here fixes the two database sizes, and the definition is extended to the corresponding projector on the ordered database space.

\begin{definition}[Bijective signed relations]
\label{def:bijective-signed-relations}
For a query pattern \(\theta=(a,b,c,d)\) of total size \(t\), let
\begin{align}
    \mathcal B_{\theta}
    &:=
    \left\{
        (\Splus,\Sminus)\in
        \mathcal R_{a+b}\times\mathcal R_{c+d}:
        \Splus\uplus \Sminus
        \text{ is an injective partial permutation}
    \right\}.
\end{align}
Equivalently, \((\Splus,\Sminus)\in\mathcal B_{\theta}\) exactly when the \(t\) domain labels in the two relations are pairwise distinct, and their \(t\) image labels are pairwise distinct. 
The set \(\mathcal B_\theta\) depends only on the
database sizes \((a+b,c+d)\). 
Different query patterns with the same sizes therefore share the same signed relations and the same database states \(|\Splus,\Sminus\rangle\), which do not include a pattern label.

Define the corresponding database projector by
\begin{align}
    \Pi_{\theta}^{\mathrm{bij}}
    &:={}
    \sum_{(\Splus,\Sminus)\in\mathcal B_{\theta}}
    |\Splus,\Sminus\rangle\!\langle \Splus,\Sminus|.
    \label{eq:bijective-signed-database-projector}
\end{align}

On the ordered database space \((\mathcal O_{a+b})_+\otimes(\mathcal O_{c+d})_-\), define the ordered bijective projector by
\begin{align}
    \Pi_{\theta}^{\mathrm{bij,ord}}
    :=
    \sum_{\substack{
        \boldsymbol x,\boldsymbol y\in[N]^{t}\\
        x_i\neq x_j,\ y_i\neq y_j\ \text{for }i\neq j
    }}
    \bigotimes_{i=1}^{t}
    |x_i,y_i\rangle\!\langle x_i,y_i|.
\end{align}
\end{definition}

Reordering the edges within either database preserves bijectivity.
Hence, \(\Pi_{\theta}^{\mathrm{bij,ord}}\) commutes with the projector onto the symmetric subspace of each database. 
Its restriction to the symmetric database space is
\begin{align}
    \left.
        \Pi_{\theta}^{\mathrm{bij,ord}}
    \right|_{(\mathcal D_{a+b})_+
             \otimes(\mathcal D_{c+d})_-}
    =\Pi_{\theta}^{\mathrm{bij}}.
\end{align}

The notation \(\mathcal B_{\theta_\lambda}\), \(\Pi_{\theta_\lambda}^{\mathrm{bij}}\), and \(\Pi_{\theta_\lambda}^{\mathrm{bij,ord}}\) refers to Definition~\ref{def:bijective-signed-relations} with \(\theta\) replaced by \(\theta_\lambda\). 
The two database sizes are then \(p_\lambda\) and \(q_\lambda\).

\section{A Parallel Recorder for the Permutation-Phase Action}
\label{sec:recorder-and-physical-environment}

The \emph{parallel recorder} defined in this section is adapted from \cite{MH24} for parallel queries.
To analyze the diamond distance between the DPFC and Haar ensembles the parallel recorder is used within a hybrid argument.
First, start with DPFC with Clifford endpoints. 
The actual permutation and phase layers are replaced by the recorder without changing \(C,D\). 
Then, the endpoint distribution is replaced from Clifford to Haar and it is shown that this does not change the induced map.
Last, the recorder is replaced again with the physical PF layers, and this reproduces the Haar ensemble due to Haar invariance.
The two steps where the recorder and the PF layers are replaced with one another incur the approximation errors bounded in this section.

One may think of the recorder as implementing a coherent version of lazy sampling.\footnote{For quantum lazy sampling via compressed oracles, see \cite{Z19}.}
For an input \(|x\rangle\), it simulates the PF layers by generating a superposition of candidate constraints of the form \(\pi(x)=y\) as edges \((x,y)\in[N]^2\) in an auxiliary database.
Two database registers distinguish the phase contributions. 
Queries of type \(U,U^T\) contribute phases with positive sign, whereas queries of type \(U^*,U^\dagger\) contribute their complex conjugates, with negative sign. 
The corresponding edges are stored in the \(+\) and \(-\) databases, respectively. 
Each database is symmetrized, which represents the fact that the order of queries does not matter.
The answers are generated independently, so that two different inputs may have the same output with small probability.
Hence, after tracing out the database registers only an approximation of the average over all possible permutations and phases is obtained.
The recorder map is defined separately in each EPR sector on the residual queries only. 
Its comparison with the physical channel is restricted to \(\nEspace{\theta_\lambda}\).

Next, to actually bound the difference between DPFC, with the physical permutation and phase actions, and an operator where the PF layers are replaced by the recorder, a purification of the PF layers is introduced.
Its environment is mapped isometrically into the recorder's database space, placing both maps in a common environment shared by all query patterns and EPR sectors. 
Their bijective components (Subsection~\ref{sec:bijective-database-sectors}) agree up to a normalization factor. 
The subsequent analysis bounds the weight outside this subspace after averaging over the endpoint unitaries. 
Together with a bound on the normalization of the recorder, these estimates bound the diamond-norm distance between the physical DPFC channel and the map induced by the recorder.

\subsection{The Ordered Recorder}
\label{sec:ordered-four-query-recorder}

The construction begins with an \emph{ordered} version of the recorder, using the ordered database spaces from Section~\ref{sec:symmetric-relation-databases}.
For a forward query on a computational-basis state \(|x\rangle\), the recorder produces a uniform superposition over output labels \(y\), and stores the corresponding edge \((x,y)\) in the database.
For a reverse query, it similarly generates a superposition over \(x\) labels given \(|y\rangle\) and stores the corresponding edge. 
As mentioned, these answers are generated independently, allowing different inputs to receive the same output and the other way around. 
Thus, the ordered recorder does not give an exact purification of the permutation-phase average. 
The later comparison bounds the contribution of such collisions.

\begin{definition}[Ordered one-query recorders]
\label{def:ordered-one-query-recorders}
The two maps \(E_{\mathrm L}^{\mathrm{ord}},E_{\mathrm R}^{\mathrm{ord}}: \mathcal H_N\to\mathcal H_N\otimes\mathcal E\) are defined by
\begin{align}
    E_{\mathrm L}^{\mathrm{ord}}|x\rangle
    &:=
    \frac{1}{\sqrt N}
    \sum_{y\in[N]}
    |y\rangle_Q|x,y\rangle_{\mathcal E}
    =|x\rangle_{\mathcal E_{\mathrm X}}\otimes
    |\Omega_N\rangle_{Q,\mathcal E_{\mathrm Y}},
    \label{eq:ordered-left-map}\\
    E_{\mathrm R}^{\mathrm{ord}}|y\rangle
    &:=
    \frac{1}{\sqrt N}
    \sum_{x\in[N]}
    |x\rangle_Q|x,y\rangle_{\mathcal E}
    =|\Omega_N\rangle_{Q,\mathcal E_{\mathrm X}}\otimes
    |y\rangle_{\mathcal E_{\mathrm Y}}.
    \label{eq:ordered-right-map}
\end{align}
\end{definition}

The \(\mathrm L\)-map supplies the domain label and generates the image label. 
The \(\mathrm R\)-map supplies the image label and generates the domain label. 
Both output the generated label to the query register. 
Equivalently, \(E_{\mathrm L}^{\mathrm{ord}}\) places the input in \(\mathcal E_{\mathrm X}\) and prepares the normalized EPR state \(|\Omega_N\rangle_{Q,\mathcal E_{\mathrm Y}}\), while \(E_{\mathrm R}^{\mathrm{ord}}\) places the input in \(\mathcal E_{\mathrm Y}\) and prepares \(|\Omega_N\rangle_{Q,\mathcal E_{\mathrm X}}\).

\begin{definition}[Ordered parallel recorder for four query types]
\label{def:ordered-parallel-four-query-recorder}
Define the ordered recorder
\begin{align*}
    E_{\theta_\lambda}^{\mathrm{ord}}:
    \mathcal A_{\theta_\lambda}
    \longrightarrow
    \mathcal A_{\theta_\lambda}
    \otimes(\mathcal O_{p_\lambda})_+
    \otimes(\mathcal O_{q_\lambda})_-
\end{align*}
using the one-query recorders:
\begin{align}
    E_{\theta_\lambda}^{\mathrm{ord}}
    &:=
    \Gamma_{\theta_\lambda}
    \left[
        \bigl(E_{\mathrm L}^{\mathrm{ord}}\bigr)^{\otimes a_\lambda}
        \otimes
        \bigl(E_{\mathrm R}^{\mathrm{ord}}\bigr)^{\otimes b_\lambda}
        \otimes
        \bigl(E_{\mathrm L}^{\mathrm{ord}}\bigr)^{\otimes c_\lambda}
        \otimes
        \bigl(E_{\mathrm R}^{\mathrm{ord}}\bigr)^{\otimes d_\lambda}
    \right].
\end{align}
Here \(\Gamma_{\theta_\lambda}\) reorders the tensor-product output by placing all query outputs first and all edge registers second, preserving the canonical order in each group.

The edge registers associated with the \(U,U^T\) query slots form the \emph{positive ordered database}, and those associated with the \(U^*,U^\dagger\) slots form the \emph{negative ordered database}.
Within each database, the edge slots remain in the order of their creating queries. 
The phase interpretation of these signs will be explained in Section~\ref{sec:parallel-four-query-recorder}.
Empty tensor products act on the one-dimensional empty database.
\end{definition}

Notice that by construction,
\begin{align}
    \bigl(E_{\theta_\lambda}^{\mathrm{ord}}\bigr)^\dagger
    E_{\theta_\lambda}^{\mathrm{ord}}
    &=I_{\mathcal A_{\theta_\lambda}}.
    \label{eq:ordered-recorder-isometry}
\end{align}

\subsection{The Parallel Recorder}
\label{sec:parallel-four-query-recorder}

As mentioned, the parallel recorder is an object designed to simulate the permutation and phase operators parallel actions. 
Within each signed database, the permutation constraints and accumulated phase depend only on the recorded relation, not on the ordering of its edges. 
The ordered recorder nevertheless distinguishes these orderings.
Symmetrizing the databases combines all constraints coherently into relation states, allowing comparison with the purification of the random permutation and phase operations constructed in subsection~\ref{sec:physical-permutation-phase-layer}.

\begin{definition}[Parallel recorder for four query types]
\label{def:parallel-four-query-recorder}
Define the unnormalized four-query symmetric recorder
\begin{align*}
    E_{\theta_\lambda}:
    \mathcal A_{\theta_\lambda}
    \longrightarrow
    \mathcal A_{\theta_\lambda}
    \otimes(\mathcal D_{p_\lambda})_+
    \otimes(\mathcal D_{q_\lambda})_-
\end{align*}
is obtained by symmetrizing the two ordered databases:
\begin{align}
    E_{\theta_\lambda}
    &:=
    \left(
        I_{\mathcal A_{\theta_\lambda}}
        \otimes\sqrt{p_\lambda!}\,
        \Pi_{\mathrm{sym}}^{N^2,p_\lambda}
        \otimes\sqrt{q_\lambda!}\,
        \Pi_{\mathrm{sym}}^{N^2,q_\lambda}
    \right)
    E_{\theta_\lambda}^{\mathrm{ord}}.
    \label{eq:recorder-from-ordered-maps}
\end{align}

The symmetrizers act on the ordered database spaces, and their images are identified with the symmetric database spaces.
\end{definition}

For an ordered list \((e_1,\ldots,e_k)\) with underlying multiset
\(S\), Definition~\ref{def:oriented-relation-databases} gives
\begin{align}
    \sqrt{k!}\,
    \Pi_{\mathrm{sym}}^{N^2,k}|e_1,\ldots,e_k\rangle
    &=
    \sqrt{\prod_e m_e(S)!}\,|S\rangle.
    \label{eq:ordered-edges-to-relation-state}
\end{align}
The factors \(\sqrt{p_\lambda!}\) and \(\sqrt{q_\lambda!}\) in Equation~\eqref{eq:recorder-from-ordered-maps} are chosen so that each ordered list of distinct edges is mapped to its normalized relation state with coefficient one. 
Indeed, a list of \(k\) distinct edges has \(k!\) orthogonal orderings, so the state \(|S\rangle/\sqrt{k!}\) is obtained after projection onto the symmetric subspace. 
Multiplying by \(\sqrt{k!}\) cancels this reduction. 
Since bijective relations have distinct edges, this scaling preserves the amplitude contributed by each ordered history on the sector used for the later comparison. 
When edges repeat, Equation~\eqref{eq:ordered-edges-to-relation-state} instead introduces the multiplicity factor \(\sqrt{\prod_e m_e(S)!}\), so the recorder is not an isometry.

The database signs record the phase contributions of the middle-layer query actions. 
For an edge \((x,y)\), the \(U,U^T\) slots contribute \(\omega_3^{f(x)}\), while the \(U^*,U^\dagger\) slots contribute \(\omega_3^{-f(x)}\). 
Their edges are stored in \(\mathcal D_+\) and \(\mathcal D_-\), respectively. 
The resulting signed relation will later identify the recorder database with the physical permutation-phase environment on collision-free histories.
In \cite{MH24} the two databases are labeled \(\mathsf L,\mathsf R\).
Here \(+,-\) indicate the phase sign, while \(\mathrm L,\mathrm R\) indicate the supplied label.\footnote{In \cite{MH24} they are the same thing, they consider only forward and inverse queries.}

The symmetric recorder acts explicitly as follows. On input
\(
    |\boldsymbol x\rangle_U
    |\boldsymbol y\rangle_T
    |\boldsymbol x'\rangle_*
    |\boldsymbol y'\rangle_\dagger
\), sum over
\(
    \boldsymbol u\in[N]^{a_\lambda}\), \(
    \boldsymbol v\in[N]^{b_\lambda}\), \(
    \boldsymbol u'\in[N]^{c_\lambda}\), \(
    \boldsymbol v'\in[N]^{d_\lambda}
\).
Define the two recorded relations
\begin{align*}
    \Splus
    &:=
    \{(x_i,u_i):1\leq i\leq a_\lambda\}
    \uplus
    \{(v_i,y_i):1\leq i\leq b_\lambda\},\\
    \Sminus
    &:=
    \{(x_i',u_i'):1\leq i\leq c_\lambda\}
    \uplus
    \{(v_i',y_i'):1\leq i\leq d_\lambda\}.
\end{align*}
Expanding the ordered recorder in \eqref{eq:recorder-from-ordered-maps} and applying \eqref{eq:ordered-edges-to-relation-state} to each database gives
\begin{align}
    & E_{\theta_\lambda}
    \bigl(
        |\boldsymbol x\rangle_U
        |\boldsymbol y\rangle_T
        |\boldsymbol x'\rangle_*
        |\boldsymbol y'\rangle_\dagger
    \bigr)
    = \nonumber\\
    & \qquad
    \frac{1}{N^{t_\lambda/2}}
    \sum_{\boldsymbol u,\boldsymbol v,\boldsymbol u',\boldsymbol v'}
    \sqrt{
        \prod_e m_e(\Splus)!\,
        \prod_e m_e(\Sminus)!
    }\,
    |\boldsymbol u\rangle_U
    |\boldsymbol v\rangle_T
    |\boldsymbol u'\rangle_*
    |\boldsymbol v'\rangle_\dagger
    |\Splus\rangle_+
    |\Sminus\rangle_-.
    \label{eq:parallel-four-query-recorder-basis-action}
\end{align}

Equation~\eqref{eq:four-query-recorder-sequential-composition} in Appendix~\ref{app:four-query-gram-analysis} shows that this parallel recorder is exactly a sequential application, starting from empty databases, of the one-query edge-appending maps of ~\cite{MH24}, restated here in Definition~\ref{def:symmetric-one-edge-recorders}. 

\subsection{Unitary Invariance}

To bound the diamond distance between the map induced by the recorder with the actual DPFC ensemble, the contribution of endpoint unitaries \(C, D\) must also be taken into account. 
The identity below shows that applying these endpoints around the recorder is equivalent to rotating only its database registers. 
After these registers are traced out, the rotation is canceled, so the recorder induces the same map on the query registers for every choice of \(C,D\). 
The recorder therefore provides an intermediate hybrid between Haar and DPFC, independent of the endpoint distribution. 
Bounding the difference to this map for Clifford and Haar endpoints, then bounds the distance between the two ensembles by the triangle inequality.

The approximation argument compares the maps before tracing out the databases, where the rotations still depend on \(C, D\). 
Averaging over these endpoint unitaries is essential for bounding the weight of inputs where the recorder fails to simulate a permutation properly.

The following definition is adapted from \cite{MH24} to fixed database lengths and the symmetric relation subspaces.

\begin{definition}[Signed database rotation]
\label{def:signed-database-rotation}
For \(C,D\in\mathrm U(N)\), define the one-edge rotations
\begin{align}
    W_+[C,D]
    :=
    C\otimes D^T,
    \qquad
    W_-[C,D]
    :=
    C^*\otimes D^\dagger.
\end{align}
Define the rotation on the ordered databases by
\begin{align}
    \mathcal Q_{\theta_\lambda}^{\mathrm{ord}}[C,D]
    &:=
    W_+[C,D]^{\otimes p_\lambda}
    \otimes W_-[C,D]^{\otimes q_\lambda}.
    \label{eq:ordered-signed-database-rotation}
\end{align}
The tensor powers preserve the symmetric database spaces. 
Their restriction defines the database rotation on the output space of \(E_{\theta_\lambda}\) by
\begin{align}
    \mathcal Q_{\theta_\lambda}[C,D]
    &:=
    \left.
        W_+[C,D]^{\otimes p_\lambda}
    \right|_{(\mathcal D_{p_\lambda})_+}
    \otimes
    \left.
        W_-[C,D]^{\otimes q_\lambda}
    \right|_{(\mathcal D_{q_\lambda})_-}.
\end{align}
\end{definition}

\begin{lemma}[Unitary invariance of the four-type recorder]
\label{lem:four-type-endpoint-covariance}
For every \(C,D\in\mathrm U(N)\),
\begin{align}
    &\left[
        \left(
            D^{\otimes a_\lambda}
            \otimes(C^T)^{\otimes b_\lambda}
            \otimes(D^*)^{\otimes c_\lambda}
            \otimes(C^\dagger)^{\otimes d_\lambda}
        \right)
        \otimes I
    \right]
    E_{\theta_\lambda}
    \nonumber\\
    &\qquad\cdot
    \left(
        C^{\otimes a_\lambda}
        \otimes(D^T)^{\otimes b_\lambda}
        \otimes(C^*)^{\otimes c_\lambda}
        \otimes(D^\dagger)^{\otimes d_\lambda}
    \right)
    \nonumber\\
    &\quad=
    \left(
        I_{\mathcal A_{\theta_\lambda}}
        \otimes\mathcal Q_{\theta_\lambda}[C,D]
    \right)
    E_{\theta_\lambda}.
    \label{eq:four-type-endpoint-covariance}
\end{align}
The same identity holds with \(E_{\theta_\lambda}\) and \(\mathcal Q_{\theta_\lambda}[C,D]\) replaced by \(E_{\theta_\lambda}^{\mathrm{ord}}\) and \(\mathcal Q_{\theta_\lambda}^{\mathrm{ord}}[C,D]\), respectively.
\end{lemma}

\begin{proof}
    The proof is given in Appendix~\ref{app:proof-endpoint-covariance}.
\end{proof}

\subsection{Gram Operator and Norm Bound}
\label{sec:four-query-gram-control}

Symmetrization can change norms, so the parallel recorder is not generally an isometry. 
The Gram operator introduced in this subsection measures this norm distortion.
The subsequent bound limits its possible increase in squared norm. 
This will be used to bound the error when comparing the recorder with the purification of the random permutation and phase operations constructed in Section~\ref{sec:physical-permutation-phase-layer}.

Most of the analysis of the Gram operator's properties is deferred to Appendix~\ref{app:four-query-gram-analysis}, where the upper bound is proved. 
In particular, Lemma~\ref{lem:four-query-common-bijective-gram} in Appendix~\ref{app:common-bijective-gram} shows that the two ordered and symmetric recorders have the same bijective contribution to their Gram operators, so their deviation from one another is determined by the non-bijective sector.

\begin{definition}[Four-query Gram operator]
\label{def:four-query-gram-operator}
The Gram operator of the residual recorder is
\begin{align}
    G_{\theta_\lambda}
    &:=
    E_{\theta_\lambda}^\dagger E_{\theta_\lambda}.
    \label{eq:four-query-gram-operator}
\end{align}
\end{definition}

\begin{lemma}[Gram upper bound]
\label{lem:four-query-gram-upper-bound}
On \(\mathcal A_{\theta_\lambda}\),
\begin{align}
    G_{\theta_\lambda}-I
    &\preceq
    \left[
        \exp\!\left(
            \frac{p_\lambda(p_\lambda-1)+q_\lambda(q_\lambda-1)}{2N}
        \right)-1
    \right]I
    \preceq \left(e^{t_\lambda^2/(2N)}-1\right)I,
    \label{eq:four-query-gram-upper-bound}\\
    \left\|E_{\theta_\lambda}\right\|_\infty^2
    &= \left\|G_{\theta_\lambda}\right\|_\infty
    \leq e^{t_\lambda^2/(2N)}.
    \label{eq:four-query-recorder-norm-bound}
\end{align}
\end{lemma}

\begin{proof}
    See Appendix~\ref{app:direct-gram-upper-bound}.
\end{proof}

\subsection{Purification of the Permutation-Phase Actions}
\label{sec:physical-permutation-phase-layer}

The physical purification stores the random choices \((\pi,f)\) coherently in an auxiliary environment. 
Each pair labels an orthogonal state and determines the permutation-phase action. 
The queries act on all slots controlled on this environment register.
Discarding this environment recovers the averaged permutation-phase channel. 
This representation allows the physical environment states corresponding to bijective relations to be matched with the recorder's database states as the following subsections show.

\begin{definition}[Physical permutation-phase environment]
\label{def:physical-permutation-phase-environment}
Let \(\mathcal E_{\mathrm{PF}}\) be the physical permutation-phase environment Hilbert space, defined by the following orthonormal
basis
\begin{align}
    \left\{
        |\pi,f\rangle:
        \pi\in S_N,\quad f:[N]\to\mathbb Z_3
    \right\}.
\end{align}
\end{definition}

The following is the four-query, fixed-sector version of the purified permutation function oracle of~\cite{MH24}.

\begin{definition}[Physical permutation-phase isometry for four query types]
\label{def:physical-four-query-pf-isometry}
Define the map \(K_{\theta_\lambda}^{\mathrm{PF}}:\mathcal A_{\theta_\lambda}\to\mathcal A_{\theta_\lambda}\otimes\mathcal E_{\mathrm{PF}}\) by
\begin{align}
    & K_{\theta_\lambda}^{\mathrm{PF}}|\psi\rangle
    := \notag \\
    &\qquad
    \frac{1}{\sqrt{N!\,3^N}}
    \sum_{\pi\in S_N}
    \sum_{f:[N]\to\mathbb Z_3}
    \Bigl(
        M_{\pi,f}^{\otimes a_\lambda}
        \otimes
        (M_{\pi,f}^{T})^{\otimes b_\lambda}
        \otimes
        (M_{\pi,f}^*)^{\otimes c_\lambda}
        \otimes
        (M_{\pi,f}^{\dagger})^{\otimes d_\lambda}
    \Bigr)
    |\psi\rangle|\pi,f\rangle.
\end{align}
\end{definition}

The environment labels are orthonormal, and the operator acting on the query registers in every summand is unitary. 
Hence,
\begin{align}
    \bigl(K_{\theta_\lambda}^{\mathrm{PF}}\bigr)^\dagger
    K_{\theta_\lambda}^{\mathrm{PF}}
    =I,
\end{align}
meaning that \(K_{\theta_\lambda}^{\mathrm{PF}}\) is an isometry. 
Discarding its environment gives the physical permutation-phase channel \(\Phi_{\theta_\lambda}^{\mathrm{PF}}:\mathsf L(\mathcal A_{\theta_\lambda})\to\mathsf L(\mathcal A_{\theta_\lambda})\), which is restricted to inputs in the residual no-EPR subspace \(\nEspace{\theta_\lambda}\) for the comparison with the recorder:
\begin{align}
    \Phi_{\theta_\lambda}^{\mathrm{PF}}
    &:={}
    \operatorname{Tr}_{\mathcal E_{\mathrm{PF}}}
    \circ
    \operatorname{Ad}_{K_{\theta_\lambda}^{\mathrm{PF}}}.
    \label{eq:physical-four-query-pf-channel}
\end{align}

\subsection{Physical Signed-Relation States}

To compare the physical purification with the recorder, this subsection identifies the physical environment states corresponding to bijective database states. 
Mapping these states to the recorder's relation states will place the bijective components of both maps in the same space.

More concretely, for any query pattern \(\theta\in\Theta_t\) and EPR sector \(\lambda\in\Lambda_\theta\), every pair in \(\mathcal B_{\theta_\lambda}\) fixes an injective map on \(t_\lambda\) points and therefore has exactly \((N-t_\lambda)!\) extensions to a permutation of \([N]\). 
The physical relation state defined below groups these extensions into a coherent superposition, with amplitudes encoding the phase contributions. 
This state is later mapped to the corresponding recorder database state.
The next definition is the signed, fixed-sector version of the physical relation states of \cite{MH24}.

\begin{definition}[Physical signed-relation state]
\label{def:physical-signed-relation-state}
For \((\Splus,\Sminus)\in\mathcal B_{\theta_\lambda}\), define
\begin{align}
    |\pf_{\Splus,\Sminus}\rangle
    &:={}
    \frac{1}{\sqrt{(N-t_\lambda)!\,3^N}}
    \sum_{\substack{
        \pi\in S_N\\
        \pi\supseteq \Splus\uplus \Sminus
    }}
    \sum_{f:[N]\to\mathbb Z_3}
    \omega_3^{
        \sum_{(x,y)\in \Splus}f(x)
        -
        \sum_{(x,y)\in \Sminus}f(x)
    }
    |\pi,f\rangle.
\end{align}
Here \(\pi\supseteq \Splus\uplus \Sminus\) means that \(\pi(x)=y\) for every edge \((x,y)\) in either relation. 
Positive-database edges carry the phase character \(+f(x)\), while negative-database edges carry \(-f(x)\). 
Since \(t_\lambda=|\Splus|+|\Sminus|\), the state depends only on the signed relation \((\Splus,\Sminus)\), not on a query pattern or EPR-sector label. 
Thus, the same relation occurring in different patterns or sectors always defines the same physical state.
\end{definition}

\begin{lemma}[Orthonormality of physical signed-relation states]
\label{lem:physical-signed-relation-orthonormality}
For any \(\theta,\theta'\in\Theta_t\), \(\lambda\in\Lambda_\theta\), \(\lambda'\in\Lambda_{\theta'}\), and relations \((\Splus,\Sminus)\in\mathcal B_{\theta_\lambda}\) and \((T_+,T_-)\in\mathcal B_{\theta'_{\lambda'}}\),
\begin{align}
    \langle\pf_{T_+,T_-}
    \mid\pf_{\Splus,\Sminus}\rangle
    &=
    \begin{cases}
        1,
        &(T_+,T_-)=(\Splus,\Sminus),\\
        0,
        &\text{otherwise}.
    \end{cases}
    \label{eq:physical-signed-relation-orthonormality}
\end{align}
\end{lemma}

\begin{proof}
    The claim was proven in \cite{MH24}.
\end{proof}

The orthonormality permits the physical environment to be compressed into the bijective sector of the two signed databases.
The following compression is adapted from \cite{MH24}, restricted to the signed relations of a fixed residual query pattern.

\begin{definition}[Physical-environment compression]
\label{def:physical-four-query-environment-compression}
Define \(\mathsf{Comp}_{\theta_\lambda}:\mathcal E_{\mathrm{PF}}\to(\mathcal D_{p_\lambda})_+\otimes(\mathcal D_{q_\lambda})_-\) by
\begin{align}
    \mathsf{Comp}_{\theta_\lambda}
    &:={}
    \sum_{(\Splus,\Sminus)\in\mathcal B_{\theta_\lambda}}
    |\Splus,\Sminus\rangle
    \langle\pf_{\Splus,\Sminus}|.
\end{align}
\end{definition}

Lemma~\ref{lem:physical-signed-relation-orthonormality} implies
\begin{align}
    \mathsf{Comp}_{\theta_\lambda}
    \mathsf{Comp}_{\theta_\lambda}^\dagger
    =
    \Pi_{\theta_\lambda}^{\mathrm{bij}}.
\end{align}

\begin{lemma}[Exact bijective-sector matching for four query types]
For all \(\lambda \in \Lambda_\theta\), 
\label{lem:exact-four-query-bijective-sector-matching}
\begin{align}
    \left(
        I_{\mathcal A_{\theta_\lambda}}
        \otimes\mathsf{Comp}_{\theta_\lambda}
    \right)
    K_{\theta_\lambda}^{\mathrm{PF}}
    &=
    \sqrt{
        \frac{
            N^{t_\lambda}(N-t_\lambda)!
        }{N!}
    }
    \left(
        I_{\mathcal A_{\theta_\lambda}}
        \otimes\Pi_{\theta_\lambda}^{\mathrm{bij}}
    \right)
    E_{\theta_\lambda}.
    \label{eq:exact-four-query-bijective-sector-matching}
\end{align}
\end{lemma}

This lemma shows that the bijective component of the physical isometry agrees exactly with the corresponding component of the unnormalized recorder, up to the extension-counting factor.

\begin{proof}
See Appendix~\ref{app:physical-signed-relation-matching}.
\end{proof}

\subsection{A Common Environment}

The permutation-phase environment \(\mathcal E_{\mathrm{PF}}\) is common to all query patterns and EPR sectors. 
To preserve coherences between them, extend the prescribed relation-state identification to one isometry on the entire physical environment, such that a given environment state is mapped to the same database state regardless of the query pattern or EPR sector with which it is entangled. 
This way, discarding the environment gives the same query state as before, even for superpositions of different patterns and sectors.

\begin{definition}[Isometric extension of physical-environment compression]
\label{def:four-query-isometric-compression-extension}
For a fixed query budget \(t\), choose and fix an isometry
\begin{align}
    \widetilde{\mathsf{Comp}}_t:
    \mathcal E_{\mathrm{PF}}
    \longrightarrow
    \mathcal D_+\otimes\mathcal D_-
\end{align}
with the following properties,
\begin{align}
    \widetilde{\mathsf{Comp}}_t
    |\pf_{\Splus,\Sminus}\rangle
    &=
    |\Splus,\Sminus\rangle,
    \qquad
    \forall (\Splus,\Sminus)\in
    \bigcup_{\theta\in\Theta_t}
    \bigcup_{\lambda\in\Lambda_\theta}
    \mathcal B_{\theta_\lambda},
    \nonumber\\
    \widetilde{\mathsf{Comp}}_t^{\dagger}
    \widetilde{\mathsf{Comp}}_t
    &=
    I_{\mathcal E_{\mathrm{PF}}}.
    \label{eq:four-query-isometric-compression-extension}
\end{align}
In addition, it maps the orthogonal complement of the span of these physical relation states into database bidegrees \((r,s)\) with \(r+s>t\). 
The same choice of \(\widetilde{\mathsf{Comp}}_t\) is used for every \(\theta\in\Theta_t\) and \(\lambda\in\Lambda_\theta\).
\end{definition}

Such an isometry exists. After identifying repeated occurrences of the same signed relation, Lemma~\ref{lem:physical-signed-relation-orthonormality} gives one orthonormal family across all query patterns and residual EPR sectors.
The corresponding database states form an orthonormal family as well, so the identification defines an isometry on their span.
It extends isometrically to all of \(\mathcal E_{\mathrm{PF}}\), with the orthogonal complement mapped into bidegrees \(r+s>t\), because the physical environment is finite-dimensional and the two databases have unbounded relation length. 
These degrees are unused by every recorder dilation, since each residual size \(t_\lambda=p_\lambda+q_\lambda\) is at most \(t\).

Applying this isometry to the discarded environment leaves its partial trace unchanged on every operator, including off-diagonal blocks between distinct query patterns or EPR sectors. This motivates the following definition,

\begin{definition}[Common-environment physical isometry]
\label{def:four-query-common-environment-physical-isometry}
Define \(\widetilde K_{\theta_\lambda}^{\mathrm{PF}}:\mathcal A_{\theta_\lambda}\to\mathcal A_{\theta_\lambda}\otimes\mathcal D_+\otimes\mathcal D_-\) by
\begin{align}
    \widetilde K_{\theta_\lambda}^{\mathrm{PF}}
    &:={}
    \left(
        I_{\mathcal A_{\theta_\lambda}}
        \otimes
        \widetilde{\mathsf{Comp}}_t
    \right)
    K_{\theta_\lambda}^{\mathrm{PF}}.
\end{align}
\end{definition}

Equation~\eqref{eq:four-query-isometric-compression-extension} implies that \(\widetilde K_{\theta_\lambda}^{\mathrm{PF}}\) is an isometry. 
Because an isometry applied only to a discarded environment does not change the induced channel,
\begin{align}
    \Phi_{\theta_\lambda}^{\mathrm{PF}} = 
    \operatorname{Tr}_{\mathcal D_+\mathcal D_-}
    \circ
    \operatorname{Ad}_{\widetilde K_{\theta_\lambda}^{\mathrm{PF}}}.
    \label{eq:four-query-common-environment-physical-channel}
\end{align}
Thus \(\widetilde K_{\theta_\lambda}^{\mathrm{PF}}\) takes values in the same query-output and database environment spaces as \(E_{\theta_\lambda}\). 
Moreover, projecting the physical isometry onto the recorder's database bidegree gives
\begin{align}
    \left(
            I_{\mathcal A_{\theta_\lambda}}
            \otimes\Pi_{\theta_\lambda}^{\mathrm{bij}}
        \right)
        \widetilde K_{\theta_\lambda}^{\mathrm{PF}}
    = \left(
            I_{\mathcal A_{\theta_\lambda}}
            \otimes\mathsf{Comp}_{\theta_\lambda}
        \right)
        K_{\theta_\lambda}^{\mathrm{PF}}
    = \sqrt{
            \frac{N^{t_\lambda}(N-t_\lambda)!}{N!}
        }
        \left(
            I_{\mathcal A_{\theta_\lambda}}
            \otimes\Pi_{\theta_\lambda}^{\mathrm{bij}}
        \right)
        E_{\theta_\lambda}.
    \label{eq:four-query-common-environment-bijective-component}
\end{align}
Indeed, by construction,
\begin{align}
    \Pi_{\theta_\lambda}^{\mathrm{bij}}
    \widetilde{\mathsf{Comp}}_t
    =
    \mathsf{Comp}_{\theta_\lambda},
\end{align}
which gives the first equality. The second uses Lemma~\ref{lem:exact-four-query-bijective-sector-matching}.

Equation~\eqref{eq:four-query-common-environment-bijective-component} shows that, for every input, the physical purification and the recorder have proportional components in the bijective database subspace.
The agreement on the bijective subspace reduces the comparison to bounding the weight outside this subspace and the normalization factor.
The following section uses the \(2\)-design property of the endpoint distributions to bound the weight outside the bijective subspace after averaging over \(C,D\).

\section{Nonadaptive Security of DPFC}
\label{sec:four-query-endpoint-averaging}
\label{sec:four-query-diamond-comparison}

This section uses the matching identity \eqref{eq:four-query-common-environment-bijective-component} to estimate the diamond-norm distance between the physical DPFC channel and the map induced by replacing the permutation phase layers with the recorder. 
The comparison is first carried out on each residual no-EPR subspace. 
The bound on the diamond norm distance is obtained by averaging over the endpoint unitaries to limit the recorder's weight outside the bijective database subspace together with the upper bound on the recorder's norm distortion from the preceding section.
The estimates for individual sectors are then combined for arbitrary coherent superpositions of query patterns and EPR sectors.
Finally, comparing both DPFC and Haar with the same recorder map gives the desired security bound.

\subsection{Endpoint-Labelled Dilations}
\label{sec:endpoint-labelled-dilations}
For a fixed EPR sector, this subsection constructs dilations of the physical DPFC channel and the recorder map with outputs in the same Hilbert space.
Fix an exact unitary \(2\)-design \(\nu\) and a residual query pattern \(\theta_\lambda\). 
The physical channel \(\Phi_{\operatorname{DPFC}(\nu)}^{\theta_\lambda}:\mathsf L(\mathcal A_{\theta_\lambda})\to\mathsf L(\mathcal A_{\theta_\lambda})\) is defined by
\begin{align}
    \Phi_{\operatorname{DPFC}(\nu)}^{\theta_\lambda}
    &:={}
    \mathbb E_{\substack{C,D\sim\nu\\ \pi,f}}
    \left[
        \operatorname{Ad}_{
            \U_{\theta_\lambda}(DP_\pi F_fC)
        }
    \right].
    \label{eq:fixed-sector-dpfc-channel}
\end{align}

The parallel recorder was constructed to simulate the random permutation-phase action. 
To compare the physical DPFC channel with the recorder map, place either the physical purification \(\widetilde K_{\theta_\lambda}^{\mathrm{PF}}\) or the recorder \(E_{\theta_\lambda}\) between the same input and output endpoint unitaries. 
For fixed \(C,D\), these two choices define the fixed-endpoint physical dilation \(V_{C,D,\lambda}\) and the fixed-endpoint recorder dilation \(W_{C,D,\lambda}\), respectively.

\begin{definition}[Fixed-endpoint physical and recorder dilations]
\label{def:four-query-fixed-endpoint-branches}
For fixed \(C,D\in\mathrm U(N)\), define the physical and recorder dilations \(V_{C,D,\lambda},W_{C,D,\lambda}:\mathcal A_{\theta_\lambda}\to\mathcal A_{\theta_\lambda}\otimes\mathcal D_+\otimes\mathcal D_-\) by
\begin{align}
    V_{C,D,\lambda}
    &:={}
    \Bigl[
        \left(
            D^{\otimes a_\lambda}
            \otimes(C^T)^{\otimes b_\lambda}
            \otimes(D^*)^{\otimes c_\lambda}
            \otimes(C^\dagger)^{\otimes d_\lambda}
        \right)
        \otimes I_{\mathcal D_+\mathcal D_-}
    \Bigr]
    \widetilde K_{\theta_\lambda}^{\mathrm{PF}}
    \nonumber\\
    &\qquad\cdot
    \left(
        C^{\otimes a_\lambda}
        \otimes(D^T)^{\otimes b_\lambda}
        \otimes(C^*)^{\otimes c_\lambda}
        \otimes(D^\dagger)^{\otimes d_\lambda}
    \right),
    \label{eq:four-query-fixed-endpoint-physical-branch}
\end{align}
and
\begin{align}
    W_{C,D,\lambda}
    &:={}
    \Bigl[
        \left(
            D^{\otimes a_\lambda}
            \otimes(C^T)^{\otimes b_\lambda}
            \otimes(D^*)^{\otimes c_\lambda}
            \otimes(C^\dagger)^{\otimes d_\lambda}
        \right)
        \otimes I_{\mathcal D_+\mathcal D_-}
    \Bigr]
    E_{\theta_\lambda}
    \nonumber\\
    &\qquad\cdot
    \left(
        C^{\otimes a_\lambda}
        \otimes(D^T)^{\otimes b_\lambda}
        \otimes(C^*)^{\otimes c_\lambda}
        \otimes(D^\dagger)^{\otimes d_\lambda}
    \right)
    \nonumber\\
    &=
    \left(
        I_{\mathcal A_{\theta_\lambda}}
        \otimes\mathcal Q_{\theta_\lambda}[C,D]
    \right)
    E_{\theta_\lambda}.
    \label{eq:four-query-fixed-endpoint-recorder-branch}
\end{align}
The last equality follows from the unitary invariance of the recorder (Lemma~\ref{lem:four-type-endpoint-covariance}), which transfers the endpoint actions to the database registers.
\end{definition}

The physical dilation is an isometry. Since the database rotation is unitary, the recorder dilation has the same Gram operator as the parallel recorder, namely \(G_{\theta_\lambda}\):
\begin{align}
    V_{C,D,\lambda}^\dagger V_{C,D,\lambda}=I,
    \qquad
    W_{C,D,\lambda}^\dagger W_{C,D,\lambda}=G_{\theta_\lambda}.
\end{align}

To purify the endpoint randomness, introduce auxiliary registers \(\mathsf C,\mathsf D\) whose labels specify the sampled unitaries \(C,D\).
For a discrete distribution \(\nu\), these registers are initialized in
\begin{align}
    \sum_{C,D}\sqrt{\nu(C)\nu(D)}
    \,|C\rangle_{\mathsf C}|D\rangle_{\mathsf D},
\end{align}
where distinct labels are orthogonal.
Applying the corresponding fixed-endpoint dilations coherently, controlled by these labels, gives the dilations below.
Tracing out \(\mathsf C,\mathsf D\) recovers the average over independently sampled endpoints.
For continuous distributions, including Haar measure, the sum is converted to an integral analogously.

\begin{definition}[Physical and Recorder Dilations with Purified Endpoint Randomness]
\label{def:four-query-endpoint-labelled-dilations}
Let \(\mathcal H_{\mathsf C}\) and \(\mathcal H_{\mathsf D}\) denote the spaces of the endpoint registers. 
Define
\begin{align}
    \mathcal V_{\nu,\lambda},\mathcal W_{\nu,\lambda}:
    \mathcal A_{\theta_\lambda}
    \longrightarrow
    \mathcal H_{\mathsf C}\otimes\mathcal H_{\mathsf D}
    \otimes\mathcal A_{\theta_\lambda}\otimes\mathcal D_+\otimes\mathcal D_-
\end{align}
by
\begin{align}
    \mathcal V_{\nu,\lambda}
    &:={}
    \sum_{C,D}
    \sqrt{\nu(C)\nu(D)}\,
    |C\rangle_{\mathsf C}|D\rangle_{\mathsf D}
    \otimes V_{C,D,\lambda},
    \label{eq:four-query-endpoint-labelled-physical-isometry}\\
    \mathcal W_{\nu,\lambda}
    &:={}
    \sum_{C,D}
    \sqrt{\nu(C)\nu(D)}\,
    |C\rangle_{\mathsf C}|D\rangle_{\mathsf D}
    \otimes W_{C,D,\lambda}.
    \label{eq:four-query-endpoint-labelled-recorder-isometry}
\end{align}
\end{definition}

Orthogonality of the endpoint labels gives
\begin{align}
    \mathcal V_{\nu,\lambda}^\dagger\mathcal V_{\nu,\lambda}
    &=I,
    &
    \mathcal W_{\nu,\lambda}^\dagger\mathcal W_{\nu,\lambda}
    &=G_{\theta_\lambda}.
    \label{eq:four-query-endpoint-labelled-grams}
\end{align}

Define the recorder map \(\mathcal R_{\theta_\lambda}:\mathsf L(\mathcal A_{\theta_\lambda})\to\mathsf L(\mathcal A_{\theta_\lambda})\) by
\begin{align}
    \mathcal R_{\theta_\lambda}
    &:=\operatorname{Tr}_{\mathcal D_+\mathcal D_-}
       \circ\operatorname{Ad}_{E_{\theta_\lambda}}.
    \label{eq:unnormalized-four-query-recorder-map}
\end{align}

Tracing out \(\mathsf C,\mathsf D\) averages the fixed-endpoint dilations over \(C,D\sim\nu\). 
For the physical dilation, subsequently tracing out the databases recovers the DPFC channel, by \eqref{eq:four-query-common-environment-physical-channel} and \eqref{eq:four-query-fixed-endpoint-physical-branch}.
For the recorder dilation, the database rotation \(\mathcal Q_{\theta_\lambda}[C,D]\) is unitary and therefore does not affect the partial trace over the databases. 
The resulting recorder map \(\mathcal R_{\theta_\lambda}\) is independent of the endpoint pair \(C,D\), and hence of the distribution from which the endpoints are sampled.
Consequently,
\begin{align}
    \Phi_{\operatorname{DPFC}(\nu)}^{\theta_\lambda}
    &=
    \operatorname{Tr}_{\mathsf C\mathsf D\mathcal D_+\mathcal D_-}
    \circ
    \operatorname{Ad}_{\mathcal V_{\nu,\lambda}},
    \label{eq:four-query-endpoint-labelled-physical-channel}\\
    \mathcal R_{\theta_\lambda}
    &=
    \operatorname{Tr}_{\mathsf C\mathsf D\mathcal D_+\mathcal D_-}
    \circ
    \operatorname{Ad}_{\mathcal W_{\nu,\lambda}}.
    \label{eq:four-query-endpoint-labelled-recorder-channel}
\end{align}

\subsection{Collision Bound and Bijective Matching}

The diamond distance between the DPFC channel and the recorder map \(\mathcal R_{\theta_\lambda}\) is bounded through their dilations.
An upper bound on the operator norm distance between the dilations bounds the diamond norm distance up to a constant by Lemma~\ref{lem:common-environment-stinespring-bound}.
In this subsection we show that the two dilations agree on the bijective subspace following Equation~\eqref{eq:four-query-common-environment-bijective-component}, and bound the weight outside this subspace.

Define the fixed-endpoint ordered recorder dilation \(W_{C,D,\lambda}^{\mathrm{ord}}:\mathcal A_{\theta_\lambda}\to\mathcal A_{\theta_\lambda}\otimes(\mathcal O_{p_\lambda})_+\otimes(\mathcal O_{q_\lambda})_-\) by
\begin{align}
    W_{C,D,\lambda}^{\mathrm{ord}}
    &:=(I_{\mathcal A_{\theta_\lambda}}
        \otimes\mathcal Q_{\theta_\lambda}^{\mathrm{ord}}[C,D])
       E_{\theta_\lambda}^{\mathrm{ord}}.
    \label{eq:four-query-fixed-endpoint-ordered-branch}
\end{align}
This dilation is an isometry by Equation~\eqref{eq:ordered-recorder-isometry} and unitarity of \(\mathcal Q_{\theta_\lambda}^{\mathrm{ord}}[C,D]\).
Database projectors below act as the identity on query-output registers, and \(C,D\sim\nu\) are independent.

\begin{lemma}[Endpoint-averaged ordered collision bound]
\label{lem:four-query-endpoint-averaged-collision-bound}
Let \(\nu\) be an exact unitary \(2\)-design. Then,
\begin{align}
    0
    & \preceq \mathbb E_{C,D\sim\nu}\!\left[
            \Pi_{\theta_\lambda}^{\mathrm{nE}}
            (W_{C,D,\lambda}^{\mathrm{ord}})^\dagger
            (I-\Pi_{\theta_\lambda}^{\mathrm{bij,ord}})
            W_{C,D,\lambda}^{\mathrm{ord}}
            \Pi_{\theta_\lambda}^{\mathrm{nE}}
        \right]
    \preceq \frac{2t_\lambda^2}{N}
            \Pi_{\theta_\lambda}^{\mathrm{nE}}.
    \label{eq:four-query-endpoint-averaged-ordered-collision-bound}
\end{align}
\end{lemma}

\begin{proof}
See Appendix~\ref{app:four-query-endpoint-averaged-collision-bound}.
\end{proof}

For the ordered recorder, averaging over the endpoint unitaries bounds each pairwise domain or image collision by \(O(1/N)\) on no-EPR inputs.
Summing over all pairs gives the \(O(t_\lambda^2/N)\) bound in \eqref{eq:four-query-endpoint-averaged-ordered-collision-bound}.
The parallel recorder is obtained by symmetrizing the ordered recorder's databases.
On the bijective subspace, this symmetrization preserves inner products.
Therefore, the ordered and the symmetrized recorder have the same norm on their bijective components for every input (Appendix~\ref{app:common-bijective-gram}).
Thus, Equation~\eqref{eq:four-query-endpoint-averaged-ordered-collision-bound} is used in the dilation comparison by replacing the symmetrized recorder's bijective contribution by the ordered recorder's.

The matching identity~\eqref{eq:four-query-common-environment-bijective-component} now extends to the physical and recorder dilations, including the endpoint unitaries \(C,D\). 
Their bijective components agree up to the same normalization factor for every fixed endpoint pair.

\begin{lemma}[Bijective Matching with Fixed Endpoint Unitaries]
\label{lem:four-query-projected-fixed-endpoint-matching}
For every \(C,D\in\mathrm U(N)\),
\begin{align}
    \left(
        I_{\mathcal A_{\theta_\lambda}}
        \otimes\Pi_{\theta_\lambda}^{\mathrm{bij}}
    \right)
    V_{C,D,\lambda}
    &=
    \alpha_{t_\lambda}
    \left(
        I_{\mathcal A_{\theta_\lambda}}
        \otimes\Pi_{\theta_\lambda}^{\mathrm{bij}}
    \right)
    W_{C,D,\lambda},
    \qquad
    \alpha_k:=\sqrt{\frac{N^k(N-k)!}{N!}}.
    \label{eq:four-query-fixed-endpoint-bijective-matching}
\end{align}
\end{lemma}

\begin{proof}
See Appendix~\ref{app:four-query-projected-fixed-endpoint-matching}.
\end{proof}

\subsection{Recorder Approximation on a Fixed EPR Sector}
\label{sec:four-query-fixed-sector-comparison}

This subsection compares the physical DPFC channel with the completely positive recorder map \(\mathcal R_{\theta_\lambda}\) on each residual no-EPR subspace. 
As mentioned, the bijective matching determines the overlap of their dilations, while the ordered collision bound and the Gram bound yield the upper bound on the approximation error. 
Discarding the common database environment gives a diamond-norm estimate for the two maps. 
The next subsection combines these estimates across all query patterns and EPR sectors.

\begin{lemma}[Fixed-sector comparison with the recorder]
\label{lem:four-query-fixed-sector-recorder-comparison}
Let \(\nu\) be an exact unitary \(2\)-design. The endpoint-labelled
dilations satisfy
\begin{align}
    \left\|
        \left(
            \mathcal V_{\nu,\lambda}
            -\mathcal W_{\nu,\lambda}
        \right)
        \Pi_{\theta_\lambda}^{\mathrm{nE}}
    \right\|_\infty^2
    &\leq\frac{4t_\lambda^2}{N}+e^{t_\lambda^2/(2N)}-1.
    \label{eq:four-query-fixed-sector-dilation-comparison}
\end{align}
Consequently,
\begin{align}
    \left\|
        \left.
        \left(
            \Phi_{\operatorname{DPFC}(\nu)}^{\theta_\lambda}
            -\mathcal R_{\theta_\lambda}
        \right)
        \right|_{\mathsf L(
            \nEspace{\theta_\lambda}
        )}
    \right\|_\diamond
    \leq
    \left(1+e^{t_\lambda^2/(4N)}\right)
    \sqrt{\frac{4t_\lambda^2}{N}+e^{t_\lambda^2/(2N)}-1}.
    \label{eq:four-query-fixed-sector-recorder-comparison}
\end{align}
\end{lemma}

The application to the security bound will assume that \(t_\lambda^2/N\) is bounded by an absolute constant. 
In this regime, the squared dilation distance is \(O(t_\lambda^2/N)\), so the diamond-norm distance is \(O(t_\lambda/\sqrt N)\), with constants depending only on the chosen bound.

\begin{proof}
The physical dilation has Gram operator \(I\), and the recorder dilation has Gram operator \(G_{\theta_\lambda}\), by \eqref{eq:four-query-endpoint-labelled-grams}. 
Consequently,
\begin{align}
    \left\|
        (\mathcal V_{\nu,\lambda}-\mathcal W_{\nu,\lambda})
        \Pi_{\theta_\lambda}^{\mathrm{nE}}
    \right\|_\infty^2
    =
    \left\|
        \Pi_{\theta_\lambda}^{\mathrm{nE}}
        \left[
            I+G_{\theta_\lambda}
            -\mathcal V_{\nu,\lambda}^\dagger\mathcal W_{\nu,\lambda}
            -\mathcal W_{\nu,\lambda}^\dagger\mathcal V_{\nu,\lambda}
        \right]
        \Pi_{\theta_\lambda}^{\mathrm{nE}}
    \right\|_\infty.
    \label{eq:four-query-fixed-sector-dilation-distance-squared}
\end{align}

To evaluate the overlap, fix \(C,D\) and consider the database registers of the two dilations. 
The recorder dilation \(W_{C,D,\lambda}\) always has database lengths \((p_\lambda,q_\lambda)\). 
By construction, \(\widetilde{\mathsf{Comp}}_t\) has no output in the non-bijective database subspace at these lengths. 
Indeed, \eqref{eq:four-query-isometric-compression-extension} maps the prescribed physical relation states to bijective database states, and the orthogonal complement of their span to database lengths whose total exceeds \(t\geq p_\lambda+q_\lambda\). 
The endpoint unitaries in \eqref{eq:four-query-fixed-endpoint-physical-branch} act only on the query registers, so they do not change these database lengths.
Therefore, applying \(I-\Pi_{\theta_\lambda}^{\mathrm{bij}}\) to the database registers of the embedded physical dilation \(V_{C,D,\lambda}\) removes its entire component with database lengths \((p_\lambda,q_\lambda)\).
The remaining components have different database lengths and are consequently orthogonal to the recorder output. 
Hence,
\begin{align}
    \bigl[(I-\Pi_{\theta_\lambda}^{\mathrm{bij}})
        V_{C,D,\lambda}\bigr]^\dagger W_{C,D,\lambda}
    &=0.
    \label{eq:four-query-physical-complement-recorder-orthogonality}
\end{align}
Recall \(\alpha_k\) from \eqref{eq:four-query-fixed-endpoint-bijective-matching}.
Decomposing the physical dilation into its bijective and complementary components, applying \eqref{eq:four-query-physical-complement-recorder-orthogonality}, and substituting \eqref{eq:four-query-fixed-endpoint-bijective-matching} gives
\begin{align}
    V_{C,D,\lambda}^\dagger W_{C,D,\lambda}
    =\bigl[\Pi_{\theta_\lambda}^{\mathrm{bij}}
        V_{C,D,\lambda}\bigr]^\dagger W_{C,D,\lambda}
    =\alpha_{t_\lambda} W_{C,D,\lambda}^\dagger
        \Pi_{\theta_\lambda}^{\mathrm{bij}}W_{C,D,\lambda}.
    \label{eq:four-query-fixed-sector-bijective-overlap}
\end{align}

The two dilations use the same orthogonal endpoint labels.
Thus, their overlap is the average of the fixed-endpoint overlaps in \eqref{eq:four-query-fixed-sector-bijective-overlap}.
Using \eqref{eq:four-query-common-bijective-gram} the parallel recorder's bijective contribution is replaced with that of the ordered recorder. 
Therefore,
\begin{align}
    \Pi_{\theta_\lambda}^{\mathrm{nE}}
      \mathcal V_{\nu,\lambda}^\dagger\mathcal W_{\nu,\lambda}
      \Pi_{\theta_\lambda}^{\mathrm{nE}}
    &=\mathbb E_{C,D\sim\nu}\!\left[
        \Pi_{\theta_\lambda}^{\mathrm{nE}}
        V_{C,D,\lambda}^\dagger W_{C,D,\lambda}
        \Pi_{\theta_\lambda}^{\mathrm{nE}}
    \right]
    \nonumber\\
    &=\alpha_{t_\lambda}\mathbb E_{C,D\sim\nu}\!\left[
        \Pi_{\theta_\lambda}^{\mathrm{nE}}
        W_{C,D,\lambda}^\dagger
        \Pi_{\theta_\lambda}^{\mathrm{bij}}W_{C,D,\lambda}
        \Pi_{\theta_\lambda}^{\mathrm{nE}}
    \right]
    \nonumber\\
    &=\alpha_{t_\lambda}\mathbb E_{C,D\sim\nu}\!\left[
        \Pi_{\theta_\lambda}^{\mathrm{nE}}
        (W_{C,D,\lambda}^{\mathrm{ord}})^\dagger
        \Pi_{\theta_\lambda}^{\mathrm{bij,ord}}
        W_{C,D,\lambda}^{\mathrm{ord}}
        \Pi_{\theta_\lambda}^{\mathrm{nE}}
    \right].
    \label{eq:four-query-fixed-sector-ordered-overlap}
\end{align}
Since \(\Pi_{\theta_\lambda}^{\mathrm{nE}}\mathcal V_{\nu,\lambda}^\dagger\mathcal W_{\nu,\lambda}\Pi_{\theta_\lambda}^{\mathrm{nE}}\) is Hermitian, the two cross terms in \eqref{eq:four-query-fixed-sector-dilation-distance-squared} sum to \(-2\Pi_{\theta_\lambda}^{\mathrm{nE}}\mathcal V_{\nu,\lambda}^\dagger\mathcal W_{\nu,\lambda}\Pi_{\theta_\lambda}^{\mathrm{nE}}\).
Moreover, since \(\alpha_{t_\lambda} \geq 1\), substituting the bijective form of the overlap from \eqref{eq:four-query-fixed-sector-ordered-overlap} into \eqref{eq:four-query-fixed-sector-dilation-distance-squared} gives
\begin{align}
    0&\preceq
        \Pi_{\theta_\lambda}^{\mathrm{nE}}
        (\mathcal V_{\nu,\lambda}-\mathcal W_{\nu,\lambda})^\dagger
        (\mathcal V_{\nu,\lambda}-\mathcal W_{\nu,\lambda})
        \Pi_{\theta_\lambda}^{\mathrm{nE}}
    \nonumber\\
    &=\Pi_{\theta_\lambda}^{\mathrm{nE}}
        (I+G_{\theta_\lambda})\Pi_{\theta_\lambda}^{\mathrm{nE}}
        -2\alpha_{t_\lambda}\mathbb E_{C,D\sim\nu}\!\left[
            \Pi_{\theta_\lambda}^{\mathrm{nE}}
            (W_{C,D,\lambda}^{\mathrm{ord}})^\dagger
            \Pi_{\theta_\lambda}^{\mathrm{bij,ord}}
            W_{C,D,\lambda}^{\mathrm{ord}}
            \Pi_{\theta_\lambda}^{\mathrm{nE}}
        \right]
    \nonumber\\
    &\preceq\Pi_{\theta_\lambda}^{\mathrm{nE}}
        (I+G_{\theta_\lambda})\Pi_{\theta_\lambda}^{\mathrm{nE}}
        -2\mathbb E_{C,D\sim\nu}\!\left[
            \Pi_{\theta_\lambda}^{\mathrm{nE}}
            (W_{C,D,\lambda}^{\mathrm{ord}})^\dagger
            \Pi_{\theta_\lambda}^{\mathrm{bij,ord}}
            W_{C,D,\lambda}^{\mathrm{ord}}
            \Pi_{\theta_\lambda}^{\mathrm{nE}}
        \right]
    \nonumber\\
    &=\Pi_{\theta_\lambda}^{\mathrm{nE}}
        (G_{\theta_\lambda}-I)\Pi_{\theta_\lambda}^{\mathrm{nE}}
        +2\mathbb E_{C,D\sim\nu}\!\left[
            \Pi_{\theta_\lambda}^{\mathrm{nE}}
            (W_{C,D,\lambda}^{\mathrm{ord}})^\dagger
            (I-\Pi_{\theta_\lambda}^{\mathrm{bij,ord}})
            W_{C,D,\lambda}^{\mathrm{ord}}
            \Pi_{\theta_\lambda}^{\mathrm{nE}}
        \right]
    \nonumber\\
    &\preceq\left(e^{t_\lambda^2/(2N)}-1
            +2\frac{2t_\lambda^2}{N}\right)
        \Pi_{\theta_\lambda}^{\mathrm{nE}}
    =\left(\frac{4t_\lambda^2}{N}+e^{t_\lambda^2/(2N)}-1\right)
        \Pi_{\theta_\lambda}^{\mathrm{nE}}.
    \label{eq:four-query-fixed-sector-gram-collision-distance}
\end{align}
Replacing \(\alpha_{t_\lambda}\) by \(1\) gives an upper bound because the averaged bijective contribution is positive semidefinite and is subtracted with coefficient \(2\alpha_{t_\lambda}\).
The next equality uses the ordered recorder's isometry to express its bijective contribution as the identity minus its collision contribution. 
The last inequality applies the Gram bound \eqref{eq:four-query-gram-upper-bound} and the ordered collision bound
\eqref{eq:four-query-endpoint-averaged-ordered-collision-bound}.
Taking the operator norm proves \eqref{eq:four-query-fixed-sector-dilation-comparison}.

Apply \eqref{eq:four-query-common-environment-diamond-bound} relating the dilation operator norm distance with the induced maps diamond norm distance, in addition to \eqref{eq:four-query-endpoint-labelled-grams} and \eqref{eq:four-query-recorder-norm-bound}, to the restrictions of the two dilations to \(\nEspace{\theta_\lambda}\). 
By \eqref{eq:four-query-endpoint-labelled-physical-channel} and \eqref{eq:four-query-endpoint-labelled-recorder-channel}, their induced maps are the restrictions of the physical channel and \(\mathcal R_{\theta_\lambda}\), respectively. 
Hence,
\begin{align}
    \left\|
            \left.
            \left(
                \Phi_{\operatorname{DPFC}(\nu)}^{\theta_\lambda}
                -\mathcal R_{\theta_\lambda}
            \right)
            \right|_{\mathsf L(
                \nEspace{\theta_\lambda}
            )}
        \right\|_\diamond
        &\leq
        \bigl(1+\|\mathcal W_{\nu,\lambda}\|_\infty\bigr)
        \left\|
            (\mathcal V_{\nu,\lambda}-\mathcal W_{\nu,\lambda})
            \Pi_{\theta_\lambda}^{\mathrm{nE}}
        \right\|_\infty \notag\\
        &\leq
        \left(1+e^{t_\lambda^2/(4N)}\right)
        \sqrt{\frac{4t_\lambda^2}{N}+e^{t_\lambda^2/(2N)}-1}.
\end{align}
\end{proof}

\subsection{Coherent Assembly of Query Patterns and EPR Sectors}
\label{sec:four-query-coherent-sector-assembly}

As per the security model described in Section~\ref{sec:informal-our-results}, inputs may be coherent superpositions of both query patterns and residual EPR sectors. 
The distance bound from the preceding subsection must therefore be extended to account for cross terms between these components. 
This subsection extends the physical and recorder dilations to all EPR sectors and query patterns.
Orthogonality of the output sectors bounds the norm of the difference between the assembled dilations by the largest sector bound, with no loss from the number of patterns or sectors.

For each \(\theta\in\Theta_t\), the unitary \(\mathcal I_\theta\) from \eqref{eq:global-sector-unitary} maps the query space \(\mathcal A_t\) to the sector-coordinate space \(\widetilde{\mathcal A}_\theta\). 
Define the enlarged output space for this pattern by
\begin{align}
    \mathcal Y_\theta
    &:={}
    \bigoplus_{\lambda\in\Lambda_\theta}
    \left(
        \mathcal A_{\theta_\lambda}
        \otimes\mathcal M_\lambda
    \right).
    \label{eq:four-query-enlarged-sector-output-space}
\end{align}
The sector-coordinate space \(\widetilde{\mathcal A}_\theta\) is identified with its natural subspace of \(\mathcal Y_\theta\), and physical sector outputs are regarded as vectors in this enlarged output space. The recorder output does not generally lie in the residual no-EPR subspace, whereas the physical output does.

When several query patterns are considered together, write \(\mathcal M_{\theta,\lambda}\) for the multiplicity space \(\mathcal M_\lambda\) associated with \(\theta\) in \eqref{eq:double-sector-multiplicity-space}. 
For a fixed pattern, keep the shorthand \(\mathcal M_\lambda\).
Define the coherent sector-coordinate and enlarged output spaces by
\begin{align}
    \widetilde{\mathcal A}^{[t]}
    :=
    \bigoplus_{\theta\in\Theta_t}\widetilde{\mathcal A}_\theta
    =
    \bigoplus_{\substack{
        \theta\in\Theta_t\\
        \lambda\in\Lambda_\theta
    }}
    \left(
        \nEspace{\theta_\lambda}
        \otimes\mathcal M_{\theta,\lambda}
    \right), \quad
    \mathcal Y^{[t]}
    :=
    \bigoplus_{\theta\in\Theta_t}\mathcal Y_\theta
    =
    \bigoplus_{\substack{
        \theta\in\Theta_t\\
        \lambda\in\Lambda_\theta
    }}
    \left(
        \mathcal A_{\theta_\lambda}
        \otimes\mathcal M_{\theta,\lambda}
    \right).
    \label{eq:coherent-enlarged-sector-output-space}
\end{align}

Under the canonical identification \(\mathcal Q_t\otimes\mathcal A_t \simeq\bigoplus_{\theta\in\Theta_t}\mathcal A_t\), define
\begin{align}
    \mathcal I^{[t]}
    :=
    \bigoplus_{\theta\in\Theta_t}\mathcal I_\theta
    :
    \mathcal Q_t\otimes\mathcal A_t
    \longrightarrow
    \widetilde{\mathcal A}^{[t]}.
    \label{eq:coherent-sector-output-inclusion}
\end{align}
The map \(\mathcal I^{[t]}\) is unitary onto \(\widetilde{\mathcal A}^{[t]}\), since each summand has this property.
Throughout, \(\widetilde{\mathcal A}^{[t]}\) is identified with its natural subspace of \(\mathcal Y^{[t]}\). Physical outputs in sector coordinates are regarded as vectors in \(\mathcal Y^{[t]}\), and output operators on \(\widetilde{\mathcal A}^{[t]}\) are extended by zero on its orthogonal complement in \(\mathcal Y^{[t]}\). These inclusions are left implicit.
The equivariance identity \eqref{eq:double-sector-intertwining} gives, for every \(U\in\mathrm U(N)\),
\begin{align}
    \mathcal I^{[t]}\U^{[t]}(U)
    &=
    \left[\bigoplus_{\substack{
        \theta\in\Theta_t\\
        \lambda\in\Lambda_\theta
    }}
    \left(
        \left.\U_{\theta_\lambda}(U)
        \right|_{\nEspace{\theta_\lambda}}
        \otimes I_{\mathcal M_{\theta,\lambda}}
    \right)\right]\mathcal I^{[t]}.
    \label{eq:coherent-sector-intertwining}
\end{align}

The objects used in the single-sector comparison are now extended to all query patterns and EPR sectors, starting with the database rotation, followed by the coherent recorder and the physical and recorder dilations.
The coherent recorder and the physical and recorder dilations first apply \(\mathcal I^{[t]}\) to express the input in the EPR-sector decomposition coordinates.

For fixed \(C,D\), the unitary invariance identity specifies the same one-edge rotations \(W_+[C,D]\) and \(W_-[C,D]\) for every query pattern and EPR sector. 
The shared databases are variable-length registers: states with different pairs of lengths \((r,s)\) lie in orthogonal subspaces \((\mathcal D_r)_+\otimes(\mathcal D_s)_-\).
On each such subspace, apply \(W_+[C,D]\) once to each positive edge and \(W_-[C,D]\) once to each negative edge. 
These tensor powers preserve the symmetric database spaces. 
Taking the direct sum of their restrictions defines a single unitary on \(\mathcal D_+\otimes\mathcal D_-\):
\begin{align}
    \mathcal Q[C,D]
    &:={}
    \bigoplus_{r,s\geq0}
    \left(
        \left.
        W_+[C,D]^{\otimes r}
        \right|_{(\mathcal D_r)_+}
        \otimes
        \left.
        W_-[C,D]^{\otimes s}
        \right|_{(\mathcal D_s)_-}
    \right).
    \label{eq:four-query-global-database-rotation}
\end{align}
This unitary preserves both database lengths and acts on superpositions of different lengths.
In particular, \(\mathcal Q_{\theta_\lambda}[C,D]\) is the restriction of \(\mathcal Q[C,D]\) to lengths \((p_\lambda,q_\lambda)\).
Summands with the same lengths therefore use exactly the same rotation, while summands with different lengths use different orthogonal blocks of this one unitary.

\begin{definition}[Coherent recorder]
\label{def:coherent-recorder}
The coherent unnormalized recorder \(\mathbf E^{[t]}: \mathcal Q_t\otimes\mathcal A_t\to\mathcal Y^{[t]} \otimes\mathcal D_+\otimes\mathcal D_-\) is the map defined by
\begin{align}
    \mathbf E^{[t]}
    &:={}
    \left[\bigoplus_{\substack{
        \theta\in\Theta_t\\
        \lambda\in\Lambda_\theta
    }}
    \left(
        \left.E_{\theta_\lambda}
        \right|_{\nEspace{\theta_\lambda}}
        \otimes I_{\mathcal M_{\theta,\lambda}}
    \right)\right]\mathcal I^{[t]}.
\end{align}
The output tensor factors are ordered with the query and multiplicity registers first, followed by the two database registers.
The direct sums are over the output sectors \((\theta,\lambda)\), with one shared database environment \(\mathcal D_+\otimes\mathcal D_-\).
The \((\theta,\lambda)\)-summand has database bidegree \((p_\lambda,q_\lambda)\), as determined by its residual pattern.
Summands with the same bidegree use the same environment subspace, even when their query patterns differ. 
Different bidegrees give orthogonal database outputs.
Define the induced completely positive recorder map \(\mathbf R^{[t]}:\mathsf L(\mathcal Q_t\otimes\mathcal A_t)\to\mathsf L(\mathcal Y^{[t]})\) by
\begin{align}
    \mathbf R^{[t]}
    &:={}
    \operatorname{Tr}_{\mathcal D_+\mathcal D_-}
    \circ\operatorname{Ad}_{\mathbf E^{[t]}}.
    \label{eq:coherent-joint-recorder-cp-map}
\end{align}
\end{definition}

Write \(V_{C,D,\theta,\lambda}\) and \(W_{C,D,\theta,\lambda}\) for the maps in \eqref{eq:four-query-fixed-endpoint-physical-branch} and \eqref{eq:four-query-fixed-endpoint-recorder-branch}, with the dependence on the query pattern made explicit.
Both coherent maps below use the same tensor-factor ordering as \(\mathbf E^{[t]}\).

\begin{definition}[Coherent physical and recorder dilations with fixed endpoints]
\label{def:coherent-physical-and-recorder-branches}
For fixed \(C,D\in\mathrm U(N)\), define \(\mathbf V_{C,D}^{[t]},\mathbf W_{C,D}^{[t]}:\mathcal Q_t\otimes\mathcal A_t\to\mathcal Y^{[t]}\otimes\mathcal D_+\otimes\mathcal D_-\) by
\begin{align}
    \mathbf V_{C,D}^{[t]}
    &:={}
    \left[\bigoplus_{\substack{
        \theta\in\Theta_t\\
        \lambda\in\Lambda_\theta
    }}
    \left(
        \left.V_{C,D,\theta,\lambda}
        \right|_{\nEspace{\theta_\lambda}}
        \otimes I_{\mathcal M_{\theta,\lambda}}
    \right)\right]\mathcal I^{[t]},
    \\
    \mathbf W_{C,D}^{[t]}
    &:={}
    \left[\bigoplus_{\substack{
        \theta\in\Theta_t\\
        \lambda\in\Lambda_\theta
    }}
    \left(
        \left.W_{C,D,\theta,\lambda}
        \right|_{\nEspace{\theta_\lambda}}
        \otimes I_{\mathcal M_{\theta,\lambda}}
    \right)\right]\mathcal I^{[t]}.
\end{align}
Every physical summand uses the same compression \(\widetilde{\mathsf{Comp}}_t\) of the common permutation-phase environment.
\end{definition}

The single database unitary \(\mathcal Q[C,D]\) from \eqref{eq:four-query-global-database-rotation} restricts to \(\mathcal Q_{\theta_\lambda}[C,D]\) on the database output space of each recorder summand. 
It can therefore be factored out of the direct sum over query patterns and EPR sectors, giving the following form again
\begin{align}
    \mathbf W_{C,D}^{[t]}
    &=
    \left(
        I_{\mathcal Y^{[t]}}
        \otimes\mathcal Q[C,D]
    \right)\mathbf E^{[t]}.
    \label{eq:coherent-global-recorder-rotation}
\end{align}

\begin{definition}[Coherent global dilations]
\label{def:coherent-global-dilations}
For a probability measure \(\nu\) on \(\mathrm U(N)\), define \(\mathbf V_\nu^{[t]},\mathbf W_\nu^{[t]}: \mathcal Q_t\otimes\mathcal A_t \longrightarrow \mathcal H_{\mathsf C}\otimes\mathcal H_{\mathsf D} \otimes\mathcal Y^{[t]}\otimes\mathcal D_+\otimes\mathcal D_-\) by
\begin{align}
    \mathbf V_\nu^{[t]}
    &:={}
    \sum_{C,D}
    \sqrt{\nu(C)\nu(D)}\,
    |C\rangle_{\mathsf C}|D\rangle_{\mathsf D}
    \otimes\mathbf V_{C,D}^{[t]},
    \\
    \mathbf W_\nu^{[t]}
    &:={}
    \sum_{C,D}
    \sqrt{\nu(C)\nu(D)}\,
    |C\rangle_{\mathsf C}|D\rangle_{\mathsf D}
    \otimes\mathbf W_{C,D}^{[t]}.
\end{align}
\end{definition}

Unitarity of \(\mathcal I^{[t]}\), orthogonality of the output sectors, and the isometry of each physical dilation imply that \(\mathbf V_\nu^{[t]}\) is an isometry.
The recorder dilation \(\mathbf W_\nu^{[t]}\) is generally not an isometry, since the recorder is unnormalized.
The coordinate transformation \(\mathcal I^{[t]}\) expresses the physical channel output in sector coordinates, regarded as lying in \(\mathcal Y^{[t]}\), the output space of the recorder map \(\mathbf R^{[t]}\). 
By the equivariance identity~\eqref{eq:coherent-sector-intertwining}, tracing out the environment of the physical dilation yields the physical channel with its output expressed in these coordinates.
The maps induced by these dilations satisfy
\begin{align}
    \operatorname{Ad}_{\mathcal I^{[t]}}
    \circ
    \Phi_{\operatorname{DPFC}(\nu)}^{[t]}
    &=
    \operatorname{Tr}_{\mathsf C\mathsf D\mathcal D_+\mathcal D_-}
    \circ\operatorname{Ad}_{\mathbf V_\nu^{[t]}},
    \label{eq:coherent-global-physical-channel}\\
    \mathbf R^{[t]}
    &=
    \operatorname{Tr}_{\mathsf C\mathsf D\mathcal D_+\mathcal D_-}
    \circ\operatorname{Ad}_{\mathbf W_\nu^{[t]}}.
    \label{eq:coherent-global-recorder-cp-map}
\end{align}

The following lemma extends Lemma~\ref{lem:four-query-fixed-sector-recorder-comparison} to inputs that are coherent superpositions of query patterns and EPR sectors. 
The common environment permits comparison of the assembled dilations, and the orthogonal \((\theta,\lambda)\)-summands of their output space ensure that the bound does not change with the number of patterns or sectors.

\begin{lemma}[Coherent sector assembly]
\label{lem:four-query-coherent-sector-assembly}
Let \(\nu\) be an exact unitary \(2\)-design. Then,
\begin{align}
    \left\|
        \operatorname{Ad}_{\mathcal I^{[t]}}
        \circ
        \Phi_{\operatorname{DPFC}(\nu)}^{[t]}
        -
        \mathbf R^{[t]}
    \right\|_\diamond
    &\leq
    \left(1+e^{t^2/(4N)}\right)
    \sqrt{\frac{4t^2}{N}+e^{t^2/(2N)}-1}.
    \label{eq:four-query-global-recorder-comparison}
\end{align}
\end{lemma}

\begin{proof}
The query outputs of distinct \((\theta,\lambda)\)-summands lie in orthogonal subspaces of \(\mathcal Y^{[t]}\). 
Hence their cross terms vanish when forming a Gram operator, even when the database bidegrees coincide.
Applying \eqref{eq:four-query-endpoint-labelled-grams} in each summand gives
\begin{align}
    \bigl(\mathbf W_\nu^{[t]}\bigr)^\dagger
    \mathbf W_\nu^{[t]}
    =
    \bigl(\mathbf E^{[t]}\bigr)^\dagger
    \mathbf E^{[t]}
    =
    (\mathcal I^{[t]})^\dagger\left[
    \bigoplus_{\substack{
        \theta\in\Theta_t\\
        \lambda\in\Lambda_\theta
    }}
    \left[
        \left.
        \left(
            \Pi_{\theta_\lambda}^{\mathrm{nE}}
            G_{\theta_\lambda}
            \Pi_{\theta_\lambda}^{\mathrm{nE}}
        \right)
        \right|_{\nEspace{\theta_\lambda}}
        \otimes I_{\mathcal M_{\theta,\lambda}}
    \right]\right]\mathcal I^{[t]}.
    \label{eq:four-query-global-dilation-grams}
\end{align}
The compression by \(\Pi_{\theta_\lambda}^{\mathrm{nE}}\) comes from restricting each recorder's input to its residual no-EPR subspace.
Unitary conjugation by \(\mathcal I^{[t]}\) preserves the operator norm, and the operator norm of a direct sum is the largest norm of its blocks.
Since the physical dilation is an isometry, \eqref{eq:four-query-global-dilation-grams} and \eqref{eq:four-query-recorder-norm-bound} give
\begin{align}
    \|\mathbf V_\nu^{[t]}\|_\infty
    =1, \quad
    \|\mathbf W_\nu^{[t]}\|_\infty
    =
    \|\mathbf E^{[t]}\|_\infty
    =
    \max_{\substack{
        \theta\in\Theta_t\\
        \lambda\in\Lambda_\theta
    }}
    \left\|
        E_{\theta_\lambda}\Pi_{\theta_\lambda}^{\mathrm{nE}}
    \right\|_\infty
    \leq e^{t^2/(4N)}.
    \label{eq:four-query-global-dilation-norms}
\end{align}

Write \(\mathcal V_{\nu,\theta,\lambda}\) and \(\mathcal W_{\nu,\theta,\lambda}\) for the physical and recorder dilations in \eqref{eq:four-query-endpoint-labelled-physical-isometry} and \eqref{eq:four-query-endpoint-labelled-recorder-isometry}, with their dependence on the query pattern \(\theta\) made explicit. 
Similarly, write \(t_{\theta,\lambda}:=|\theta_\lambda|=t-2(\ell+m)\) for the residual size when \(\lambda=(\ell,m)\).
The common endpoint and database environments identify each global dilation with the direct sum of the sector dilations, tensored with the multiplicity identities, followed on the right by \(\mathcal I^{[t]}\).
Their difference therefore has the same form. Right multiplication by the unitary \(\mathcal I^{[t]}\) preserves its operator norm, and the \((\theta,\lambda)\)-summands of \(\mathcal Y^{[t]}\) are orthogonal. 
Thus, applying~\eqref{eq:four-query-fixed-sector-dilation-comparison} to each block and taking the maximum over all orthogonal sectors gives,
\begin{align}
    \left\|
        \mathbf V_\nu^{[t]}
        -
        \mathbf W_\nu^{[t]}
    \right\|_\infty^2
    &=
    \max_{\substack{
        \theta\in\Theta_t\\
        \lambda\in\Lambda_\theta
    }}
    \left\|
        \left(
            \mathcal V_{\nu,\theta,\lambda}
            -
            \mathcal W_{\nu,\theta,\lambda}
        \right)
        \Pi_{\theta_\lambda}^{\mathrm{nE}}
    \right\|_\infty^2
    \nonumber\\
    & \leq
    \max_{\substack{
        \theta\in\Theta_t\\
        \lambda\in\Lambda_\theta
    }}
    \left(
        \frac{4t_{\theta,\lambda}^2}{N}
        +e^{t_{\theta,\lambda}^2/(2N)}-1
    \right)
    \nonumber\\
    &\leq\frac{4t^2}{N}+e^{t^2/(2N)}-1.
    \label{eq:four-query-global-dilation-distance}
\end{align}

Together with Lemma~\ref{lem:common-environment-stinespring-bound}, \eqref{eq:four-query-global-dilation-distance} bounds the cross-sector terms as well: the lemma converts the distance between the full dilations into a diamond-norm bound for their induced maps, including their action on off-diagonal input blocks between distinct query patterns or EPR sectors.
It uses the fact that the operator norm equals the maximum of norms of all of its orthogonal blocks. 
Thus, no separate estimates of these blocks are needed. 
Applying the lemma to the maps in \eqref{eq:coherent-global-physical-channel} and \eqref{eq:coherent-global-recorder-cp-map}, and substituting \eqref{eq:four-query-global-dilation-norms} and \eqref{eq:four-query-global-dilation-distance}, yields
\begin{align}
    \left\|
        \operatorname{Ad}_{\mathcal I^{[t]}}
        \circ
        \Phi_{\operatorname{DPFC}(\nu)}^{[t]}
        -
        \mathbf R^{[t]}
    \right\|_\diamond
    & \leq
    \left(
        \|\mathbf V_\nu^{[t]}\|_\infty
        +
        \|\mathbf W_\nu^{[t]}\|_\infty
    \right)
    \left\|
        \mathbf V_\nu^{[t]}
        -
        \mathbf W_\nu^{[t]}
    \right\|_\infty \nonumber \\
    & \leq
    \left(1+e^{t^2/(4N)}\right)
    \sqrt{\frac{4t^2}{N}+e^{t^2/(2N)}-1}.
\end{align}
\end{proof}

\subsection{Main Result Assembly}

In this section we put the parts together to assemble the final security bound. 
It follows by applying Lemma~\ref{lem:four-query-coherent-sector-assembly} to two endpoint distributions: the chosen unitary \(2\)-design and Haar measure.
Both resulting channels are compared with the same recorder map \(\mathbf R^{[t]}\), which is independent of the endpoint distribution.
With Haar endpoints, the DPFC unitary is itself Haar distributed, so the triangle inequality combines the two recorder comparisons into a bound on the distance between DPFC and Haar. 
The theorem states this comparison in a general form, followed by a concrete parameter choice giving a simpler numerical bound.

\begin{theorem}[Coherent nonadaptive four-query security of DPFC]
\label{thm:main-four-query-dpfc-security}
Let \(\mathfrak D\) be an exact unitary \(2\)-design on \(\mathrm U(N)\), and let \(0\leq t\leq N\) be an integer. Then,
\begin{align}
    \left\|
        \Phi_{\operatorname{DPFC}(\mathfrak D)}^{[t]}
        -
        \Phi_{\mathrm H}^{[t]}
    \right\|_\diamond
    &\leq
    2\left(1+e^{t^2/(4N)}\right)
    \sqrt{\frac{4t^2}{N}+e^{t^2/(2N)}-1}.
    \label{eq:main-four-query-dpfc-security}
\end{align}
Specifically, if \(t^2\leq N/16\), then
\begin{align}
    \left\|
        \Phi_{\operatorname{DPFC}(\mathfrak D)}^{[t]}
        -
        \Phi_{\mathrm H}^{[t]}
    \right\|_\diamond
    &\leq\frac{10t}{\sqrt N}.
    \label{eq:main-four-query-dpfc-security-example}
\end{align}
\end{theorem}

\begin{proof}
If the endpoint distribution is Haar, then, after conditioning on \(D,\pi,f\), the unitary \(DP_\pi F_f\) is fixed and \(C\sim\mu_{\mathrm H}\). 
Haar invariance implies that \(DP_\pi F_fC\) is Haar distributed, including its global phase.
Consequently,
\begin{align}
    \Phi_{\operatorname{DPFC}(\mu_{\mathrm H})}^{[t]}
    &=
    \Phi_{\mathrm H}^{[t]}.
    \label{eq:four-query-haar-endpoint-absorbs-dpfc}
\end{align}

Since \(t\leq N\), the unitary coordinate transformation \(\mathcal I^{[t]}\) is available. 
The recorder map \(\mathbf R^{[t]}\) in \eqref{eq:coherent-joint-recorder-cp-map} is the same for \(\nu=\mathfrak D\) and \(\nu=\mu_{\mathrm H}\).
Using the triangle inequality,
\begin{align}
    \left\|
        \Phi_{\operatorname{DPFC}(\mathfrak D)}^{[t]}
        -
        \Phi_{\mathrm H}^{[t]}
    \right\|_\diamond
    &=
    \left\|
        \operatorname{Ad}_{\mathcal I^{[t]}}
        \circ
        \left(
            \Phi_{\operatorname{DPFC}(\mathfrak D)}^{[t]}
            -
            \Phi_{\operatorname{DPFC}(\mu_{\mathrm H})}^{[t]}
        \right)
    \right\|_\diamond
    \nonumber\\
    & \leq
    \left\|
        \operatorname{Ad}_{\mathcal I^{[t]}}
        \circ
        \Phi_{\operatorname{DPFC}(\mathfrak D)}^{[t]}
        -
        \mathbf R^{[t]}
    \right\|_\diamond
    +
    \left\|
        \mathbf R^{[t]}
        -
        \operatorname{Ad}_{\mathcal I^{[t]}}
        \circ
        \Phi_{\operatorname{DPFC}(\mu_{\mathrm H})}^{[t]}
    \right\|_\diamond
    \nonumber \\
    &\leq
    2\left(1+e^{t^2/(4N)}\right)
    \sqrt{\frac{4t^2}{N}+e^{t^2/(2N)}-1}.
    \label{eq:main-four-query-dpfc-security-triangle}
\end{align}
The final inequality applies Lemma~\ref{lem:four-query-coherent-sector-assembly} to \(\nu=\mathfrak D\) and \(\nu=\mu_{\mathrm H}\), both exact unitary \(2\)-designs.

Choosing \(t^2\leq N/16\) gives
\begin{align}
    2\left(1+e^{t^2/(4N)}\right)
    \sqrt{\frac{4t^2}{N}+e^{t^2/(2N)}-1}
    &\leq
    2\left(1+e^{1/64}\right)\sqrt5\,\frac{t}{\sqrt N}
    \leq\frac{10t}{\sqrt N},
\end{align}
which proves \eqref{eq:main-four-query-dpfc-security-example}.
\end{proof}

\section{An Adaptive Birthday-Scale Attack on DPFC}
\label{sec:cba-attack}

The proof of Theorem~\ref{thm:main-four-query-dpfc-security} establishes non-adaptive security of DPFC, allowing arbitrary entangled inputs and coherent superpositions of query types. 
A natural follow-up question is whether the same security parameters can be achieved in an adaptive model, where arbitrary known operations may be performed between oracle queries. 
In this section, we show that, for DPFC with Clifford endpoints, an adaptive adversary can distinguish the DPFC and Haar-random ensembles with constant advantage using \(O(\sqrt N)\) queries to \(U\) and \(U^\dagger\).

Clifford endpoints provide an efficiently implementable way to randomize the input and output of the phase-permutation layer.
However, as shown in this section, since Clifford operators normalize the Pauli group, for a fraction of at least \(1/(N+1)\) nonidentity Pauli operators \(Q\), the operator \(UQU^\dagger\) has a Pauli expansion supported on mutually commuting Paulis. 
This signature is exploited using a constant-query commutation test and coherent search over Pauli labels.
The attack reaches constant advantage at the birthday scale, complementing the vanishing non-adaptive advantage for \(t=o(\sqrt N)\) established in Theorem~\ref{thm:main-four-query-dpfc-security}.

\begin{theorem}[Adaptive attack on DPFC with Clifford endpoints]
\label{thm:cba-main}
Set \(N_0=2^{15}\) and \(\varepsilon=9/100\).
Let \(N=2^n\geq N_0\). 
There exists an adaptive quantum adversary \(\mathcal A_N\) making \(O(\sqrt N)\) queries to \(U\) and \(U^\dagger\) such that
\begin{align}
    \Pr_{U\sim\mathsf{DPFC}(\text{Clifford}(n))}
        [\mathcal A_N^U=1]
    -
    \Pr_{U\sim\mu_{\mathrm H}}
        [\mathcal A_N^U=1]
    \geq \varepsilon.
\end{align}
The choice of \(N_0\) and \(\varepsilon\) is explained in the proof.
The query directions are fixed in advance, with known operations between queries.
\end{theorem}

The proof proceeds in three steps. 
First, a four-query commutation test checks whether two Pauli terms sampled from the expansion of \(UQU^\dagger\) commute where \(Q\) is a Pauli itself. 
Second, the test is shown to accept with certainty on a hidden fraction \(1/(N+1)\) of the Pauli labels for DPFC, while satisfying a soundness bound simultaneously for all nonidentity Pauli labels in the Haar case with probability close to \(1\). 
Finally, a coherent search Grover's-style algorithm uses this test to find a suitable Pauli candidate. 
Repeated testing of the candidate then yields the claimed distinguishing advantage.

\subsection{A Four-Query Pauli Commutation Test}
\label{subsec:cba-verifier}

This subsection introduces a test for the commutation property exploited by the attack. 
Let \(Q\) be a nonidentity Hermitian Pauli representative.
Given oracle access to a fixed unknown unitary \(U\) and its inverse \(U^\dagger\), the test draws two independent Pauli samples from the expansion of \(UQU^\dagger\), with probabilities given by the squared coefficients, and accepts if the sampled Paulis commute. 
Each sample uses one forward and one inverse query, for a total of four queries.
More concretly, expanding \(UQU^\dagger\) in the Pauli basis,
\begin{align}
    \begin{aligned}
    UQU^\dagger=\sum_{a\in\mathbb F_2^{2n}}r_a(U,Q)P_a,\quad
    r_a(U,Q)=\frac1N\operatorname{Tr}(P_aUQU^\dagger).
    \end{aligned}
    \label{eq:cba-expansion}
\end{align}
Since \(UQU^\dagger\) is Hermitian, traceless, and squares to the identity, the coefficients are real and
\begin{align}
    r_0(U,Q)=0,\qquad
    \sum_{a\in\mathbb F_2^{2n}}r_a(U,Q)^2
       =\frac1N\operatorname{Tr}\bigl((UQU^\dagger)^2\bigr)=1.
    \label{eq:cba-coefficients}
\end{align}

Define the Bell basis by
\begin{align}
    |B_a\rangle:=(P_a\otimes I)|\Omega_N\rangle,
\end{align}
where \(|\Omega_N\rangle\) is the normalized EPR state.
Equation~\eqref{eq:cba-paulis} implies that the states \(|B_a\rangle\) form an orthonormal basis. 
The test performs the following steps:

\par\medskip
\noindent\begin{minipage}{\linewidth}
\hrule\smallskip
\refstepcounter{algorithm}\label{alg:cba-verifier}
\noindent\textbf{Algorithm~\thealgorithm.} Four-query Pauli commutation test
\par\smallskip\hrule\smallskip
\noindent\textbf{Input:} A nonidentity Hermitian Pauli \(Q\) and
oracle access to \(U,U^\dagger\).
\begin{enumerate}
    \item \label{step-1}Prepare \(|\Omega_N\rangle\) and apply \(UQU^\dagger\) to its first register to obtain \(\sum_a r_a(U,Q)|B_a\rangle\).
    \item \label{step-2}Measure both registers of the resulting state in the Bell basis to obtain \(a\in\mathbb F_2^{2n}\) with probability \(r_a(U,Q)^2\).
    \item Repeat steps \ref{step-1},\ref{step-2} on fresh registers to obtain \(b\in\mathbb F_2^{2n}\).
    \item Accept if \([a,b]=0\). Otherwise, reject.
\end{enumerate}
\hrule
\end{minipage}
\par\medskip

The acceptance probability is therefore
\begin{align}
    \begin{aligned}
    q_Q(U)
    =\sum_{\substack{a,b\in\mathbb F_2^{2n}\\{[a,b]=0}}}r_a(U,Q)^2r_b(U,Q)^2
    =\frac{1+\sum_{a,b\in\mathbb F_2^{2n}}(-1)^{[a,b]}r_a(U,Q)^2r_b(U,Q)^2}{2}.
    \end{aligned}
    \label{eq:cba-verifier-probability}
\end{align}
The test uses four oracle queries, in the fixed chronological order \((U^\dagger,U,U^\dagger,U)\).

\begin{lemma}[Hidden perfect witnesses]
\label{lem:cba-witnesses}
Let \(U=DPFC\) where \(C,D\) are sampled from the Clifford group.
Then,
\begin{align}
    \left|\left\{
        a\in\mathbb F_2^{2n}\setminus\{0\}:q_{P_a}(U)=1
    \right\}\right|
    \geq N-1.
\end{align}
Consequently, when the unitary in the test above is sampled from the DPFC ensemble with Clifford endpoints, the test accepts with certainty on at least a \(1/(N+1)\) fraction of the nonidentity Pauli labels. 
\end{lemma}

The identity operator is not considered, as the test will always pass when \(Q=I\), regardless of the sampled unitary.

\begin{proof}
For each nonzero \(s\in\mathbb F_2^n\), let \(Q_s\) be the Hermitian Pauli representative equal to \(C^\dagger Z^sC\) up to sign.\footnote{Notice that for \(U=PFC\), the Pauli witnesses \(C^\dagger Z^sC\) are, up to sign, exactly the stabilizers of the states \(U^\dagger|0\rangle \propto C^\dagger|\pi^{-1}(0)\rangle\) used in the PFC inverse-query attack in \cite{MPSY24}. 
The additional Clifford in the DPFC construction prevents the state \(U^\dagger|0\rangle\) from always being a stabilizer state, so this test cannot be used directly.
Instead, the operators \(C^\dagger Z^sC\) serve as inputs to an operator commutation test.}
Since \(C\) is Clifford and conjugation preserves orthogonality, these are \(N-1\) distinct nonidentity representatives.

The diagonal operators \(F\) and \(Z^s\) commute, meaning that
\begin{align}
    UQ_sU^\dagger = (DPFC)(C^\dagger Z^s C)( C^\dagger F^\dagger P^\dagger D^\dagger)
    =\pm DP Z^sP^\dagger D^\dagger.
\end{align}
The operator \(P Z^s P^\dagger\) is diagonal and is hence a linear combination of the mutually commuting Paulis \(Z^v\).
Since \(D\) is a Clifford, each \(DZ^vD^\dagger\) is a Pauli up to sign. 
These Paulis commute pairwise, as conjugation preserves commutation. 
Thus, the Pauli expansion of \(UQ_sU^\dagger\) has mutually commuting support.
Both Bell outcomes sampled by the test lie in this support, so they always commute. 
Therefore \(q_{Q_s}(U)=1\) for each \(s\), showing the claimed \(N-1\) witnesses.
\end{proof}

Distinguishing the two ensembles also requires ruling out high acceptance in the Haar case.
The following lemma shows that with high probability over a Haar-random \(U\), \emph{every} nonidentity Pauli has a bounded acceptance probability. 

\begin{lemma}[Uniform Haar acceptance bound]
\label{lem:cba-haar}
For \(N\geq2\) and \(0<\delta<1\), define
\begin{align*}
    h_N(\delta):=\frac12+\frac{24}{N}\ln\frac{2N^4}{\delta}.
\end{align*}
Then
\begin{align}
    \Pr_{U\sim\mu_{\mathrm H}}
      \bigl[q_Q(U)\leq h_N(\delta)
            \text{ for every nonidentity Pauli }Q\bigr]
       \geq1-\delta.
\end{align}
\end{lemma}

\begin{proof}
    See Appendix~\ref{app:cba-haar}.
\end{proof}

\subsection{Searching for a Pauli with High Acceptance Probability}
\label{subsec:cba-search}

The commutation test separates the two ensembles when a suitable witness is known: some Pauli candidates have acceptance probability one in the case of DPFC, whereas Lemma~\ref{lem:cba-haar} guarantees that with probability at least \(1-\delta\) over Haar-random \(U\), every nonidentity Pauli candidate has acceptance probability at most \(h_N(\delta)\). 
The witnesses, however, depend on the Clifford endpoint \(C\) which is unknown to the distinguisher. 
As shown in Lemma~\ref{lem:cba-witnesses}, among the \(N^2-1\) nonidentity Pauli labels, the DPFC case contains at least \(N-1\) perfect witnesses, giving uniform density at least \( p_{\text{witness}} = 1/(N+1)\).
The algorithm presented here uses a Grover-style search to find a suitable witness.
A variation is required because the oracle is probabilistic on the candidates that should be rejected.
For a standard Grover search, constant success probability can be obtained by performing \(O(\log N)\) coherent repetitions of the commutation test in each Grover iteration, leading to a total query budget of \(O(\sqrt{N}\log N)\).\footnote{Prepare a uniform superposition over non-identity Pauli candidates. Controlled on the candidate, apply the coherent implementation of the four-query commutation test \(r\) times, using a separate initially zero workspace for each application, and compute an acceptance flag that equals one only if all trials accept. Leave all registers unmeasured. Candidate \(i\) passes with probability \(q_i^r\). Perfect witnesses always pass, so the total acceptance probability is at least \(p_{\mathrm{witness}}\). Candidates with \(q_i\le\tau\) contribute at most \(\tau^r\) to this probability. Thus, conditioned on acceptance, the probability to obtain a candidate with \(q_i < \tau\) is at most \(\tau^r/p_{\mathrm{witness}}\). Choosing \(r\sim \log N\) makes this probability constant for a constant \(\tau\). Standard amplitude amplification preserves this ratio within the accepted candidates, using \(O(p_{\mathrm{witness}}^{-1/2})\) calls to the preparation circuit and its inverse.}
Instead, a modification with a gradually increasing number of coherent repetitions of the commutation test (and hence queries to the oracle) requiring only \(O(\sqrt{N})\) queries overall is used.
The algorithm adapts the recursive search construction of \cite{HMW03}.

\begin{lemma}[Perfect-witness search]
\label{lem:cba-search}
Let \(\mathcal I\) be a finite candidate set and let
\begin{align*}
    W=\sum_{i\in\mathcal I}|i\rangle\langle i|\otimes W_i
\end{align*}
be a unitary called \emph{coherent verifier} acting on a candidate register and a workspace, such that
\begin{align}
    W_i|0\rangle
    =\sqrt{q_i}\,|\psi_{i,1}\rangle|1\rangle
     +\sqrt{1-q_i}\,|\psi_{i,0}\rangle|0\rangle,
\end{align}
where \(q_i\in[0,1]\) is the acceptance probability of candidate \(i\), and the states \(|\psi_{i,b}\rangle\) are normalized.
Assume access to \(W,W^\dagger\) and to a known unitary preparing \(\sum_{i\in\mathcal I}\sqrt{w_i}|i\rangle\), where \(w_i\geq0\), and \(\sum_iw_i=1\).
Suppose an unknown set \(S\subseteq\mathcal I\) satisfies
\begin{align}
    q_i=1\quad(i\in S),\qquad
    \sum_{i\in S}w_i\geq p_{\mathrm{witness}},
\end{align}
where \(p_{\mathrm{witness}}\in(0,1]\) is known in advance.
Fix \(0<\tau<1\) and choose a sufficiently large integer \(m\geq1\), depending only on \(\tau\).
There is a quantum algorithm returning a candidate \(i\) with \(q_i>\tau\) with constant probability at least \(1-e^{-1/8}-\tau^m/(1-\tau)\), using \(O(p_{\mathrm{witness}}^{-1/2})\) applications of \(W\) or \(W^\dagger\).
Explicitly, the number of applications is less than
\begin{align}
    C_\tau p_{\mathrm{witness}}^{-1/2},
    \qquad C_\tau=\frac{27(2m+1)}8.
\end{align}
\end{lemma}

\begin{proof}
    See Appendix~\ref{app:cba-search}.
\end{proof}

\subsection{An Adaptive Distinguisher Between DPFC with Clifford Endpoints and Haar}

The distinguisher searches for a candidate Pauli and then estimates its acceptance probability through repeated commutation tests. 
The full description of the algorithm is shown first, and then the numerical parameter choice.

Let \(W\) be the coherent implementation of the commutation test from Algorithm~\ref{alg:cba-verifier}, conditioned on the candidate label and leaving both the Bell-sample registers and the acceptance flag unmeasured.
Explicitly,
\begin{align}
W:=\sum_{a\ne0}|a\rangle\langle a|\otimes W_a,
\qquad
W_a|0\rangle
:=
\sum_{b,c\in\mathbb F_2^{2n}}
r_b(U,P_a)r_c(U,P_a)
|b\rangle|c\rangle|\mathbf1_{[b,c]=0}\rangle.
\end{align}
Here \(W_a\) is the coherent verifier for candidate \(P_a\), and \(|0\rangle\) denotes its initially zero workspace.

In the following algorithm, \(\delta\) bounds failure of the Haar guarantee, \(\alpha\) bounds the final test error, \(\tau\) is the search threshold and in \(r:=\left\lceil\frac{2\ln(1/\alpha)}{(\tau -h_N(\delta))^2}\right\rceil\) counts commutation-test trials on the selected candidate.

\par\medskip
\noindent\begin{minipage}{\linewidth}
\hrule\smallskip
\refstepcounter{algorithm}\label{alg:cba-distinguisher}
\noindent\textbf{Algorithm~\thealgorithm.} Adaptive Distinguishing of DPFC from Haar
\par\smallskip\hrule\smallskip
\noindent\textbf{Input:} Oracle access to \(U,U^\dagger\), the coherent implementation \(W\) of the commutation test, and parameters \(\delta,\alpha\in(0,1)\), \(h_N(\delta)<\tau<1\).
\begin{enumerate}
    \item Prepare the uniform superposition \(\frac1{\sqrt{N^2-1}}\sum_{a\ne0}|a\rangle\) and run the search of Lemma~\ref{lem:cba-search} (Algorithm~\ref{alg:cba-search}) using \(W\), for acceptance threshold \(\tau\) and with \(p_{\mathrm{witness}}=1/(N+1)\).
    Obtain a candidate Pauli \(Q\) or \(\perp\).
    \item If the search returns \(\perp\), choose an arbitrary fixed nonidentity Pauli \(Q\). 
    Run \(r\) fresh trials of the four-query commutation test on \(Q\).
    \item Accept if at least a \((\tau+h_N(\delta))/2\) fraction of the test trials accept.
    Otherwise, reject.
\end{enumerate}
\hrule
\end{minipage}
\par\medskip

\begin{proof}[Proof of Theorem~\ref{thm:cba-main}]

Conditioned on a fixed \(U\) and the Pauli \(Q\) selected in the Algorithm~\ref{alg:cba-distinguisher}, let \(X_\ell\in\{0,1\}\) indicate acceptance of commutation-test trial \(\ell\). 
These variables are independent with mean \(q_Q(U)\).
Define their average \(\overline X=r^{-1}\sum_{\ell=1}^rX_\ell\). 
For every \(t>0\), the Chernoff-Hoeffding inequality~\cite{Hoeffding63} gives
\begin{align}
    \Pr[\overline X-q_Q(U)\geq t]\leq e^{-2rt^2},
    \qquad
    \Pr[q_Q(U)-\overline X\geq t]\leq e^{-2rt^2}.
\end{align}
The final test accepts when \(\overline X\geq(\tau+h_N(\delta))/2\). 
If \(q_Q(U)>\tau\), rejection requires a downward deviation from the mean of at least \((\tau-h_N(\delta))/2\). 
If \(q_Q(U)\leq h_N(\delta)\), acceptance requires an upward deviation of at least the same size.
Applying these bounds with \(t=(\tau-h_N(\delta))/2\) and using the definition of \(r\), each error probability is at most \(\alpha\).
Thus, the final test accepts with probability at least \(1-\alpha\) when \(q_Q(U)>\tau\), and at most \(\alpha\) when \(q_Q(U)\leq h_N(\delta)\).

By Lemmas~\ref{lem:cba-witnesses} and~\ref{lem:cba-search}, the DPFC case has witness density at least \(1/(N+1)\), and the search returns \(q_Q(U)>\tau\) with probability at least \(s:=1-e^{-1/8}-\tau^m/(1-\tau)>0\), using at most \(C_\tau\sqrt{N+1}\) applications of \(W\) or \(W^\dagger\).
Hence the distinguisher accepts DPFC with probability at least \(s(1-\alpha)\). 
In the Haar case, Lemma~\ref{lem:cba-haar} bounds every nonidentity Pauli's acceptance by \(h_N(\delta)\), except with probability \(\delta\), so the acceptance probability is at most \(\alpha+\delta\). 
Therefore, the distinguishing advantage satisfies
\begin{align}
    \operatorname{Adv}(\mathcal A_N) = \Pr_{U\sim\mathsf{DPFC}(\text{Clifford}(n))}
        [\mathcal A_N^U=1]
    -
    \Pr_{U\sim\mu_{\mathrm H}}
        [\mathcal A_N^U=1]
       \geq s(1-\alpha)-\alpha-\delta.
    \label{eq:cba-combined-advantage}
\end{align}
Each application of \(W\) or \(W^\dagger\) costs four oracle queries, giving a total query count of
\begin{align}
    Q_{\mathrm{tot}}\leq4C_\tau\sqrt{N+1}+4r.
    \label{eq:cba-combined-cost}
\end{align}
It is satisfied that \(r \leq \lceil\frac{2 \ln(1/\alpha)}{c^2}\rceil \), for a constant \(0<c \leq \tau -h_N(\delta)\), and thus \(Q_{\mathrm{tot}}=O(\sqrt N)\) for fixed \(\tau,\alpha\).

For the choice of parameters, take
\begin{align*}
    \delta=2N^{-6},\qquad\tau=\frac34,\qquad\alpha=10^{-3}.
\end{align*}
For the search in Appendix~\ref{app:cba-search}, choose \(\eta=1/8\) and \(m=21\): \(\eta\) is an analysis parameter, while \(m\) fixes the initial number of applications of \(W\). 
Since \(\ln N/N\) is decreasing for every \(N\geq2^{15}\),
\begin{align}
    h_N(\delta)-\frac12
    =\frac{240\ln N}{N}
    \leq\frac{240\cdot15\ln2}{2^{15}}
    <\frac{240\cdot15}{2^{15}}
    <\frac18,
\end{align}
Thus \(h_N(\delta)\leq5/8\) and \(\tau-h_N(\delta)\geq1/8\).
The choice of \(m\) satisfies \(\tau^m/(1-\tau)<1-e^{-1/8}\), since
\begin{align}
    1-e^{-1/8}\geq\frac19,
    \qquad \frac{(3/4)^{21}}{1-3/4}<\frac1{100},
\end{align}
Equations~\eqref{eq:cba-search-success} and~\eqref{eq:cba-search-cost} give \(s>1/10\). 
In addition, \(\delta\leq10^{-3}\).
Substituting into \eqref{eq:cba-combined-advantage} gives
\begin{align}
    \begin{aligned}
    \operatorname{Adv}(\mathcal A_N)
       &>\frac1{10}(1-10^{-3})-2\cdot10^{-3}
       =\frac{979}{10000}>\frac9{100},\\
    \end{aligned}
\end{align}
Thus \(N_0=2^{15}\) and \(\varepsilon=9/100\) satisfy the theorem.
\end{proof}

\appendix

\section{Proof of the Common-Environment Dilation Bound}
\label{app:common-environment-stinespring-bound}

\begin{proof}[Proof of Lemma~\ref{lem:common-environment-stinespring-bound}]
Fix a reference space \(\mathcal R\), a density operator \(X\in\mathsf L(\mathcal H\otimes\mathcal R)\), and set
\begin{align*}
    \widetilde V:=V\otimes I_{\mathcal R},
    \qquad
    \widetilde W:=W\otimes I_{\mathcal R}.
\end{align*}
Contractivity of the trace norm under the partial trace and the triangle inequality give
\begin{align}
    \left\|
        \left(
            (\Phi_V-\Phi_W)
            \otimes\operatorname{id}_{\mathcal R}
        \right)(X)
    \right\|_1
    &\leq
    \left\|
        \widetilde V X\widetilde V^\dagger
        -
        \widetilde W X\widetilde W^\dagger
    \right\|_1
    \nonumber\\
    &=
    \left\|
        (\widetilde V-\widetilde W)X\widetilde V^\dagger
        +
        \widetilde W X(\widetilde V^\dagger-\widetilde W^\dagger)
    \right\|_1
    \nonumber\\
    &\leq
    \|\widetilde V-\widetilde W\|_\infty
    \|X\|_1
    \|\widetilde V^\dagger\|_\infty
    +
    \|\widetilde W\|_\infty
    \|X\|_1
    \|\widetilde V^\dagger-\widetilde W^\dagger\|_\infty
    \nonumber\\
    &=
    \bigl(\|V\|_\infty+\|W\|_\infty\bigr)
    \|V-W\|_\infty.
    \label{eq:four-query-common-stinespring-calculation}
\end{align}
For each summand, applying H\"older's inequality twice gives \(\|AXB\|_1\leq\|A\|_\infty\|XB\|_1 \leq\|A\|_\infty\|X\|_1\|B\|_\infty\).
The last equality uses \(\|X\|_1=1\), invariance of the operator norm under taking adjoints, and invariance under tensoring with an identity. 
Taking the supremum over \(\mathcal R\) and all density operators \(X\) proves \eqref{eq:four-query-common-environment-diamond-bound}.
If \(V\) and \(W\) are isometries, their operator norms are one, which gives the stated factor-two bound.
\end{proof}

\section{Proof of the Double EPR-Sector Decomposition}
\label{app:proof-double-epr-sector-coordinates}

\begin{proof}[Proof of Lemma~\ref{prop:exact-double-epr-sector-coordinates}]
Apply Lemma~\ref{lemma:orthogonal-epr-sector-coordinates} to the \(U,U^*\) registers with \((u,v)=(a,c)\), and independently to the \(U^T,U^\dagger\) registers with \((u,v)=(b,d)\).

First, for \(\lambda=(\ell,m)\) and \(\lambda'=(\ell',m')\), then
\begin{align}
    P_\lambda P_{\lambda'}
    &=
    \left(P_\ell^{(a,c)}P_{\ell'}^{(a,c)}\right)
    \left(P_m^{(b,d)}P_{m'}^{(b,d)}\right) =
    \delta_{\ell,\ell'}\delta_{m,m'}
    P_\ell^{(a,c)}P_m^{(b,d)}=
    \delta_{\lambda,\lambda'}P_\lambda,
\end{align}
which proves~\eqref{eq:double-sector-orthogonality}. 
In addition, 
\begin{align}
    I_\lambda^\dagger I_\lambda
    = P_\ell^{(a,c)}P_m^{(b,d)} = P_\lambda, \quad
    I_\lambda I_\lambda^\dagger
    =
    \Pi_{a-\ell,c-\ell}^{U,U^*}
    \Pi_{b-m,d-m}^{U^T,U^\dagger}
    \otimes I_{\mathcal M_\ell^{(a,c)}}
    \otimes I_{\mathcal M_m^{(b,d)}} =
    \Pi_{\theta_\lambda}^{\mathrm{nE}}
    \otimes I_{\mathcal M_\lambda}.
\end{align}
This proves~\eqref{eq:double-sector-isometry-relations}.

For the equivariance identity, Equation~\eqref{eq:epr-sector-intertwining} gives
\begin{align}
    I_\ell^{(a,c)}
    \bigl(U^{\otimes a}\otimes(U^*)^{\otimes c}\bigr)
    &=
    \left(
        U^{\otimes(a-\ell)}
        \otimes(U^*)^{\otimes(c-\ell)}
        \otimes I_{\mathcal M_\ell^{(a,c)}}
    \right)I_\ell^{(a,c)},\\
    I_m^{(b,d)}
    \bigl((U^T)^{\otimes b}\otimes(U^\dagger)^{\otimes d}\bigr)
    &=
    \left(
        (U^T)^{\otimes(b-m)}
        \otimes(U^\dagger)^{\otimes(d-m)}
        \otimes I_{\mathcal M_m^{(b,d)}}
    \right)I_m^{(b,d)}.
\end{align}
Combining these identities in canonical query order gives
\begin{align}
    I_\lambda\U_\theta(U)
    &=
    I_\lambda
    \left[
        U^{\otimes a}
        \otimes(U^T)^{\otimes b}
        \otimes(U^*)^{\otimes c}
        \otimes(U^\dagger)^{\otimes d}
    \right] \notag\\
    &=
    \left(
        U^{\otimes(a-\ell)}
        \otimes(U^T)^{\otimes(b-m)}
        \otimes(U^*)^{\otimes(c-\ell)}
        \otimes(U^\dagger)^{\otimes(d-m)}
        \otimes I_{\mathcal M_\ell^{(a,c)}}
        \otimes I_{\mathcal M_m^{(b,d)}}
    \right)I_\lambda \notag\\
    &=
    \left(
        \U_{\theta_\lambda}(U)
        \otimes I_{\mathcal M_\lambda}
    \right)I_\lambda,
\end{align}
which is~\eqref{eq:double-sector-intertwining}. 
Furthermore,
\begin{align}
    \sum_{\lambda=(\ell,m)}P_\lambda
    =
    \left(\sum_\ell P_\ell^{(a,c)}\right)
    \left(\sum_m P_m^{(b,d)}\right)=
    I_{\mathcal A_{a,b,c,d}},
\end{align}
which proves~\eqref{eq:double-sector-resolution}. 
Tensoring the multiplicity spaces gives the dimension formula. 
Finally, \eqref{eq:double-sector-orthogonality}, \eqref{eq:double-sector-isometry-relations}, and \eqref{eq:double-sector-resolution} show that \(\mathcal I_\theta\) is an isometry onto the mutually orthogonal summands of \(\widetilde{\mathcal A}_\theta\), and hence is unitary.
\end{proof}

\section{Proof of Unitary Invariance}
\label{app:proof-endpoint-covariance}

\begin{proof}[Proof of Lemma~\ref{lem:four-type-endpoint-covariance}]
For every operator \(M\) on \(\mathcal H_N\), the normalized Bell pair of Definition~\ref{def:normalized-epr-state} satisfies
\begin{align}
    (M\otimes I)|\Omega_N\rangle
    =(I\otimes M^T)|\Omega_N\rangle.
\end{align}
By \eqref{eq:ordered-left-map}-\eqref{eq:ordered-right-map}, \(E_{\mathrm L}^{\mathrm{ord}}\) places the input in \(\mathcal E_{\mathrm X}\) and prepares a Bell pair on \(Q\mathcal E_{\mathrm Y}\), while \(E_{\mathrm R}^{\mathrm{ord}}\) places the input in \(\mathcal E_{\mathrm Y}\) and prepares a Bell pair on \(Q\mathcal E_{\mathrm X}\).
Thus, an input operator acts on the stored endpoint, and an output operator on \(Q\) moves to the other endpoint with a transpose.
Applying the Bell-pair identity with \(M=D,C^T,D^*,C^\dagger\), respectively, gives
\begin{align}
    (D\otimes I_{\mathcal E})E_{\mathrm L}^{\mathrm{ord}}C
    &=
    (I_Q\otimes C\otimes D^T)
    E_{\mathrm L}^{\mathrm{ord}},
    \nonumber \\ 
    (C^T\otimes I_{\mathcal E})E_{\mathrm R}^{\mathrm{ord}}D^T
    &=
    (I_Q\otimes C\otimes D^T)
    E_{\mathrm R}^{\mathrm{ord}},
    \nonumber \\ 
    (D^*\otimes I_{\mathcal E})
    E_{\mathrm L}^{\mathrm{ord}}C^*
    &=
    (I_Q\otimes C^*\otimes D^\dagger)
    E_{\mathrm L}^{\mathrm{ord}},
    \nonumber \\ 
    (C^\dagger\otimes I_{\mathcal E})
    E_{\mathrm R}^{\mathrm{ord}}D^\dagger
    &=
    (I_Q\otimes C^*\otimes D^\dagger)
    E_{\mathrm R}^{\mathrm{ord}}.
\end{align}

Tensoring these identities over the query registers, with the output ordering in Definition~\ref{def:ordered-parallel-four-query-recorder}, gives
\begin{align}
    &\left[
        \left(
            D^{\otimes a_\lambda}
            \otimes(C^T)^{\otimes b_\lambda}
            \otimes(D^*)^{\otimes c_\lambda}
            \otimes(C^\dagger)^{\otimes d_\lambda}
        \right)
        \otimes I_{\mathcal E^{\otimes t_\lambda}}
    \right]
    E_{\theta_\lambda}^{\mathrm{ord}}
    \nonumber\\
    &\qquad\cdot
    \left(
        C^{\otimes a_\lambda}
        \otimes(D^T)^{\otimes b_\lambda}
        \otimes(C^*)^{\otimes c_\lambda}
        \otimes(D^\dagger)^{\otimes d_\lambda}
    \right)
    \nonumber\\
    &\quad=
    \left[
        I_{\mathcal A_{\theta_\lambda}}
        \otimes\mathcal Q_{\theta_\lambda}^{\mathrm{ord}}[C,D]
    \right]
    E_{\theta_\lambda}^{\mathrm{ord}}.
    \label{eq:ordered-four-query-endpoint-covariance}
\end{align}
This is the required identity for the ordered recorder. 
To pass to the symmetric recorder, observe that for every one-edge operator \(W\),
\begin{align}
    \Pi_{\mathrm{sym}}^{N^2,k}W^{\otimes k}
    &=
    W^{\otimes k}\Pi_{\mathrm{sym}}^{N^2,k},
    \label{eq:symmetric-projector-tensor-power-commutation}
\end{align}
because \(W^{\otimes k}\) commutes with every permutation of the edge slots. 
Substitute \eqref{eq:recorder-from-ordered-maps} into the left-hand side of \eqref{eq:four-type-endpoint-covariance}.
Apply \eqref{eq:ordered-four-query-endpoint-covariance}, then commute the tensor-power database rotations past the symmetrizers using \eqref{eq:symmetric-projector-tensor-power-commutation}.
Their restrictions are \(\mathcal Q_{\theta_\lambda}[C,D]\), giving the right-hand side of \eqref{eq:four-type-endpoint-covariance}.
\end{proof}

\section{Gram Analysis of the Four-Query Recorder}
\label{app:four-query-gram-analysis}

This appendix analyzes the norm distortion introduced by symmetrizing the ordered recorder. 
Section~\ref{app:common-bijective-gram} shows that the ordered and symmetric recorders assign the same weight to the bijective subspace, so their difference in squared norm is determined by the non-bijective components. 
Section~\ref{app:direct-gram-upper-bound} expresses the parallel recorder as a composition of one-edge maps and bounds the norm increase at each step. 
Combining these bounds gives the general Gram upper bound stated in Section~\ref{sec:four-query-gram-control}.

\subsection{Equality of the Recorder Gram Operators on the Bijective Database Subspace}
\label{app:common-bijective-gram}

Database operators in recorder expressions implicitly act as the identity on the query-output registers.

\begin{lemma}[Equality of the Recorder Gram Operators on the Bijective Database Subspace]
\label{lem:four-query-common-bijective-gram}
For the two bijective projectors in Definition~\ref{def:bijective-signed-relations}, the following identities hold on the residual query space \(\mathcal A_{\theta_\lambda}\):
\begin{align}
    E_{\theta_\lambda}^\dagger
    \Pi_{\theta_\lambda}^{\mathrm{bij}}
    E_{\theta_\lambda}
    &=
    \bigl(E_{\theta_\lambda}^{\mathrm{ord}}\bigr)^\dagger
    \Pi_{\theta_\lambda}^{\mathrm{bij,ord}}
    E_{\theta_\lambda}^{\mathrm{ord}},
    \label{eq:four-query-common-bijective-gram}\\
    G_{\theta_\lambda}-I
    &=
    E_{\theta_\lambda}^\dagger
    \bigl(I-\Pi_{\theta_\lambda}^{\mathrm{bij}}\bigr)
    E_{\theta_\lambda}
    -
    \bigl(E_{\theta_\lambda}^{\mathrm{ord}}\bigr)^\dagger
    \bigl(I-\Pi_{\theta_\lambda}^{\mathrm{bij,ord}}\bigr)
    E_{\theta_\lambda}^{\mathrm{ord}}.
    \label{eq:four-query-gram-collision-identity}
\end{align}
\end{lemma}

The Gram operators agree on the bijective subspace, but both recorders can place weight outside this component.
Equation~\eqref{eq:four-query-gram-collision-identity} expresses the Gram defect as the difference of their non-bijective weights.

\begin{proof}
Identify the symmetric databases with their subspaces in the ordered database spaces, extending \(\Pi_{\theta_\lambda}^{\mathrm{bij}}\) by zero on the orthogonal complement. 
Each relation state in the defining sum \eqref{eq:bijective-signed-database-projector} is fixed by the symmetric-subspace projectors of both databases. 
Substituting \eqref{eq:recorder-from-ordered-maps} therefore gives
\begin{align}
    E_{\theta_\lambda}^\dagger
    \Pi_{\theta_\lambda}^{\mathrm{bij}}E_{\theta_\lambda}
    =p_\lambda!q_\lambda!\,
    \bigl(E_{\theta_\lambda}^{\mathrm{ord}}\bigr)^\dagger
    \Pi_{\theta_\lambda}^{\mathrm{bij}}
    E_{\theta_\lambda}^{\mathrm{ord}}.
\end{align}
Thus, it remains to prove
\begin{align}
    p_\lambda!q_\lambda!\,
    \bigl(E_{\theta_\lambda}^{\mathrm{ord}}\bigr)^\dagger
    \Pi_{\theta_\lambda}^{\mathrm{bij}}
    E_{\theta_\lambda}^{\mathrm{ord}}
    =\bigl(E_{\theta_\lambda}^{\mathrm{ord}}\bigr)^\dagger
    \Pi_{\theta_\lambda}^{\mathrm{bij,ord}}
    E_{\theta_\lambda}^{\mathrm{ord}}.
\end{align}
Fix computational-basis inputs \(|\boldsymbol\xi\rangle\) and \(|\boldsymbol\xi'\rangle\). 
In each input, the labels in the \(U\) and \(U^*\) query slots must be pairwise distinct, and likewise for the labels in the \(U^T\) and \(U^\dagger\) query slots. 
Otherwise, every ordered history produced from that input has a repeated domain or image label and is annihilated by both bijective projectors.
The corresponding other matrix elements on both sides are therefore zero.
Assume henceforth that these distinctness conditions hold for both inputs.

Expand the ordered recorders in the computational basis. 
The query-output inner product forces the two generated output strings to coincide. 
Denote their common value by \(\boldsymbol w\). 
Let \(\boldsymbol e\) and \(\boldsymbol e'\) be the resulting ordered edge lists. In slot \(i\),
\begin{align}
    e_i=
    \begin{cases}
        (\xi_i,w_i),&\text{if the slot uses }E_{\mathrm L}^{\mathrm{ord}},\\
        (w_i,\xi_i),&\text{if the slot uses }E_{\mathrm R}^{\mathrm{ord}},
    \end{cases}
\end{align}
and \(e_i'\) is defined identically with \(\xi_i'\) in place of \(\xi_i\). 
Denote their underlying signed relations by \((\Splus,\Sminus)\) and \((T_+,T_-)\), respectively. 
Every edge in a bijective relation has multiplicity one.
\begin{align}
    p_\lambda!q_\lambda!\,
    \langle\boldsymbol e'|
    \Pi_{\theta_\lambda}^{\mathrm{bij}}
    |\boldsymbol e\rangle
    =
    \begin{cases}
        1,&(\Splus,\Sminus)=(T_+,T_-)\in\mathcal B_{\theta_\lambda},\\
        0,&\text{otherwise}.
    \end{cases}
\end{align}
By definition, the ordered projector satisfies
\begin{align}
    \langle\boldsymbol e'|
    \Pi_{\theta_\lambda}^{\mathrm{bij,ord}}
    |\boldsymbol e\rangle
    =
    \begin{cases}
        1,&\boldsymbol e'=\boldsymbol e
        \text{ and }(\Splus,\Sminus)\in\mathcal B_{\theta_\lambda},\\
        0,&\text{otherwise}.
    \end{cases}
\end{align}

For a fixed bijective signed relation, the common query-output string determines the edge in every slot. 
Indeed, suppose \((\Splus,\Sminus)=(T_+,T_-)\in\mathcal B_{\theta_\lambda}\).
In a slot using \(E_{\mathrm L}^{\mathrm{ord}}\), both \(e_i\) and \(e_i'\) have image label \(w_i\). 
Since the common relation contains only one edge with that image label, these edges must coincide. 
In a slot using \(E_{\mathrm R}^{\mathrm{ord}}\), the same argument applies to their common domain label \(w_i\). 
Thus, the ordered edge lists agree slot by slot. 
Conversely, equal ordered edge lists clearly give equal signed relations, so the two displayed conditions are equivalent.

Consequently, the two displayed matrix elements agree for every common output string \(\boldsymbol w\). 
Summing over these strings, with the common recorder factor \(N^{-t_\lambda}\), gives
\begin{align}
    p_\lambda!q_\lambda!\,
    \langle\boldsymbol\xi'|
    \bigl(E_{\theta_\lambda}^{\mathrm{ord}}\bigr)^\dagger
    \Pi_{\theta_\lambda}^{\mathrm{bij}}
    E_{\theta_\lambda}^{\mathrm{ord}}
    |\boldsymbol\xi\rangle=
    \langle\boldsymbol\xi'|
    \bigl(E_{\theta_\lambda}^{\mathrm{ord}}\bigr)^\dagger
    \Pi_{\theta_\lambda}^{\mathrm{bij,ord}}
    E_{\theta_\lambda}^{\mathrm{ord}}
    |\boldsymbol\xi\rangle.
\end{align}

Subtract this common contribution from the total Grams in \eqref{eq:four-query-gram-operator} and \eqref{eq:ordered-recorder-isometry} to obtain \eqref{eq:four-query-gram-collision-identity}.
\end{proof}

\subsection{A Direct Gram Upper Bound}
\label{app:direct-gram-upper-bound}

Following~\cite{MH24}, define the one-edge recorders below.

\begin{definition}[One-edge recorders]
\label{def:symmetric-one-edge-recorders}
Let \(\mathcal D\) be the space of finite linear combinations of symmetric relation states of arbitrary size, with different database sizes orthogonal. 
For the query register \(Q\), define
\begin{align*}
    E_{\mathrm L},E_{\mathrm R}:
    (\mathcal H_N)_Q\otimes\mathcal D
    \longrightarrow
    (\mathcal H_N)_Q\otimes\mathcal D
\end{align*}
on relation states by
\begin{align}
    E_{\mathrm L}|x\rangle_Q|S\rangle_{\mathcal D}
    &:=\frac1{\sqrt N}\sum_{y\in[N]}
      \sqrt{m_{(x,y)}(S)+1}
      |y\rangle_Q|S\uplus\{(x,y)\}\rangle_{\mathcal D},
    \;
    \nonumber \\
    E_{\mathrm R}|y\rangle_Q|S\rangle_{\mathcal D}
    &:=\frac1{\sqrt N}\sum_{x\in[N]}
      \sqrt{m_{(x,y)}(S)+1}
      |x\rangle_Q|S\uplus\{(x,y)\}\rangle_{\mathcal D}.
    \label{eq:symmetric-one-edge-recorder}
\end{align}
\end{definition}

By construction, the parallel recorder is a composition of these one-edge maps.
Label the query registers in the four canonical blocks by \(Q_{U,i},Q_{T,i},Q_{*,i},Q_{\dagger,i}\). 
Then, for \(|\psi\rangle\in\mathcal A_{\theta_\lambda}\),
\begin{align}
    E_{\theta_\lambda}|\psi\rangle
    &=\left(\prod_{i=1}^{d_\lambda}
        (E_{\mathrm R})_{Q_{\dagger,i},-}\right)
      \left(\prod_{i=1}^{c_\lambda}
        (E_{\mathrm L})_{Q_{*,i},-}\right)
      \left(\prod_{i=1}^{b_\lambda}
        (E_{\mathrm R})_{Q_{T,i},+}\right)
      \left(\prod_{i=1}^{a_\lambda}
        (E_{\mathrm L})_{Q_{U,i},+}\right)
      |\psi\rangle|\varnothing\rangle_+|\varnothing\rangle_-.
    \label{eq:four-query-recorder-sequential-composition}
\end{align}
Each product applies slots in increasing index order. 
Empty products are identities.
This identity follows directly from \eqref{eq:parallel-four-query-recorder-basis-action}: each step contributes \(N^{-1/2}\), and the factors for \(m\) occurrences of an edge in either database multiply to \(\sqrt{m!}\).

\begin{lemma}[One-edge Gram identities and norm bounds]
\label{lem:symmetric-one-edge-gram-and-norm}
For \(j\geq0\), identify \(\mathcal D_j\) with its symmetric subspace in edge registers \(\mathcal E_1,\ldots,\mathcal E_j\), and write \(X_i,Y_i\) for the domain and image registers of \(\mathcal E_i\). 
The operator \(\operatorname{SWAP}_{A,B}\) exchanges two equally sized registers \(A,B\) and acts as the identity on every other register. 
Then,
\begin{align}
    \left.(E_{\mathrm L}^\dagger E_{\mathrm L})
    \right|_{(\mathcal H_N)_Q\otimes\mathcal D_j}
    &=\left.
      \left(I+\frac1N\sum_{i=1}^j
          \operatorname{SWAP}_{Q,X_i}\right)
      \right|_{(\mathcal H_N)_Q\otimes\mathcal D_j},
    \label{eq:symmetric-one-edge-left-gram}\\
    \left.(E_{\mathrm R}^\dagger E_{\mathrm R})
    \right|_{(\mathcal H_N)_Q\otimes\mathcal D_j}
    &=\left.
      \left(I+\frac1N\sum_{i=1}^j
          \operatorname{SWAP}_{Q,Y_i}\right)
      \right|_{(\mathcal H_N)_Q\otimes\mathcal D_j}.
    \label{eq:symmetric-one-edge-right-gram}
\end{align}
Consequently, every \(|\psi\rangle\in(\mathcal H_N)_Q\otimes\mathcal D_j\) satisfies
\begin{align}
    \bigl\|E_{\mathrm L}|\psi\rangle\bigr\|_2^2
    &\leq\left(1+\frac jN\right)\bigl\||\psi\rangle\bigr\|_2^2,
    &
    \bigl\|E_{\mathrm R}|\psi\rangle\bigr\|_2^2
    &\leq\left(1+\frac jN\right)\bigl\||\psi\rangle\bigr\|_2^2.
    \label{eq:symmetric-one-edge-norm-bound}
\end{align}
The squared operator norm of either fixed-size restriction is therefore at most \(1+j/N\).
\end{lemma}

\begin{proof}
Fix the input sector \((\mathcal H_N)_Q\otimes\mathcal D_j\).
To compute \(E_{\mathrm L}^\dagger E_{\mathrm L}\), first express the symmetric recorder in terms of the ordered recorder:
\begin{align}
    E_{\mathrm L}
    =\sqrt{j+1}\,
      \Pi_{\mathrm{sym}}^{N^2,j+1}E_{\mathrm L}^{\mathrm{ord}}.
\end{align}
To justify this, consider a relation \(S\in\mathcal R_j\), and let \(\mathcal E_0\) hold a newly appended edge \(e\). 
On this fixed-size input sector, the relation-state normalization gives
\begin{align}
    \sqrt{j+1}\,
    \Pi_{\mathrm{sym}}^{N^2,j+1}
    \bigl(|e\rangle\otimes|S\rangle\bigr)
    &=\sqrt{m_e(S)+1}\,|S\uplus\{e\}\rangle.
    \label{eq:symmetric-one-edge-multiplicity}
\end{align}
The identity follows from the expansion of \(|S\rangle\) using \eqref{eq:oriented-normalized-relation-state}. 
Applying \eqref{eq:symmetric-one-edge-multiplicity} to the ordered recorder in \eqref{eq:ordered-left-map} and comparing with \eqref{eq:symmetric-one-edge-recorder} therefore proves the displayed representation of \(E_{\mathrm L}\). 
Consequently,
\begin{align}
    E_{\mathrm L}^\dagger E_{\mathrm L}
    = (j+1)(E_{\mathrm L}^{\mathrm{ord}})^\dagger
          \bigl(\Pi_{\mathrm{sym}}^{N^2,j+1}\bigr)^\dagger
          \Pi_{\mathrm{sym}}^{N^2,j+1}
          E_{\mathrm L}^{\mathrm{ord}}
    = (j+1)(E_{\mathrm L}^{\mathrm{ord}})^\dagger
          \Pi_{\mathrm{sym}}^{N^2,j+1}
          E_{\mathrm L}^{\mathrm{ord}}.
\end{align}

Next, expand the symmetrizer on all \(j+1\) edge registers in this expression.
The operator \(\operatorname{SWAP}_{\mathcal E_0,\mathcal E_i}\) exchanges both labels of the two edges.
Decomposing an edge permutation according to the destination of slot \(0\) gives
\begin{align}
    (j+1)\Pi_{\mathrm{sym}}^{N^2,j+1}
    &=\left(I+\sum_{i=1}^j
        \operatorname{SWAP}_{\mathcal E_0,\mathcal E_i}\right)
      \bigl(I_{\mathcal E_0}\otimes
        \Pi_{\mathrm{sym}}^{N^2,j}\bigr).
    \label{eq:one-edge-symmetrizer-expansion}
\end{align}
Indeed, full symmetrization can be performed in two stages. 
First, average over all permutations of the existing \(j\) edge registers, leaving the newly appended edge in slot \(0\). 
Then, average over the \(j+1\) possible destinations of the newly appended edge: either leave it in slot \(0\), or swap it with one of the existing edges. 
Together, these choices generate every permutation of the \(j+1\) registers exactly once. 
The symmetrizer on the existing \(j\) edge registers already includes the normalization \(1/j!\), so the second average requires a factor \(1/(j+1)\). 
Moving this factor to the left-hand side gives \eqref{eq:one-edge-symmetrizer-expansion}.

Substituting \eqref{eq:one-edge-symmetrizer-expansion} into the Gram calculation, the symmetrizer \(I_{\mathcal E_0}\otimes\Pi_{\mathrm{sym}}^{N^2,j}\) on the existing \(j\) edge registers acts as the identity, because their state lies in \(\mathcal D_j\) and the ordered recorder leaves these registers unchanged.
It follows that
\begin{align}
    E_{\mathrm L}^\dagger E_{\mathrm L}
    = (E_{\mathrm L}^{\mathrm{ord}})^\dagger
          \left(I+\sum_{i=1}^j
              \operatorname{SWAP}_{\mathcal E_0,\mathcal E_i}\right)
          E_{\mathrm L}^{\mathrm{ord}} 
        = I+\sum_{i=1}^j
          (E_{\mathrm L}^{\mathrm{ord}})^\dagger
          \operatorname{SWAP}_{\mathcal E_0,\mathcal E_i}
          E_{\mathrm L}^{\mathrm{ord}}.
\end{align}

It remains to evaluate each swap term in this sum.
Applying an ordered one-query recorder, swapping the newly appended edge with an existing edge, and then applying the recorder's adjoint gives
\begin{align}
    &\bigl((E_{\mathrm L}^{\mathrm{ord}})^\dagger
        \otimes I_{\mathcal O_j}\bigr)
      \bigl(I_Q\otimes
        \operatorname{SWAP}_{\mathcal E_0,\mathcal E_i}\bigr)
      \bigl(E_{\mathrm L}^{\mathrm{ord}}
        \otimes I_{\mathcal O_j}\bigr)
      =\frac1N\operatorname{SWAP}_{Q,X_i},
    \label{eq:one-edge-left-swap-pullback}\\
    &\bigl((E_{\mathrm R}^{\mathrm{ord}})^\dagger
        \otimes I_{\mathcal O_j}\bigr)
      \bigl(I_Q\otimes
        \operatorname{SWAP}_{\mathcal E_0,\mathcal E_i}\bigr)
      \bigl(E_{\mathrm R}^{\mathrm{ord}}
        \otimes I_{\mathcal O_j}\bigr)
      =\frac1N\operatorname{SWAP}_{Q,Y_i}.
    \label{eq:one-edge-right-swap-pullback}
\end{align}
To verify this, consider a basis vector \(|a\rangle_Q|x,y\rangle_{\mathcal E_i}\), suppressing identities on untouched registers. 
Expanding the ordered recorder and applying the swap gives
\begin{align}
    (E_{\mathrm L}^{\mathrm{ord}})^\dagger
      \operatorname{SWAP}_{\mathcal E_0,\mathcal E_i}
      E_{\mathrm L}^{\mathrm{ord}}
      |a\rangle_Q|x,y\rangle_{\mathcal E_i}
    &=
      \frac1{\sqrt N}\sum_{b\in[N]}
      \left[(E_{\mathrm L}^{\mathrm{ord}})^\dagger
      |b\rangle_Q|x,y\rangle_{\mathcal E_0}\right]
      \otimes|a,b\rangle_{\mathcal E_i} \notag\\
    &=
      \frac1N\sum_{b\in[N]}\delta_{b,y}
      |x\rangle_Q|a,b\rangle_{\mathcal E_i} \notag\\
    &=
      \frac1N|x\rangle_Q|a,y\rangle_{\mathcal E_i}
      =\frac1N\operatorname{SWAP}_{Q,X_i}
      |a\rangle_Q|x,y\rangle_{\mathcal E_i}.
\end{align}
The adjoint enforces \(b=y\), leaving precisely a swap between the query register and the existing edge's domain label. 
Exchanging the roles of the domain and image labels gives \eqref{eq:one-edge-right-swap-pullback}, with \(\operatorname{SWAP}_{Q,Y_i}\) instead.

By \eqref{eq:one-edge-left-swap-pullback}, each full-edge swap in the Gram expansion contributes \(N^{-1}\operatorname{SWAP}_{Q,X_i}\).
Substituting these terms gives
\begin{align}
    E_{\mathrm L}^\dagger E_{\mathrm L}
    = I+\frac1N\sum_{i=1}^j\operatorname{SWAP}_{Q,X_i}.
\end{align}
This proves \eqref{eq:symmetric-one-edge-left-gram}.
The same calculation for \(E_{\mathrm R}\), using \eqref{eq:one-edge-right-swap-pullback}, gives \eqref{eq:symmetric-one-edge-right-gram} with \(Y_i\) in place of \(X_i\).

Finally, every swap is a self-adjoint unitary, with eigenvalues in \(\{-1,1\}\), and hence is at most \(I\) in operator order.
Thus, on this input sector,
\begin{align}
    E_{\mathrm L}^\dagger E_{\mathrm L}
    \preceq\left(1+\frac jN\right)I,
    \qquad
    E_{\mathrm R}^\dagger E_{\mathrm R}
    \preceq\left(1+\frac jN\right)I.
\end{align}
Taking expectation values in \(|\psi\rangle\) proves \eqref{eq:symmetric-one-edge-norm-bound}.
\end{proof}

\begin{proof}[Proof of Lemma~\ref{lem:four-query-gram-upper-bound}]
In the sequential composition \eqref{eq:four-query-recorder-sequential-composition}, each one-edge map increases its database size by one. 
The database sizes before these applications are therefore \(j=0,\ldots,p_\lambda-1\) for the positive database and \(j=0,\ldots,q_\lambda-1\) for the negative database.
The norm bounds of Lemma~\ref{lem:symmetric-one-edge-gram-and-norm} also hold after tensoring with the identity on all other registers.
Submultiplicativity of the operator norm consequently gives
\begin{align}
    G_{\theta_\lambda}
    \preceq \prod_{j=0}^{p_\lambda-1}\left(1+\frac jN\right)
        \prod_{j=0}^{q_\lambda-1}\left(1+\frac jN\right)I
    \preceq \exp\!\left(
            \frac{p_\lambda(p_\lambda-1)+q_\lambda(q_\lambda-1)}{2N}
        \right)I.
    \label{eq:four-query-gram-product-upper-bound}
\end{align}
The last inequality uses \(1+z\leq e^z\) for \(z\geq0\). 
Since \(p_\lambda+q_\lambda=t_\lambda\),
\begin{align}
    p_\lambda(p_\lambda-1)+q_\lambda(q_\lambda-1)
    \leq t_\lambda^2.
\end{align}
Subtracting \(I\) from the upper bound in \eqref{eq:four-query-gram-product-upper-bound} and using this inequality proves \eqref{eq:four-query-gram-upper-bound}.
Since \(G_{\theta_\lambda}=E_{\theta_\lambda}^\dagger E_{\theta_\lambda}\) is positive semidefinite, the same upper bound gives
\begin{align}
    \|E_{\theta_\lambda}\|_\infty^2
    =\|G_{\theta_\lambda}\|_\infty
    \leq e^{t_\lambda^2/(2N)},
\end{align}
which is \eqref{eq:four-query-recorder-norm-bound}.
\end{proof}

\section{Proof of Exact Bijective-Sector Matching}
\label{app:physical-signed-relation-matching}

\begin{proof}[Proof of Lemma~\ref{lem:exact-four-query-bijective-sector-matching}]
Fix a computational-basis input \(
    |\boldsymbol x\rangle_U
    |\boldsymbol y\rangle_T
    |\boldsymbol x'\rangle_*
    |\boldsymbol y'\rangle_\dagger
\).
For output strings
\(
\boldsymbol u\in[N]^{a_\lambda} \), \(
\boldsymbol v\in[N]^{b_\lambda}\), \(
\boldsymbol u'\in[N]^{c_\lambda}\), \(
\boldsymbol v'\in[N]^{d_\lambda}
\), denote the database relations
\begin{align*}
    \Splus
    &:={}
    \{(x_i,u_i):1\leq i\leq a_\lambda\}
    \uplus
    \{(v_j,y_j):1\leq j\leq b_\lambda\},\\
    \Sminus
    &:={}
    \{(x'_i,u'_i):1\leq i\leq c_\lambda\}
    \uplus
    \{(v'_j,y'_j):1\leq j\leq d_\lambda\}.
\end{align*}
For fixed \((\pi,f)\), the four basis actions above send the chosen input to
\begin{align}
    \begin{aligned}
        \omega_3^{
            \sum_i f(x_i)
            +\sum_j f(\pi^{-1}(y_j))
            -\sum_i f(x'_i)
            -\sum_j f(\pi^{-1}(y'_j))
        } \cdot
        |\pi(\boldsymbol x)\rangle_U
        |\pi^{-1}(\boldsymbol y)\rangle_T
        |\pi(\boldsymbol x')\rangle_*
        |\pi^{-1}(\boldsymbol y')\rangle_\dagger,
    \end{aligned}
\end{align}
where permutations act componentwise on the strings.
Definition~\ref{def:physical-four-query-pf-isometry} sums these vectors over \((\pi,f)\), by acting conditioned on the environment label \(|\pi,f\rangle\) in each summand. 
To express this sum in terms of the output strings, collect all summands producing the same tuple \((\boldsymbol u,\boldsymbol v,\boldsymbol u',\boldsymbol v')\).
A permutation produces this tuple precisely when
\begin{align}
    \pi(x_i)=u_i,\qquad
    \pi(v_j)=y_j,\qquad
    \pi(x'_i)=u'_i,\qquad
    \pi(v'_j)=y'_j,
\end{align}
which is equivalent to \(\pi\supseteq \Splus\uplus \Sminus\).
Under these constraints, the phase exponent becomes exactly
\( \sum_{(x,y)\in \Splus}f(x) - \sum_{(x,y)\in \Sminus}f(x)\), with edge multiplicities counted. 
This regrouping gives
\begin{align}
    &K_{\theta_\lambda}^{\mathrm{PF}}
    \left(
        |\boldsymbol x\rangle_U
        |\boldsymbol y\rangle_T
        |\boldsymbol x'\rangle_*
        |\boldsymbol y'\rangle_\dagger
    \right)
    \nonumber\\
    &\quad={}
    \sum_{\boldsymbol u\in[N]^{a_\lambda}}
    \sum_{\boldsymbol v\in[N]^{b_\lambda}}
    \sum_{\boldsymbol u'\in[N]^{c_\lambda}}
    \sum_{\boldsymbol v'\in[N]^{d_\lambda}}
    |\boldsymbol u\rangle_U
    |\boldsymbol v\rangle_T
    |\boldsymbol u'\rangle_*
    |\boldsymbol v'\rangle_\dagger
    \nonumber\\
    &\qquad\quad{}\otimes
    \left[
    \frac{1}{\sqrt{N!\,3^N}}
    \sum_{\substack{
        \pi\in S_N\\
        \pi\supseteq \Splus\uplus \Sminus
    }}
    \sum_{f:[N]\to\mathbb Z_3}
    \omega_3^{
        \sum_{(x,y)\in \Splus}f(x)
        -
        \sum_{(x,y)\in \Sminus}f(x)
    }
    |\pi,f\rangle
    \right].
    \label{eq:four-query-fixed-output-physical-vector}
\end{align}

To match the physical purification with the recorder on the bijective sector, it remains to show that physical-environment compression eliminates the contributions of non-bijective output tuples, just as the bijective projector does on the recorder side.
Consider the bracketed environment vector in \eqref{eq:four-query-fixed-output-physical-vector} for a fixed output tuple, with associated signed multirelation \((\Splus,\Sminus)\).
There are three cases.
\begin{enumerate}
    \item No permutation extends the relation.
    The sum over permutations is empty, so this output tuple contributes zero.

    \item The relation is bijective.
    In this case the bracketed term is exactly \(|\pf_{\Splus,\Sminus}\rangle\) up to the correct factor. 
    Compression therefore maps it to the database state \(|\Splus,\Sminus\rangle\), with the same coefficient.

    \item The relation has a permutation extension but is not bijective.
    Its \(t_\lambda\) edge slots involve fewer than \(t_\lambda\) distinct domains: if all domains were distinct, their images under an extending permutation would also be distinct, making the relation bijective.
    Define the signed phase character of a signed multirelation \((\Splus,\Sminus)\) by
    \begin{align}
        \varepsilon_{\Splus,\Sminus}(x)
        :=\sum_{y\in[N]}
            \bigl(m_{(x,y)}(\Splus)-m_{(x,y)}(\Sminus)\bigr)
            \pmod 3,
        \qquad x\in[N].
    \end{align}
    For a bijective pair, this is \(1\) on positive domains, \(-1\) on negative domains, and \(0\) elsewhere.

    Denote the bracketed environment vector in \eqref{eq:four-query-fixed-output-physical-vector} by \(|v_{\Splus,\Sminus}\rangle\).
    Compression computes its overlap with each bijective physical relation state \(|\pf_{T_+,T_-}\rangle\), where \((T_+,T_-)\in\mathcal B_{\theta_\lambda}\). 
    Using
    \begin{align}
        \langle\pi',f'|\pi,f\rangle
        =\delta_{\pi',\pi}\delta_{f',f},
    \end{align}
    this overlap becomes the scalar sum
    \begin{align}
        \langle\pf_{T_+,T_-}|v_{\Splus,\Sminus}\rangle
        &=
        \frac{1}{3^N\sqrt{N!(N-t_\lambda)!}}
        \sum_{\substack{
            \pi\in S_N\\
            \pi\supseteq \Splus\uplus \Sminus\\
            \pi\supseteq T_+\uplus T_-
        }}
        \sum_{f:[N]\to\mathbb Z_3}
        \omega_3^{
            \sum_{x\in[N]}
            (\varepsilon_{\Splus,\Sminus}(x)-\varepsilon_{T_+,T_-}(x))f(x)
        }
        \nonumber\\
        &=
        \frac{1}{3^N\sqrt{N!(N-t_\lambda)!}}
        \sum_{\substack{
            \pi\in S_N\\
            \pi\supseteq \Splus\uplus \Sminus\\
            \pi\supseteq T_+\uplus T_-
        }}
        \prod_{x\in[N]}
        \left[
            \sum_{r\in\mathbb Z_3}
            \omega_3^{
                (\varepsilon_{\Splus,\Sminus}(x)-\varepsilon_{T_+,T_-}(x))r
            }
        \right].
        \label{eq:physical-signed-relation-phase-factorization}
    \end{align}
    The factorization holds because the values \(f(x)\) range independently over \(\mathbb Z_3\).
    If there is no common permutation extension, this sum is empty.
    Otherwise, \(T_+\uplus T_-\) contains a domain \(x_0\) absent from \(\Splus\uplus \Sminus\), because it has \(t_\lambda\) distinct domains while \(\Splus\uplus \Sminus\) has fewer. At this domain, the two phase characters are \(0\) and \(\pm1\). 
    The \(x_0\)-factor in the product is therefore \( \sum_{r\in\mathbb Z_3}\omega_3^{\pm r}=0 \).
    Thus every such overlap vanishes, and the compression map sends the non-bijective contribution to zero.
\end{enumerate}
Consequently, only the bijective output tuples survive compression, and
\begin{align}
    \left(
        I_{\mathcal A_{\theta_\lambda}}
        \otimes\mathsf{Comp}_{\theta_\lambda}
    \right)
    K_{\theta_\lambda}^{\mathrm{PF}}
    \left(
        |\boldsymbol x\rangle_U
        |\boldsymbol y\rangle_T
        |\boldsymbol x'\rangle_*
        |\boldsymbol y'\rangle_\dagger
    \right)
    =
    \sqrt{\frac{(N-t_\lambda)!}{N!}}
    \sum_{\substack{
        \boldsymbol u,\boldsymbol v,\boldsymbol u',\boldsymbol v'\\
        (\Splus,\Sminus)\in\mathcal B_{\theta_\lambda}
    }}
    |\boldsymbol u\rangle_U
    |\boldsymbol v\rangle_T
    |\boldsymbol u'\rangle_*
    |\boldsymbol v'\rangle_\dagger
    |\Splus,\Sminus\rangle.
    \label{eq:compressed-four-query-physical-action}
\end{align}
Here each output string ranges over the tensor power specified above.

On the same bijective sector every edge multiplicity is one.
Projecting Equation~\eqref{eq:parallel-four-query-recorder-basis-action} therefore gives
\begin{align}
    \left(
        I_{\mathcal A_{\theta_\lambda}}
        \otimes\Pi_{\theta_\lambda}^{\mathrm{bij}}
    \right)
    E_{\theta_\lambda}
    \left(
        |\boldsymbol x\rangle_U
        |\boldsymbol y\rangle_T
        |\boldsymbol x'\rangle_*
        |\boldsymbol y'\rangle_\dagger
    \right)
    =
    \frac{1}{N^{t_\lambda/2}}
    \sum_{\substack{
        \boldsymbol u,\boldsymbol v,\boldsymbol u',\boldsymbol v'\\
        (\Splus,\Sminus)\in\mathcal B_{\theta_\lambda}
    }}
    |\boldsymbol u\rangle_U
    |\boldsymbol v\rangle_T
    |\boldsymbol u'\rangle_*
    |\boldsymbol v'\rangle_\dagger
    |\Splus,\Sminus\rangle.
    \label{eq:bijective-four-query-recorder-action}
\end{align}
The two coefficients agree because
\begin{align}
    \sqrt{
        \frac{N^{t_\lambda}(N-t_\lambda)!}{N!}
    }
    \frac{1}{N^{t_\lambda/2}}
    =
    \sqrt{\frac{(N-t_\lambda)!}{N!}}.
\end{align}
Thus multiplying \eqref{eq:bijective-four-query-recorder-action} by the factor in Lemma~\ref{lem:exact-four-query-bijective-sector-matching} gives \eqref{eq:compressed-four-query-physical-action}. 
The identity holds on every computational-basis input and hence on every input by linearity, proving \eqref{eq:exact-four-query-bijective-sector-matching}.
\end{proof}

\section{Proof of the Endpoint-Averaged Collision Bound}
\label{app:four-query-endpoint-averaged-collision-bound}

Appendix~\ref{app:physical-signed-relation-matching}, established matching up to normalization between the physical purification and the parallel recorder on the bijective database subspace.
Thus, it remains to show that the recorder’s weight outside this subspace is small. 
This appendix proves the collision bound for the ordered recorder by averaging pairwise collision projectors over the endpoint unitaries. 
The result is used in Section~\ref{sec:four-query-fixed-sector-comparison} to bound the distance between the physical and recorder dilations.

\subsection{Collision Projectors and Their Endpoint Averages}
\label{app:collision-projectors-and-twirls}

\begin{definition}[Database collision projectors]
\label{def:four-query-database-collision-projectors}
Identify symmetric databases with their subspaces in the ordered database space \((\mathcal O_{p_\lambda})_+\otimes(\mathcal O_{q_\lambda})_-\). 
For two labels, define the equality projector by
\begin{align}
    \Pi^{\mathrm{eq}}
    &:=\sum_{x\in[N]}|x,x\rangle\!\langle x,x|.
\end{align}
The normalized EPR state \(|\Omega_N\rangle\) and its projector \(\Phi_N\) are as in Definition~\ref{def:normalized-epr-state}.
An edge slot has coordinate \(1\) for its domain and coordinate \(2\) for its image. 
Let \(\Pi_{r,s}^{\mathrm{eq},(\kappa)}\) apply \(\Pi^{\mathrm{eq}}\) to coordinate \(\kappa\) of edge slots \(r,s\), and let \(\Phi_{r,s}^{(\kappa)}\) apply \(\Phi_N\) to the same coordinates. 
All unspecified registers carry the identity.

\end{definition}

\begin{lemma}[Collision union bound for non-bijective relations]
\label{lem:four-query-bijective-union-bound}
On the ordered database space,
\begin{align}
    I-\Pi_{\theta_\lambda}^{\mathrm{bij,ord}}
    &\preceq
    \sum_{\kappa=1}^2
    \sum_{1\leq r<s\leq t_\lambda}
    \Pi_{r,s}^{\mathrm{eq},(\kappa)}.
    \label{eq:four-query-bijective-union-bound}
\end{align}
\end{lemma}

\begin{proof}
All projectors in \eqref{eq:four-query-bijective-union-bound} are diagonal in the ordered edge basis. 
On a bijective basis state, both sides are zero. 
On any other state, the left-hand side is one and at least one pair of domain or image labels agrees, so the right-hand side is at least one. 
This proves the operator inequality by comparing diagonal entries, as in~\cite{MH24}.
\end{proof}

\begin{lemma}[Standard and mixed equality twirls]
\label{lem:four-query-standard-and-mixed-equality-twirls}
Let \(\nu\) be an exact unitary \(2\)-design. 
Then
\begin{align}
    \mathbb E_{V\sim\nu}
    \left[
        (V\otimes V)^\dagger
        \Pi^{\mathrm{eq}}
        (V\otimes V)
    \right]
    &=
    \frac{2}{N+1}\Pi_{\mathrm{sym}}^{N,2},
    \label{eq:four-query-standard-equality-twirl}\\
    \mathbb E_{V\sim\nu}
    \left[
        (V\otimes V^*)^\dagger
        \Pi^{\mathrm{eq}}
        (V\otimes V^*)
    \right]
    &=
    \Phi_N+
    \frac{1}{N+1}(I-\Phi_N).
    \label{eq:four-query-mixed-equality-twirl}
\end{align}
The same identities hold after applying transpose, conjugation, or adjoint to the sample.
\end{lemma}

These are the twirling identities of~\cite{MH24}.

\subsection{EPR Weight of the Ordered Recorder}
\label{app:four-query-ordered-cross-sign-epr-weight}

\begin{lemma}[Ordered cross-sign EPR weight]
\label{lem:four-query-ordered-cross-sign-epr-weight}
Let \(1\leq r\leq p_\lambda<s\leq t_\lambda\) be slots in opposite databases, and let \(\kappa\in\{1,2\}\). 
Then,
\begin{align}
    0
    &\preceq \Pi_{\theta_\lambda}^{\mathrm{nE}}
        (E_{\theta_\lambda}^{\mathrm{ord}})^\dagger
        (I\otimes\Phi_{r,s}^{(\kappa)})
        E_{\theta_\lambda}^{\mathrm{ord}}
        \Pi_{\theta_\lambda}^{\mathrm{nE}}
    \preceq \frac1{N^2}\Pi_{\theta_\lambda}^{\mathrm{nE}}.
    \label{eq:four-query-ordered-cross-sign-epr-weight}
\end{align}
\end{lemma}

\begin{proof}
Equations~\eqref{eq:ordered-left-map} and \eqref{eq:ordered-right-map} place each supplied input label in one endpoint of its edge and prepare an independent normalized EPR state between the other endpoint and the query-output register.
Thus, up to a reordering of registers, the output is therefore the tensor product of the original input state on the supplied endpoints and these independent EPR pairs.

If both selected endpoints are supplied, then \(\kappa=1\) selects a \(U,U^*\) query pair, or \(\kappa=2\) selects a \(U^T,U^\dagger\) query pair. 
The ordered recorder transfers the two input registers to these database coordinates. 
Since the input is restricted to the no-EPR subspace, applying \(\Phi_{r,s}^{(\kappa)}\) gives zero.

Next, assume at least one selected endpoint is generated, call it \(B\).
Let \(Q\) be its query-output partner, and call the other selected endpoint \(A\). 
The pair \(QB\) is in \(|\Omega_N\rangle\), independently of the remaining registers. 
For every \(x\in[N]\),
\begin{align}
    (\langle\Omega_N|_{AB}\otimes I_Q)
        \left(|x\rangle_A\otimes
            \frac1{\sqrt N}\sum_y|y\rangle_Q|y\rangle_B\right)
    = \frac1N\sum_y\delta_{x,y}|y\rangle_Q = \frac1N|x\rangle_Q.
\end{align}
If \(A\) is supplied, projecting \(AB\) onto the EPR state transfers its input state, including its entanglement with other registers, to \(Q\).
If \(A\) is generated, the projection creates an EPR pair between the two query-output partners.
In either case, the projection adds a factor of \(1/N^2\) for every normalized input.
Hence \((E_{\theta_\lambda}^{\mathrm{ord}})^\dagger(I\otimes\Phi_{r,s}^{(\kappa)})E_{\theta_\lambda}^{\mathrm{ord}}=N^{-2}I\), where the identity on the right acts on all residual query-input registers.
The two cases give respectively zero and \(N^{-2}\Pi_{\theta_\lambda}^{\mathrm{nE}}\) after restriction, proving the claim.
\end{proof}

\subsection{The Ordered Collision Bound}
\label{app:four-query-ordered-collision-bound}

\begin{proof}[Proof of \eqref{eq:four-query-endpoint-averaged-ordered-collision-bound}]
Substituting \eqref{eq:four-query-fixed-endpoint-ordered-branch} into the left-hand side of \eqref{eq:four-query-endpoint-averaged-ordered-collision-bound} and then applying Lemma~\ref{lem:four-query-bijective-union-bound} gives
\begin{align}
    &\mathbb E_{C,D\sim\nu}\!\left[
        \Pi_{\theta_\lambda}^{\mathrm{nE}}
        (W_{C,D,\lambda}^{\mathrm{ord}})^\dagger
        (I-\Pi_{\theta_\lambda}^{\mathrm{bij,ord}})
        W_{C,D,\lambda}^{\mathrm{ord}}
        \Pi_{\theta_\lambda}^{\mathrm{nE}}
    \right]
    \nonumber\\
    &\quad=
    \Pi_{\theta_\lambda}^{\mathrm{nE}}
    (E_{\theta_\lambda}^{\mathrm{ord}})^\dagger
    \Bigl[I\otimes\mathbb E_{C,D\sim\nu}\!\left[
        (\mathcal Q_{\theta_\lambda}^{\mathrm{ord}}[C,D])^\dagger
        (I-\Pi_{\theta_\lambda}^{\mathrm{bij,ord}})
        \mathcal Q_{\theta_\lambda}^{\mathrm{ord}}[C,D]
    \right]\Bigr]
    E_{\theta_\lambda}^{\mathrm{ord}}
    \Pi_{\theta_\lambda}^{\mathrm{nE}}
    \nonumber\\
    &\quad\preceq
    \sum_{\kappa=1}^2\sum_{1\leq r<s\leq t_\lambda}
    \Pi_{\theta_\lambda}^{\mathrm{nE}}
    (E_{\theta_\lambda}^{\mathrm{ord}})^\dagger
    \Bigl[I\otimes\mathbb E_{C,D\sim\nu}\!\left[
        (\mathcal Q_{\theta_\lambda}^{\mathrm{ord}}[C,D])^\dagger
        \Pi_{r,s}^{\mathrm{eq},(\kappa)}
        \mathcal Q_{\theta_\lambda}^{\mathrm{ord}}[C,D]
    \right]\Bigr]
    E_{\theta_\lambda}^{\mathrm{ord}}
    \Pi_{\theta_\lambda}^{\mathrm{nE}}.
    \label{eq:four-query-ordered-pair-collision-operator}
\end{align}
The recorder \(E_{\theta_\lambda}^{\mathrm{ord}}\) and projector \(\Pi_{\theta_\lambda}^{\mathrm{nE}}\) are independent of \(C,D\), so they have been taken outside the expectation.
It therefore suffices to bound each summand in the final line. 
The database rotations in \eqref{eq:ordered-signed-database-rotation} cancel on all unselected coordinates. 
For domain coordinates, the two selected rotations are \(C,C\), \(C^*,C^*\), or \(C,C^*\), according to the database signs. 
For image coordinates they are \(D^T,D^T\), \(D^\dagger,D^\dagger\), or \(D^T,D^\dagger\). 
Same-database pairs use the standard twirl, and opposite-database pairs use the mixed twirl.

For a same-database pair, let \(\Pi_{\mathrm{sym},r,s}^{(\kappa)}\) denote the projector \(\Pi_{\mathrm{sym}}^{N,2}\) on coordinate \(\kappa\) of slots \(r,s\), acting as the identity on the other database coordinates.
Substituting \eqref{eq:four-query-standard-equality-twirl} into the corresponding summand of \eqref{eq:four-query-ordered-pair-collision-operator} gives
\begin{align}
    &\Pi_{\theta_\lambda}^{\mathrm{nE}}
    (E_{\theta_\lambda}^{\mathrm{ord}})^\dagger
    \Bigl[I\otimes\mathbb E_{C,D\sim\nu}\!\left[
        (\mathcal Q_{\theta_\lambda}^{\mathrm{ord}}[C,D])^\dagger
        \Pi_{r,s}^{\mathrm{eq},(\kappa)}
        \mathcal Q_{\theta_\lambda}^{\mathrm{ord}}[C,D]
    \right]\Bigr]
    E_{\theta_\lambda}^{\mathrm{ord}}
    \Pi_{\theta_\lambda}^{\mathrm{nE}} \notag\\
    &\quad=
    \frac{2}{N+1}
    \Pi_{\theta_\lambda}^{\mathrm{nE}}
    (E_{\theta_\lambda}^{\mathrm{ord}})^\dagger
    (I\otimes\Pi_{\mathrm{sym},r,s}^{(\kappa)})
    E_{\theta_\lambda}^{\mathrm{ord}}
    \Pi_{\theta_\lambda}^{\mathrm{nE}} \notag\\
    &\quad\preceq
    \frac{2}{N+1}
    \Pi_{\theta_\lambda}^{\mathrm{nE}}
    (E_{\theta_\lambda}^{\mathrm{ord}})^\dagger
    E_{\theta_\lambda}^{\mathrm{ord}}
    \Pi_{\theta_\lambda}^{\mathrm{nE}}
    =\frac{2}{N+1}\Pi_{\theta_\lambda}^{\mathrm{nE}}.
\end{align}
The inequality uses \(\Pi_{\mathrm{sym},r,s}^{(\kappa)}\preceq I\), and the final equality is the ordered-isometry identity \eqref{eq:ordered-recorder-isometry}.

For an opposite-database pair, substituting \eqref{eq:four-query-mixed-equality-twirl} gives
\begin{align}
    &\Pi_{\theta_\lambda}^{\mathrm{nE}}
    (E_{\theta_\lambda}^{\mathrm{ord}})^\dagger
    \Bigl[I\otimes\mathbb E_{C,D\sim\nu}\!\left[
        (\mathcal Q_{\theta_\lambda}^{\mathrm{ord}}[C,D])^\dagger
        \Pi_{r,s}^{\mathrm{eq},(\kappa)}
        \mathcal Q_{\theta_\lambda}^{\mathrm{ord}}[C,D]
    \right]\Bigr]
    E_{\theta_\lambda}^{\mathrm{ord}}
    \Pi_{\theta_\lambda}^{\mathrm{nE}} \notag\\
    &\quad=
    \Pi_{\theta_\lambda}^{\mathrm{nE}}
    (E_{\theta_\lambda}^{\mathrm{ord}})^\dagger
    \left[I\otimes\left(
        \Phi_{r,s}^{(\kappa)}
        +\frac1{N+1}(I-\Phi_{r,s}^{(\kappa)})
    \right)\right]
    E_{\theta_\lambda}^{\mathrm{ord}}
    \Pi_{\theta_\lambda}^{\mathrm{nE}} \notag\\
    &\quad=
    \frac1{N+1}\Pi_{\theta_\lambda}^{\mathrm{nE}}
    +\frac{N}{N+1}
    \Pi_{\theta_\lambda}^{\mathrm{nE}}
    (E_{\theta_\lambda}^{\mathrm{ord}})^\dagger
    (I\otimes\Phi_{r,s}^{(\kappa)})
    E_{\theta_\lambda}^{\mathrm{ord}}
    \Pi_{\theta_\lambda}^{\mathrm{nE}} \notag\\
    &\quad\preceq
    \left(\frac1{N+1}+\frac{N}{N+1}\frac1{N^2}\right)
    \Pi_{\theta_\lambda}^{\mathrm{nE}}
    =\frac1N\Pi_{\theta_\lambda}^{\mathrm{nE}}.
\end{align}
The second equality uses \eqref{eq:ordered-recorder-isometry}, and the inequality substitutes \eqref{eq:four-query-ordered-cross-sign-epr-weight} from Lemma~\ref{lem:four-query-ordered-cross-sign-epr-weight}.
There are two coordinates per pair, \(\binom{p_\lambda}{2}+\binom{q_\lambda}{2}\) same-database pairs, and \(p_\lambda q_\lambda\) opposite-database pairs.
Consequently,
\begin{align}
    \mathbb E_{C,D\sim\nu}\!\left[
            \Pi_{\theta_\lambda}^{\mathrm{nE}}
            (W_{C,D,\lambda}^{\mathrm{ord}})^\dagger
            (I-\Pi_{\theta_\lambda}^{\mathrm{bij,ord}})
            W_{C,D,\lambda}^{\mathrm{ord}}
            \Pi_{\theta_\lambda}^{\mathrm{nE}}
        \right]
    &\preceq \left(
            \frac{2p_\lambda(p_\lambda-1)+2q_\lambda(q_\lambda-1)}{N+1}
            +\frac{2p_\lambda q_\lambda}{N}
        \right)\Pi_{\theta_\lambda}^{\mathrm{nE}}
    \nonumber\\
    &\preceq \frac{2t_\lambda^2}{N}
            \Pi_{\theta_\lambda}^{\mathrm{nE}}.
    \label{eq:four-query-ordered-collision-pair-count}
\end{align}
The last inequality uses \(N+1\geq N\) and \(p_\lambda(p_\lambda-1)+q_\lambda(q_\lambda-1)+p_\lambda q_\lambda \leq(p_\lambda+q_\lambda)^2=t_\lambda^2\).
\end{proof}

\section{Proof of Bijective Matching with Fixed Endpoint Unitaries}
\label{app:four-query-projected-fixed-endpoint-matching}

\begin{proof}[Proof of Lemma~\ref{lem:four-query-projected-fixed-endpoint-matching}]
Substitute \eqref{eq:four-query-fixed-endpoint-physical-branch}.
The database projector commutes with the output endpoint unitaries, and \eqref{eq:four-query-common-environment-bijective-component} replaces the projected physical isometry by the projected recorder.
Explicitly,
\begin{align}
    &(I\otimes\Pi_{\theta_\lambda}^{\mathrm{bij}})V_{C,D,\lambda} \notag
    \\
    &\quad=
    \Bigl[
        \left(
            D^{\otimes a_\lambda}
            \otimes(C^T)^{\otimes b_\lambda}
            \otimes(D^*)^{\otimes c_\lambda}
            \otimes(C^\dagger)^{\otimes d_\lambda}
        \right)\otimes I
    \Bigr]
    (I\otimes\Pi_{\theta_\lambda}^{\mathrm{bij}})
    \widetilde K_{\theta_\lambda}^{\mathrm{PF}} \notag
    \\
    &\qquad\cdot
    \left(
        C^{\otimes a_\lambda}
        \otimes(D^T)^{\otimes b_\lambda}
        \otimes(C^*)^{\otimes c_\lambda}
        \otimes(D^\dagger)^{\otimes d_\lambda}
    \right) \notag
    \\
    &\quad=
    \alpha_{t_\lambda}(I\otimes\Pi_{\theta_\lambda}^{\mathrm{bij}})
    \Bigl[
        \left(
            D^{\otimes a_\lambda}
            \otimes(C^T)^{\otimes b_\lambda}
            \otimes(D^*)^{\otimes c_\lambda}
            \otimes(C^\dagger)^{\otimes d_\lambda}
        \right)\otimes I
    \Bigr]E_{\theta_\lambda} \notag
    \\
    &\qquad\cdot
    \left(
        C^{\otimes a_\lambda}
        \otimes(D^T)^{\otimes b_\lambda}
        \otimes(C^*)^{\otimes c_\lambda}
        \otimes(D^\dagger)^{\otimes d_\lambda}
    \right) \notag
    \\
    &\quad=
    \alpha_{t_\lambda}(I\otimes\Pi_{\theta_\lambda}^{\mathrm{bij}})
    (I\otimes\mathcal Q_{\theta_\lambda}[C,D])E_{\theta_\lambda}
    =\alpha_{t_\lambda}(I\otimes\Pi_{\theta_\lambda}^{\mathrm{bij}})
      W_{C,D,\lambda}.
\end{align}
The third equality is the unitary invariance identity \eqref{eq:four-type-endpoint-covariance}, the last is \eqref{eq:four-query-fixed-endpoint-recorder-branch}.
This proves \eqref{eq:four-query-fixed-endpoint-bijective-matching}.
\end{proof}

\section{Proofs for the Adaptive Birthday-Scale Attack}
\label{app:cba-proofs}

This appendix supplies the uniform Haar bound and the detailed perfect-witness search construction used in Section~\ref{sec:cba-attack}.

\subsection{Uniform Haar Acceptance Bound}
\label{app:cba-haar}

\begin{proof}[Proof of Lemma~\ref{lem:cba-haar}]
First, bound the coefficient of an arbitrary Pauli in the expansion of \(UQU^\dagger\). 
The proof first shows that each coefficient has mean zero, then uses concentration and a union bound to bound all squared coefficients simultaneously with high probability. 
These bounds will then imply that two Paulis sampled from the expansion commute with probability close to \(1/2\), as required to bound \(q_Q(U)\) in \eqref{eq:cba-verifier-probability}. 

Fix a nonidentity Pauli representative \(Q\) and a nonzero label \(a\in\mathbb F_2^{2n}\), and consider the coefficient \(r_a(U,Q)\) defined in \eqref{eq:cba-expansion}.
The coefficient is real because \(P_a\) and \(UQU^\dagger\) are Hermitian. 
Haar invariance implies that \(\mathbb E_U[UQU^\dagger]\) commutes with every unitary, so it is a scalar multiple of the identity. 
Its trace equals \(\operatorname{Tr}(Q)=0\), since \(Q\) is a nonidentity Pauli, so \(\mathbb E_U[UQU^\dagger]=0\). 
By linearity,
\begin{align}
    \mathbb E_U[r_a(U,Q)]
       =\frac1N\operatorname{Tr}\!\left(
           P_a\,\mathbb E_U[UQU^\dagger]\right)=0,
\end{align}
as mentioned.

To bound deviations from this average, the following concentration inequality from~\cite{MM13} applies.
If a real function \(f\) on \(\mathrm U(N)\) is \(L\)-Lipschitz for the unnormalized Hilbert-Schmidt metric \(\|U-V\|_2\), then
\begin{align}
    \Pr[f(U)\geq\mathbb Ef(U)+s]
       \leq \exp\!\left(-\frac{Ns^2}{12L^2}\right)
       \quad(s>0).
    \label{eq:cba-haar-concentration}
\end{align}

For the coefficient function \(U\mapsto r_a(U,Q)\), use the Hilbert-Schmidt Cauchy-Schwarz inequality and invariance of that norm under unitary multiplication,
\begin{align}
    |r_a(U,Q)-r_a(V,Q)|
       \leq\frac{\|P_a\|_2}{N}
              \|UQU^\dagger-VQV^\dagger\|_2 \leq\frac2{\sqrt N}\|U-V\|_2.
\end{align}

Thus, \(r_a(\cdot,Q)\) is \(2/\sqrt N\)-Lipschitz with respect to the Hilbert-Schmidt metric.
Applying \eqref{eq:cba-haar-concentration} to \(r_a(\cdot,Q)\) and \(-r_a(\cdot,Q)\), both of which have mean zero and the same Lipschitz constant, bounds the two tails. 
Adding these bounds gives
\begin{align}
    \Pr[|r_a(U,Q)|\geq s]
       \leq2\exp(-N^2s^2/48).
\end{align}
Taking \(s^2=48\ln(2N^4/\delta)/N^2\) and a union bound over fewer than \(N^4\) choices of nonidentity \(Q\) and nonzero label \(a\) shows that, except with probability \(\delta\), the bound
\begin{align}
    r_a(U,Q)^2
       \leq\frac{48}{N^2}\ln\frac{2N^4}{\delta}
    \label{eq:cba-transfer-bound}
\end{align}
holds simultaneously for every such \(Q\) and \(a\).

The remaining goal is to bound the test acceptance probability \(q_Q(U)\), which is the probability that two independently sampled Paulis commute.
Consider an event in \eqref{eq:cba-transfer-bound}, which has probability at least \(1-\delta\) over Haar-random \(U\).
Fix \(U\) on this event and any \(Q\ne I\).
The following calculation shows that the small sampling weights force this deviation from \(1/2\) in Equation~\eqref{eq:cba-verifier-probability} to be small.

For \(a,b\in\mathbb F_2^{2n}\) the Cauchy-Schwarz inequality yields,
\begin{align}
    \begin{aligned}
    \left|\sum_{a,b}(-1)^{[a,b]}r_a(U,Q)^2r_b(U,Q)^2\right|
       &\leq\left(\sum_a r_a(U,Q)^4\right)^{1/2}
              \left[\sum_a\left(\sum_b(-1)^{[a,b]}r_b(U,Q)^2\right)^2
              \right]^{1/2}\\
        &=\left(\sum_a r_a(U,Q)^4\right)^{1/2}
        \left[\sum_a\sum_{b,c} (-1)^{[a,b]+[a,c]} r_b(U,Q)^2r_c(U,Q)^2
              \right]^{1/2} \\
     &=\left(\sum_a r_a(U,Q)^4\right)^{1/2}
              \left[\sum_{b,c}r_b(U,Q)^2r_c(U,Q)^2\sum_a(-1)^{[a,b+c]}
              \right]^{1/2} \\
    &=\left(\sum_a r_a(U,Q)^4\right)^{1/2}
              \left[\sum_{b,c}r_b(U,Q)^2r_c(U,Q)^2 \cdot N^2 \cdot \delta_{b,c}
              \right]^{1/2} \\
       &=N\sum_a r_a(U,Q)^4
       \leq N\max_a r_a(U,Q)^2\sum_a r_a(U,Q)^2\\
       &\leq\frac{48}{N}\ln\frac{2N^4}{\delta}.
    \end{aligned}
    \label{eq:cba-sign-bound}
\end{align}

Combining \eqref{eq:cba-verifier-probability} with this bound gives \(q_Q(U)\leq h_N(\delta)\), as required.
\end{proof}

\subsection{Perfect-Witness Search}
\label{app:cba-search}

The algorithm follows the recursive Grover-style construction of \cite{HMW03}.
Their search guarantee assumes that every candidate has acceptance probability either at most \(1/10\) or at least \(9/10\). 
If a fraction at least \(p>0\) of the candidates has acceptance probability at least \(9/10\), their algorithm can find such a candidate with constant success probability using \(O(p^{-1/2})\) expected verifier calls.

In the application to distinguishing DPFC from Haar, the idea is to choose \(h_N(\delta)<\tau<1\), where \(h_N(\delta)\) is the parameter from Lemma~\ref{lem:cba-haar}.
For Haar-random \(U\), Lemma~\ref{lem:cba-haar} guarantees that except with probability \(\delta\), all candidate acceptance probabilities are at most \(h_N(\delta)\).
For DPFC, even though perfect witnesses exist, candidates with \(\tau<q_i<1\) are also acceptable outputs: fresh trials of the four-query commutation test distinguishes them from Haar candidates using the gap \(\tau-h_N(\delta)\).

\begin{proof}[Proof of Lemma~\ref{lem:cba-search}]
Set
\begin{align}
    J:=\left\lceil\log_3(p_{\mathrm{witness}}^{-1/2})\right\rceil.
\end{align}
The circuits \(A_0,\ldots,A_J\) describe the algorithm steps.
Let \(R_j\) denote all registers used by \(A_j\), including its candidate register, workspaces, and stage success flag.
The circuit \(A_{j+1}\) uses a copy of \(R_j\), together with \(m+j+1\) new verifier workspaces and a new stage success flag.
Each stage is executed on a separate, initially zero registers. 
calls to \(A_j\) within \(A_{j+1}\) act on the copy of \(R_j\) belonging to the current stage.

\par\medskip
\noindent\begin{minipage}{\linewidth}
\hrule\smallskip
\refstepcounter{algorithm}\label{alg:cba-search}
\noindent\textbf{Algorithm~\thealgorithm.} Perfect-witness search
\par\smallskip\hrule\smallskip
\noindent\textbf{Input:} Access to \(W,W^\dagger\), a unitary preparing \(\sum_i\sqrt{w_i}|i\rangle\) and its inverse, a threshold \(0<\tau<1\), and \(J\) as defined above.
In addition, choose an integer \(m\geq1\), such that \(\tau^m/(1-\tau)<1-e^{-1/8}\)
\begin{enumerate}
    \item Apply \(A_0\) on registers \(R_0\): Prepare the candidate register in \(\sum_i\sqrt{w_i}|i\rangle\) and initialize \(m\) separate verifier workspaces to zero.
    For each workspace, apply \(W\) to the candidate register together with that workspace.
    Reversibly compute a stage success flag, initially zero, that equals one if and only if all \(m\) acceptance flags equal one. 
    Do not measure any of the registers.
    \item For \(j=0,\ldots,J-1\), apply \(A_{j+1}\) on a separate, initially zero registers denoted by \(R_{j+1}\), as follows:
    \begin{enumerate}
        \item \textbf{Amplification.} On an \(R_j\) portion of the current stage's registers, apply \(A_j\), then \(Z\) to its success flag, followed by \(A_j^\dagger\), then the reflection \(2|0\rangle\langle0|_{R_j}-I_{R_j}\), and then \(A_j\).
        \item \textbf{Verification.} On \(m+j+1\) new all-zero workspaces, apply \(W\) to the candidate register within \(R_j\) together with that workspace.
        Reversibly compute the new stage success flag, that equals one if and only if the success flag within \(R_j\) and all \(m+j+1\) new acceptance flags equal one. 
    \end{enumerate}
    \item \textbf{Output.} Measure the success flags of all stages and return the candidate label from the first successful stage.
    If no stage succeeds, return \(\perp\).
\end{enumerate}
\hrule
\end{minipage}
\par\medskip

For the analysis, denote
\begin{align*}
    p_j=\Pr[A_j\text{ succeeds}],\qquad
    g_j=\Pr[A_j\text{ succeeds with a candidate in }S].
\end{align*}

The construction taken from \cite{HMW03} recursively combines amplitude-amplification step with verifier trials whose number grows with the stage index. 
Then, they use majority voting on the acceptance flags of the individual trials. 
Here, perfect witnesses always accept, so requiring all additional trials to accept suppresses candidates with \(q_i\leq\tau\) without rejecting any candidate in \(S\).\footnote{Ordinary majority voting would not suppress candidates with \(1/2<q_i\leq\tau\) when \(\tau>1/2\). Requiring an acceptance rate above \((1+\tau)/2\) would work, but would require a concentration bound on the probability that the observed acceptance rate exceeds this threshold when \(q_i\leq\tau\). Let \(r\) be the number of trials. Since \(q_i=1\) for \(i\in S\), requiring all \(r\) trials to accept preserves these candidates exactly and gives the simpler acceptance probability \(q_i^r\leq\tau^r\) for \(q_i\leq\tau\).}

The amplification step itself is standard Grover amplitude amplification.
As explained in \cite{HMW03} as well, at every step \(A_j\), every amplitude of a basis state whose stage success flag equals one is multiplied by \(3-4p_j\). 
Every candidate in \(S\) passes all additional verifier trials with certainty. 
Therefore the probability of success with a candidate in \(S\) is unchanged by those trials. 
Applying the amplification identity to this part of the accepting component gives
\begin{align}
    g_0=\sum_{i\in S}w_i\geq p_{\mathrm{witness}},
    \qquad
    g_{j+1}=(3-4p_j)^2g_j.
    \label{eq:cba-simplified-perfect-recurrence}
\end{align}

The remaining argument differs from the completeness analysis in \cite{HMW03}.
Their analysis tracks the successive amplification steps by bounding the probabilities of success separately for candidates with acceptance probability at least \(9/10\) and those with acceptance probability at most \(1/10\). 
Here, candidates outside \(S\) do not necessarily have low acceptance probability, and candidates with \(q_i>\tau\) are acceptable outputs even when they are outside \(S\).
Since the stages are executed independently, a lower bound on their sum gives a lower bound on the probability that at least one stage's success flag measures one, as shown below.
The argument shows that if \(\sum_{j=0}^{J}p_j\) were too small, the recurrence would force \(g_J\) to exceed \(p_J\), which is impossible.
After bounding the probability that some stage succeeds, it remains to exclude outputs with \(q_i\leq\tau\).

Fix an analysis parameter \(0<\eta\leq1/4\). 
The claim is that
\begin{align}
    \sum_{j=0}^{J}p_j\geq\eta.
    \label{eq:cba-simplified-total-weight}
\end{align}
Suppose instead that this sum is less than \(\eta\).
\begin{align}
    \begin{aligned}
        g_J
        &=9^Jg_0\left[\prod_{j=0}^{J-1}
            \left(1-\frac43p_j\right)\right]^2 \geq 
            9^Jp_{\mathrm{witness}}
            \left(1-\frac43\sum_{j=0}^{J-1}p_j\right)^2
        \geq\left(1-\frac43\eta\right)^2
        >\eta.
    \end{aligned}
\end{align}
The first inequality uses \(g_0\geq p_{\mathrm{witness}}\) and \(\prod_j(1-x_j)\geq1-\sum_jx_j\) for \(x_j\in[0,1]\).
The second inequality uses \(9^Jp_{\mathrm{witness}}\geq1\) and \(\sum_{j=0}^{J-1}p_j<\eta\), and the last follows from \(0<\eta\leq1/4\).
But \(g_J\leq p_J\leq\sum_{j=0}^{J}p_j<\eta\), which is a contradiction.

For a fixed verifier \(W\), the separate stage executions act on disjoint, initially zero register collections, so their success events are independent. 
Hence
\begin{align}
    \Pr[\text{some stage succeeds}]
    =1-\prod_{j=0}^{J}(1-p_j)
    \geq1-\exp\!\left(-\sum_{j=0}^{J}p_j\right)
    \geq1-e^{-\eta}.
\end{align}
Whenever some stage succeeds, the algorithm returns a candidate.
For any candidate \(i\) with \(q_i\leq\tau\), the \(m+j\) new trials at the end of stage \(j\) all accept with probability \(q_i^{m+j}\leq\tau^{m+j}\).
A union bound therefore gives
\begin{align}
    \Pr[\text{return an }i\text{ with }q_i\leq\tau]
    \leq\sum_{j=0}^{J}\tau^{m+j}
    \leq\frac{\tau^m}{1-\tau}.
    \label{eq:cba-simplified-low-output}
\end{align}
Subtracting \eqref{eq:cba-simplified-low-output} therefore gives
\begin{align}
    \Pr[\text{return an }i\text{ with }q_i>\tau]
    \geq1-e^{-\eta}-\frac{\tau^m}{1-\tau}.
    \label{eq:cba-search-success}
\end{align}
Taking \(\eta=1/8\) gives the success bound in the lemma.

It remains to calculate the query complexity. 
The count follows the analysis used in \cite{HMW03} and provided here for completeness.
Let \(T_j\) count verifier calls, meaning applications of \(W\) or \(W^\dagger\), in \(A_j\).
The initial stage uses \(m\) verifier calls.
Constructing \(A_{j+1}\) uses \(A_j\) twice and \(A_j^\dagger\) once for amplification, followed by \(m+j+1\) new verifier calls. 
Thus,
\begin{align}
    T_0=m,\qquad T_{j+1}=3T_j+m+j+1.
\end{align}
Solving this recurrence by induction gives
\begin{align}
    T_j=3^j\sum_{k=0}^{j}\frac{m+k}{3^k}
    \leq3^j\sum_{k=0}^{\infty}\frac{m+k}{3^k}
    =3^j\frac{6m+3}{4}.
\end{align}
The algorithm executes each stage once on separate registers.
Using \(3^J<3/\sqrt{p_{\mathrm{witness}}}\), its total number of verifier calls is therefore bounded by
\begin{align}
    \sum_{j=0}^{J}T_j
    &\leq\sum_{j=0}^{J}3^j\frac{6m+3}{4}
        =\frac{3(2m+1)}8\left(3^{J+1}-1\right)
    <\frac{27(2m+1)}{8\sqrt{p_{\mathrm{witness}}}}
        =O(p_{\mathrm{witness}}^{-1/2}).
        \label{eq:cba-search-cost}
\end{align}
For fixed \(\tau\), the stated choice of \(m\) depends only on \(\tau\), so \eqref{eq:cba-search-success} gives constant success probability and \eqref{eq:cba-search-cost} gives the claimed query bound with \(C_\tau=27(2m+1)/8\).
\end{proof}

\bibliographystyle{alpha}
\bibliography{references}

\end{document}